\documentclass[
  journal=pasa,
  manuscript=research-paper, 
  year=2026,
  volume=??,
]{cup-journal}

\usepackage{microtype,siunitx,booktabs}
\usepackage{natbib}
\setcitestyle{aysep={}}

\usepackage{CJKutf8}    
\usepackage{tikz,xcolor,hyperref} 
\usepackage{amsmath}    
\usepackage{amssymb}    
\usepackage{longtable}
\usepackage{multicol}
\usepackage{float}
\usepackage{booktabs}
\usepackage{siunitx}
\usepackage{array}

\newcolumntype{H}{>{\setbox0=\hbox\bgroup}c<{\egroup}@{}}

\definecolor{lime}{HTML}{A6CE39}
\DeclareRobustCommand{\orcidicon}{%
    \begin{tikzpicture}
    \draw[lime, fill=lime] (0,0) 
    circle [radius=0.16] 
    node[white] {{\fontfamily{qag}\selectfont \tiny ID}};
    \draw[white, fill=white] (-0.0625,0.095) 
    circle [radius=0.007];
    \end{tikzpicture}
    \hspace{-1mm}
}

\newcommand{\orcidChrisO}{\href{https://orcid.org/0000-0003-0017-349X}{\orcidicon}}
\newcommand{\orcidChrisW}{\href{https://orcid.org/0000-0002-4569-016X}{\orcidicon}}
\newcommand{\orcidSamuel}{\href{https://orcid.org/0000-0001-9372-4611}{\orcidicon}}
\newcommand{\orcidPatrick}{\href{https://orcid.org/0000-0003-4237-0520}{\orcidicon}}
\newcommand{\orcidJack}{\href{https://orcid.org/0000-0001-8359-2328}{\orcidicon}}
\newcommand{\orcidRachel}{\href{https://orcid.org/0000-0002-5325-2709}{\orcidicon}}

\newcommand\ion[2]{\text{#1\,\textsc{\lowercase{#2}}}}	

\title{AllBRICQS: the All-sky BRIght, Complete Quasar Survey --- Quasar Selection and Final Southern $G_{\rm RP}<16$ Sample}

\author{Christian Wolf~\orcidChrisW}
\affiliation{
Research School of Astronomy and Astrophysics, Australian National University, Canberra ACT 2611, Australia}
\alsoaffiliation{
Centre for Gravitational Astrophysics, Australian National University, Canberra ACT 2600, Australia}
\email[Christian Wolf]{christian.wolf@anu.edu.au}

\author{~{Christopher A.\ Onken}~\orcidChrisO}
\affiliation{
Research School of Astronomy and Astrophysics, Australian National University, Canberra ACT 2611, Australia}

\author{~Wei Jeat Hon~\orcidJack}
\affiliation{
Research School of Astronomy and Astrophysics, Australian National University, Canberra ACT 2611, Australia}
\alsoaffiliation{ 
School of Physics, University of Melbourne, Parkville, VIC 3010, Australia}

\author{~Samuel Lai~\begin{CJK}{UTF8}{gbsn}(赖民希)\end{CJK}~\orcidSamuel}
\affiliation{ 
Research School of Astronomy and Astrophysics, Australian National University, Canberra ACT 2611, Australia}

\author{~Patrick Tisserand~\orcidPatrick}
\affiliation{ 
Sorbonne Universit\'{e}, CNRS, UMR 7095, Institut d'Astrophysique de Paris, 98 bis bd Arago, F-75014 Paris, France}

\author{~Rachel Webster~\orcidRachel}
\affiliation{
School of Physics, University of Melbourne, Parkville, VIC 3010, Australia}

\doi{}

\received {dd Mmm YYYY}
\revised  {dd Mmm YYYY}
\accepted {dd Mmm YYYY}
\published{dd Mmm YYYY}

\keywords{galaxies: active; galaxies: Seyfert; quasars: general; surveys} 

\begin{document}

\begin{abstract}
We describe the final quasar sample from the Southern part of the All-sky BRIght, Complete Quasar Survey (AllBRICQS), which aimed to discover the remaining optically bright quasars using data from Gaia and WISE. We discuss our final selection and present a further 119 quasars in the redshift range $z=[0.03;4.35]$. This brings the sample of spectroscopically confirmed quasars with $G_{\rm RP}<16$ and $z>0.1$ at Dec$<+20^{\circ}$ and Galactic latitude $|b|>10^{\circ}$ to 745 objects. The spectra from the ANU 2.3m telescope at Siding Spring Observatory cover a wavelength range from 360 to 950~nm with $R\sim3000$. For redshifts of $z>0.12$, we expect the sample to be complete at $G_{\rm RP}<16$, a regime that is usually avoided by large spectroscopic surveys due to detector saturation. We note that at least one strongly lensed quasar has escaped our selection. 
\end{abstract}

\section{Introduction}

Quasars are supermassive black holes (SMBH) that typically appear brighter than their host galaxies due to heat released during viscous accretion of matter. While their brightness makes them easy to observe, they often hide in plain sight while they are not distinguished from the vastly more numerous Milky Way stars or unresolved galaxies in the sky. In principle, quasars allow us to tackle several questions of cosmological significance, including: 
\begin{enumerate}
    \item What is the origin of the cosmic SMBH population in the early Universe and its later growth throughout the cosmic epochs \citep{Ina20}? 
    \item How does accretion work at the highest rates, which are estimated to range up to a solar mass per day in the most extreme case, J0529$-$4351 \citep{Wolf24}?
    \item How does the energy released by accretion affect their host galaxies given it exceeds that of starlight and emanates at high concentration in the galaxies' cores  \citep{Fabian12}?
    \item And what can we learn about cosmic gas, heavy elements and the expansion of the Universe by observing absorption signatures of gas clouds along the line-of-sight to bright quasars \citep{Smith96,Rauch98,2008MNRAS.386.1192L}? 
\end{enumerate}

Since the first quasar, 3C~273 at redshift $z=0.157$, was discovered by \citet{1963Natur.197.1040S}, over a million further quasars have been identified \citep[e.g.][]{2020ApJS..249...17R,Flesch23,DESIQSO23} and even larger lists of quasar candidates have been proposed \citep[e.g. "Quaia";][]{Quaia24}.
However, there is particular interest in constructing complete as well as pure samples of quasars, for two reasons:
\begin{enumerate}
    \item They describe the intrinsic population and its evolution rather then merely showing our selection functions and their structure with redshift and cosmic epoch \citep[e.g.][]{W03,PDvar16}.
    \item They enable cosmological applications, e.g., maps of large-scale structure, and act as reference frames against which to measure peculiar motion of our own Milky Way Galaxy \citep[e.g.][]{Ross20eBOSS,Dipole24}.
\end{enumerate}

The distribution of quasar brightness is so steep that any apparent magnitude bin of 1~mag width contains more quasars than exist at all brighter magnitudes combined, at least in the well-studied range at optical ($B$, $r$ or $i$) magnitudes $<19$. The brightest quasars are of particular interest
because:
\begin{enumerate}
    \item they are the easiest to observe in detail and provide the best signal for most types of further analysis;
    \item they tend to point to the most massive SMBH and to the physically largest accretion discs with the best chance of future spatially resolved measurements;
    \item they resemble the most extreme accretion events in the Universe, which may be the hardest to sustain and explain.
\end{enumerate}

At the top of the apparent magnitude scale is 3C~273, which, owing to its brightness, was the first quasar to have its broad-emission line region (BLR) mapped by "BLR astrometry" with the GRAVITY instrument \citep{Gravity17} at the Very Large Telescope Interferometer (VLTI) of the European Southern Observatory (ESO) on Paranal in Chile: the results appear consistent with a neatly rotating disc of gas clouds orbiting an SMBH with 200 to 500 million solar masses \citep{2018Natur.563..657G}, consistent with a range of previous SMBH mass estimates \citep[e.g.][]{Malkan83}. 

At the top of the intrinsic luminosity scale is J0529$-$4351, whose estimated accretion rate would exceed one billion solar masses in 3~Myr if sustained \citep{Wolf24}. This was the first target studied with the upgraded GRAVITY+ instrument after Laser Guide Star Adaptive Optics \citep[LGS-AO;][]{Gravity26LGSAO} became available at the VLTI: these revealed the currently largest accretion disc in the Universe on both a physical scale and an angular scale, with a BLR radius of $\sim$1~pc or $\sim$150 micro-arcseconds ($\mu$as). The observations are most consistent with massive outflows from a large disc that orbits an SMBH of $\sim$0.8 billion solar masses \citep{Gravity26}, while canonical virial mass estimates, ignorant of biases from outflows, had initially placed the mass around 15-20 billion solar masses ($\sim$20$\times$ larger) and reveal interesting challenges for the application of virial methods in studies of cosmic SMBH evolution.

Methods for finding quasars across wide regions of sky have come a long way since the \citet{1963Natur.197.1040S} discovery of 3C~273 led to the first systematic selection of conspicuously blue star-like point sources associated with strong radio sources. They involved mostly selecting sources with star-like morphology but optical colours that are unlike stars \citep[e.g.][]{WHO91,W03,2005MNRAS.360..839R}. A major advance was enabled by the all-sky mid-infrared survey from the Wide-Field Infrared Survey Explorer (WISE), given that the red thermal emission from warm dusty tori helps quasars stand out from the blue Rayleigh-Jeans tails of hot stellar atmospheres \citep[e.g.][]{2010AJ....140.1868W,Stern12,Wu12}. 

More impactful was the advent of high-quality parallax and proper motion (PPM) information from the ESA Gaia mission: this enabled rejecting a vastly larger fraction of Galactic stars from candidate samples and pushing the boundaries of colour criteria further towards a now vastly depopulated stellar locus, thus making them more inclusive of the variety of quasar colours. The publication of Gaia Data Release~2 \citep[DR2;][]{2018A&A...616A...1G} led to the discovery of the then most luminous known quasar, J2157$-$3602, at redshift $z=4.7$ \citep{2018PASA...35...24W}, within 48 hours of the release. Several groups soon employed the Gaia PPM information as well as initially colour information from the SkyMapper Southern Survey \citep[SMSS;][]{2018PASA...35...10W} for wide-area quasar searches, especially in the historically less well studied Southern hemisphere. This included the QUBRICS survey that aimed to build a quasar sample at $z>2.5$ for the Sandage test \citep{2020ApJS..250...26B} and the higher-redshift SMSS survey to build a complete $z\sim 5$ sample \citep{2020MNRAS.491.1970W, 2022MNRAS.511..572O}. Further improvements to the quality of both WISE data \citep[with catWISE2020;][]{catWISE} and Gaia data \citep[DR3;][]{2021A&A...649A...1G} have since sharpened our ability to distinguish quasars from other sources, enabling us to perform the most complete all-sky selection yet. 

Surprisingly, the most luminous quasar of the last 9~Gyr of cosmic history was discovered without WISE or Gaia data: J1144$-$4309 is only 1~mag fainter in $i$-band than 3C~273 but resides at much higher redshift, $z=0.835$ \citep{On22}. While the object has been recorded on photographic plates since 1901, it remained unidentified six decades longer than 3C~273, until it was serendipitously found during a survey for Galactic symbiotic binaries \cite{Lucy25} in SMSS Data Release 1 \citep[DR1;][]{2018PASA...35...10W}.

Ironically, finding quasar needles in the haystack of stars seemed often easier at faint magnitudes where the quasars are relatively more common. Hence, the primary source for bright Southern quasars has long been the Hamburg-ESO Survey \citep{2000A&A...358...77W}, which used an objective prism on a wide-field Schmidt telescope. However, today the quality of colour and PPM data is so much better that quasar selection at bright magnitudes has become virtually trivial, so much that popular statistical methods such as Bayesian or machine learning algorithms are not required at redshifts of $z>0.1$ as we will emphasize in this paper. The flip side of looking specifically at the bright side of the quasar population is that these most outstanding specimens in brightness exceed the saturation limits ($\sim$16 mag) of the most massive spectroscopic surveys undertaken to collect systematic data for many objects, e.g., the Sloan Digital Sky Survey \citep[SDSS;][]{2000AJ....120.1579Y,2020ApJS..249...17R} that has produced multiple quasar catalogues already, including most recently substantial follow-up of X-ray sources from eROSITA \citep{2021A&A...647A...1P} in the Southern hemisphere \citep{sdss_dr20_2026}. 
This 
led to the All-sky BRIght, Complete Quasar Survey (AllBRICQS), 
with a first sample published by \citet{On23} covering Southern declinations of Dec $<+20 \deg$ and 
\citet{Choi25} covering 
Dec $>+20\deg$. The Southern effort was significantly larger and the initial focus of AllBRICQS, given that large surveys of the past, such as SDSS, had focussed on the Northern hemisphere, while many unique astronomical observatories and capabilities, such as ESO with its powerful VLTI, ALMA, and the new Vera C.~Rubin Observatory reside in the Southern hemisphere.

Here, we present the final 
Southern sample, where follow-up is now complete at $G_{\rm RP}<16$ and $|b|>10\deg$. In this project, we also observed previously known quasars in the selection range that do not have good-quality spectra from systematic surveys in the literature. As a final product, we will produce a catalogue and a digital spectral atlas. Some objects beyond our primary selection cuts are included as well. A complementary paper that focuses on quasars of lower luminosity, a.k.a. Seyfert-1 galaxies, in the nearby Universe at $z<0.1$ is presented by Amrutha et al. (2026b). We note for context that quasars and Seyfert-1 galaxies together are also known as type-1 AGN (Active Galactic Nuclei) in the unification scheme of \citet{An93} and \citet{UP95}, and are two traditional labels for the same physical phenomenon. 

The structure of the paper is as follws: in Section~2 we present properties of known quasars, our selection method and its likely completeness. Section~3 discusses our observations and Section~4 the properties of the final sample. Section~5 closes with a summary. Gaia magnitudes are in the AB system, while WISE magnitudes are in the Vega system. Throughout the paper, we use a flat $\Lambda$CDM cosmology with $\Omega_{\rm m}=0.3$ and a Hubble-Lema\^itre constant of $H_0=70$km~s$^{-1}$~Mpc$^{-1}$.

\section{Quasars and their selection}

\begin{figure*}
    \centering
    \includegraphics[width=0.48\columnwidth]{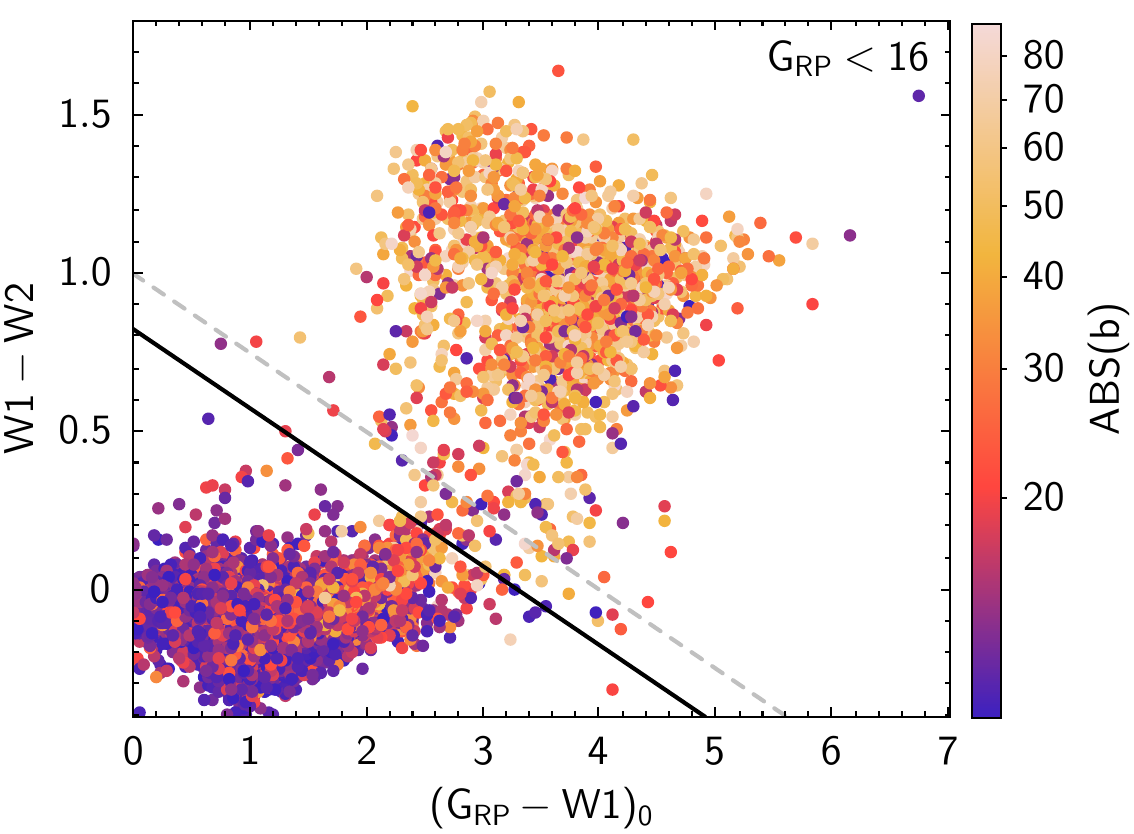} \hspace{5mm}
    \includegraphics[width=0.48\columnwidth]{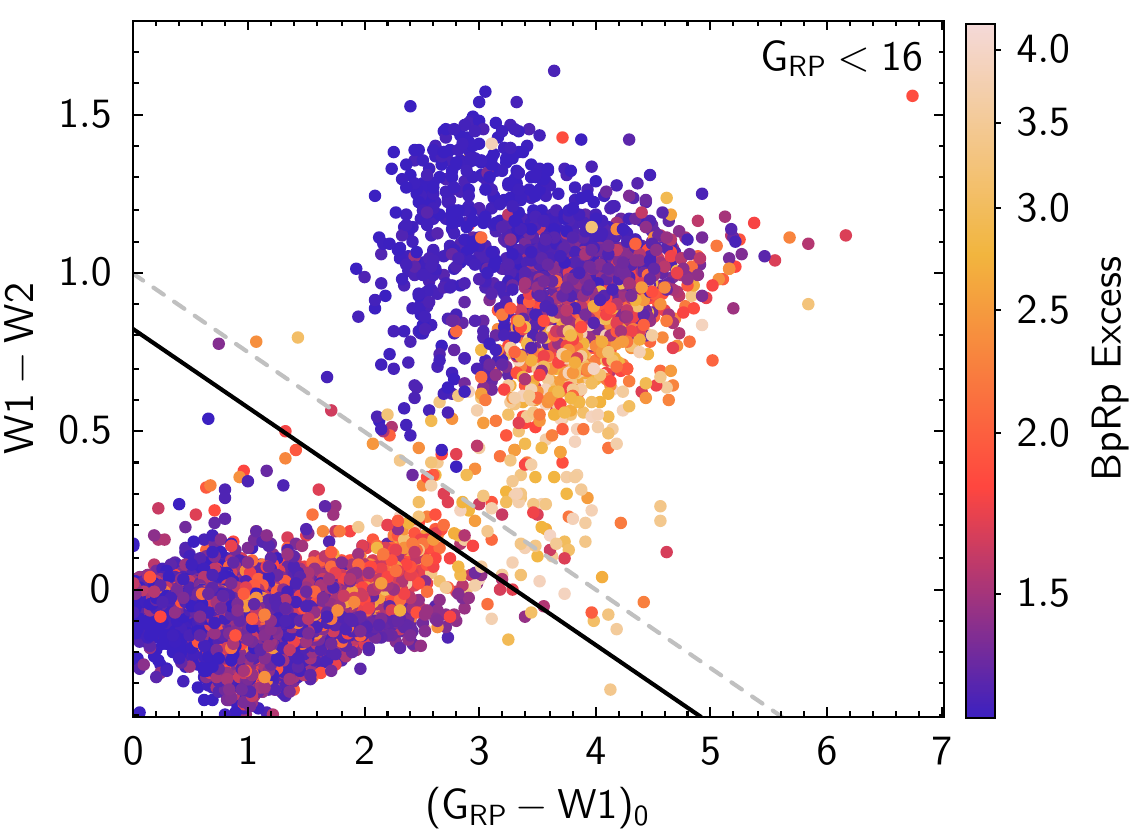} 
    \includegraphics[width=0.48\columnwidth]{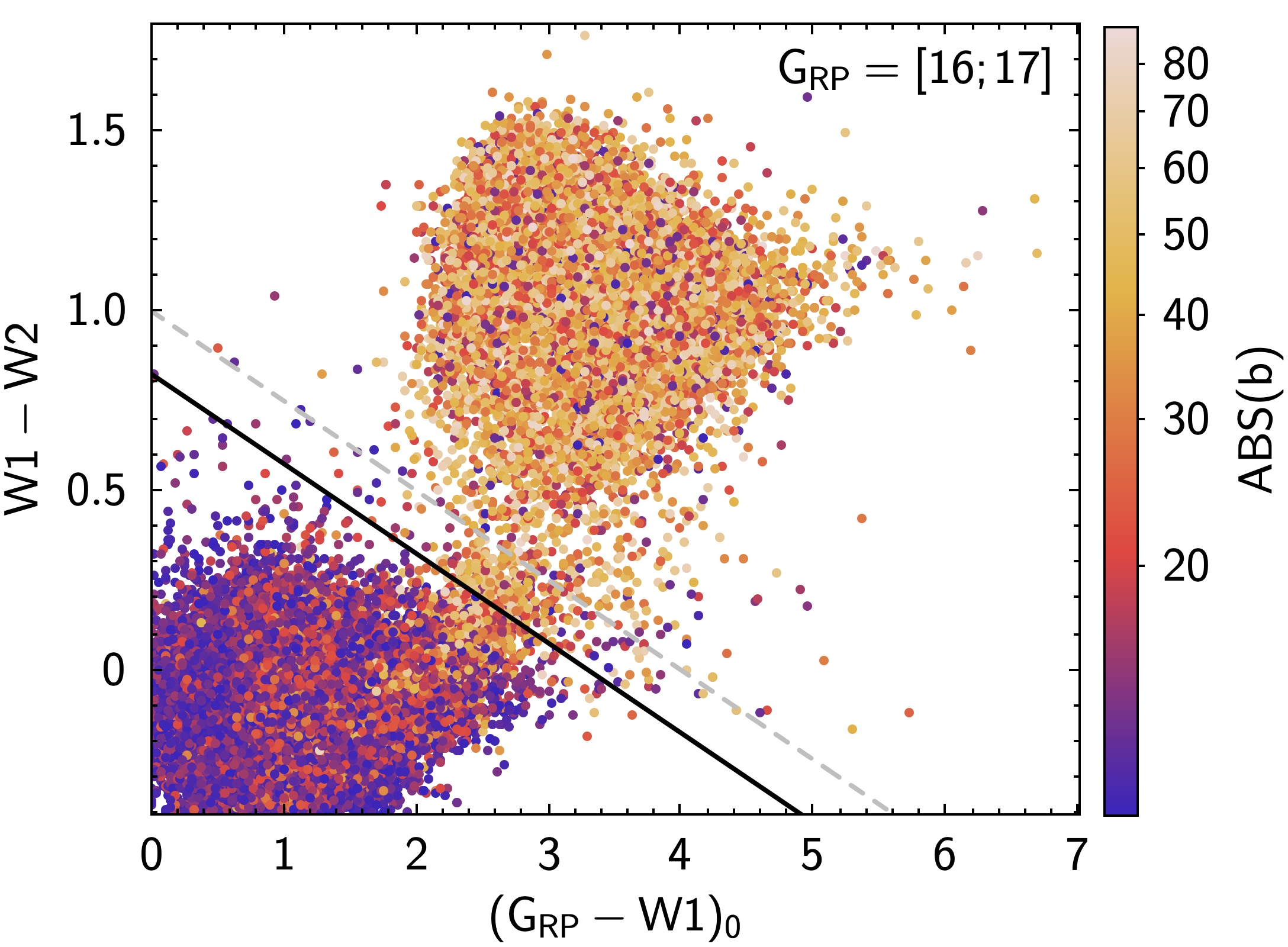} \hspace{5mm}
    \includegraphics[width=0.48\columnwidth]{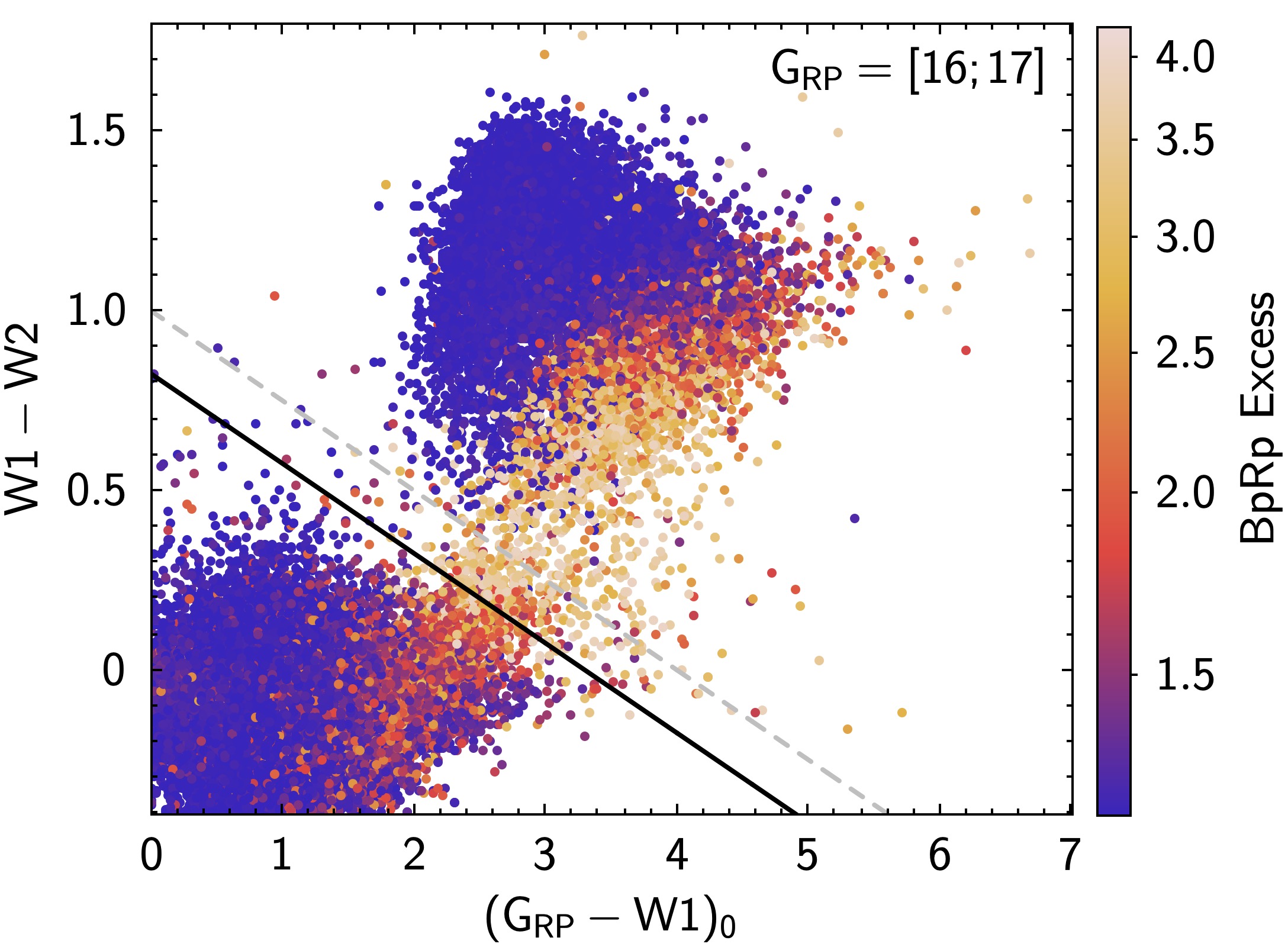}
    \includegraphics[width=0.48\columnwidth]{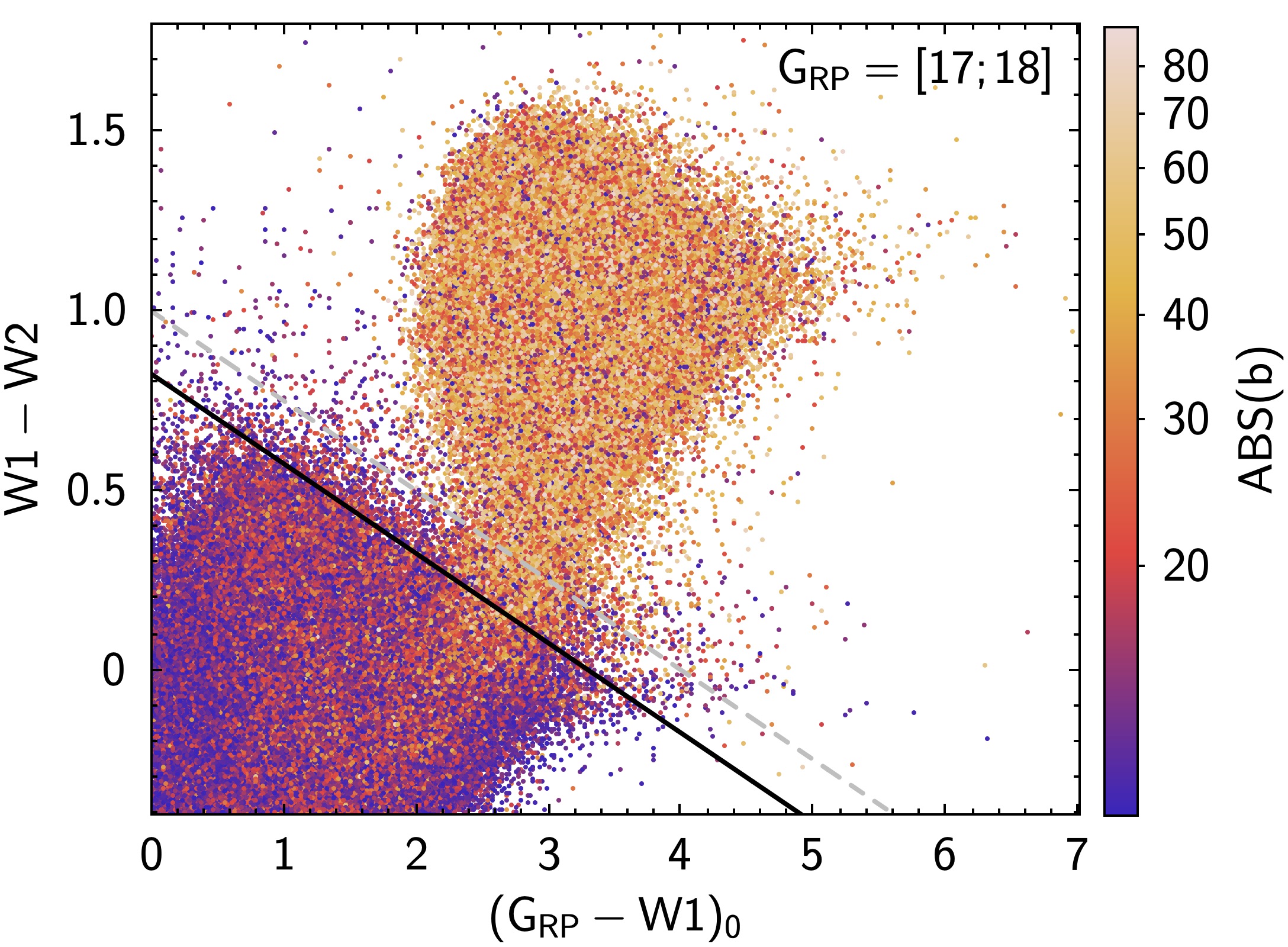} \hspace{5mm}
    \includegraphics[width=0.48\columnwidth]{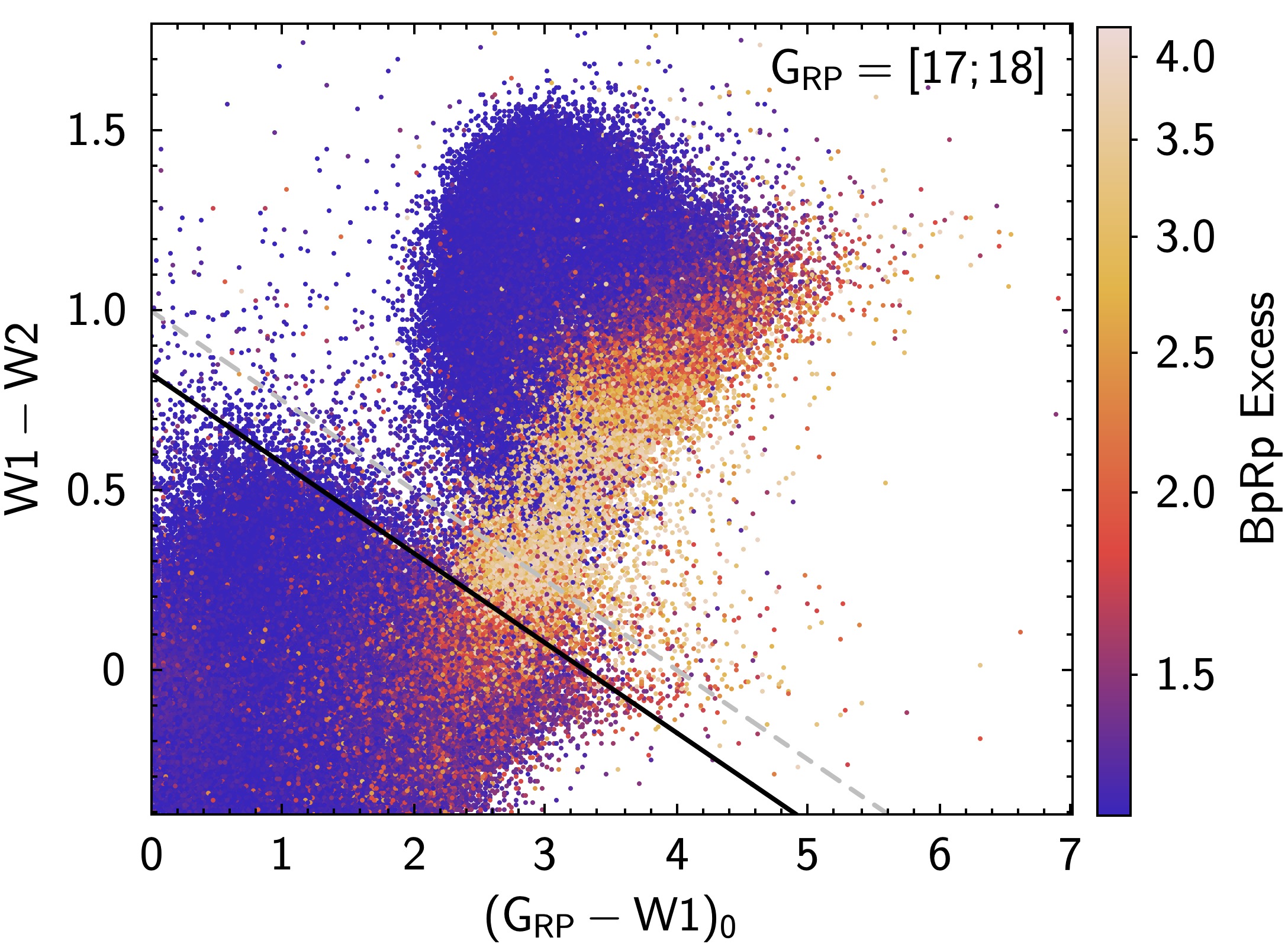} 
    \caption{Colour-colour diagram W1$-$W2 vs. $(G_{\rm RP}-$W1$)_0$ of Gaia-WISE objects at Galactic latitude $|b|>10\deg$ with negligible PPM (parallax and proper motion) signatures, colour-coded by Galactic latitude (left) and BpRp Excess Factor (right). 
    Rows step in magnitude from bright objects with $G_{\rm RP}<16$ (top) over $G_{\rm RP}=[16;17]$ (centre) to faint objects with $G_{\rm RP}=[17;18]$ (bottom). 
    The bimodal distribution represents stars on the bottom left and quasars at the top centre. Higher BpRp Excess Factors represent extended objects, which are low-redshift, low-luminosity Seyfert-1 galaxies with visible host galaxies. The grey dashed line appears to be a balanced separator between the two modes in the faintest bin, but for an as complete as possible quasar selection we adopt the black solid line to improve completeness for low-luminosity AGN at low redshift.}
    \label{fig:cands}
\end{figure*}

In this section, we discuss the selection process for our quasar sample, as well as its completeness considering the properties of known quasars. We use two different magnitude cuts: for our first exploration of potential targets on sky, we push down to a magnitude of Gaia $G_{\rm RP}=18$, although we have no intention to pursue a complete verification of quasars to this depth. As we show later, we aim to identify a complete quasar sample at 
$G_{\rm RP}\le 16$, while we also observe a handful of fainter objects. This magnitude limit is motivated by:
\begin{enumerate}
    \item our limited capacity for spectroscopic follow-up given a low sky density and a single-object spectrograph 
    \item the magnitude range of massive spectroscopic surveys of the Southern quasar population, which limit themselves to objects that are fainter than roughly $r\sim 16$ (exact values to be confirmed). 
\end{enumerate}

Quasars tend to appear point-like as their unresolved bright nucleus outshines its extended host galaxy; their equivalents at lower nuclear luminosity, Seyfert-1 galaxies, appear as regular extended galaxies with a noticeably bright nucleus; at lowest nuclear luminosities, objects appear morphologically similar to galaxies without nuclear activity.

The most abundant type of point-like object in the sky are Milky Way stars, which form a haystack hiding the quasar needles in plain sight. Hence, a technique for removing the overwhelming star population is required. Nearby stars will show parallax and proper motion (PPM) signatures, while quasars and galaxies will not.

Hence, we first select objects with little PPM signature and a point-like or modestly extended appearance. This is likely to exclude nearby Seyfert galaxies, that appear extended already in the view of backyard telescopes. It will still include plenty of stars that are too distant to reveal their motion in the Milky Way through observable PPM signatures.

\subsection{Parent sample selection}\label{sec:parent}

In the first instance, we select 1\,592\,316 objects from the Gaia DR3 source catalogue using the following selection criteria:
\begin{itemize}
    \item all Gaia sources in the magnitude range $10<G_{\rm RP}<18$ 
    \item persistent sources with at least 10 $G_{\rm RP}$ measurements, i.e.: \texttt{phot\_rp\_n\_obs>9} (the mean number is $\sim$40)
    \item Galactic latitude $|b|>10 \deg$
    \item a modest parallax signature: \texttt{parallax\_over\_error} $<5$
    \item a modest proper motion signature allowing objects with: \texttt{(pmra/pmra\_error)$^2$+(pmdec/pmdec\_error)$^2$<25 OR pmra$^2$+pmdec$^2$<0.04}
    \item a modest extended-source signature with a photometric excess ratio (hereafter, BpRp Excess Factor) $<4$ (a typical value for point sources is $\sim 1.2$)
    \item a modest level of local field crowding with fewer than 30 neighbouring sources within a 15" radius.
\end{itemize}

In a second step, we join the resulting sample with the catWISE2020 \citep{catWISE} and AllWISE \citep{2010AJ....140.1868W} catalogues using a matching radius of 2". Around 99\% of the objects have catWISE2020 matches, and we use all objects even though 12\% of them have warning flags for at least one of the bands. Where catWISE2020 provides no magnitudes for the W1 and W2 bands, we resort to AllWISE data provided the AllWISE flags for W1 and W2 are equal to "0". This eliminates 10\,650 objects and brings the sample size to 1\,581\,666.

In a third and final step, we reject sources that are in the footprint of galaxies from the Local Volume Galaxy catalogue \citep[LVG;][]{LVG13}, which we define as a circular region with a radius that is 1.3$\times$ the major axis half-length, thus eliminating 63\,115 ($\sim$4\%) of the sample, for a final sample size of 1\,513\,925. This step reduces the sky area by $\sim$227 deg$^2$ or 0.67\% 
but avoids selecting photometrically variable red giant stars and crowded areas that bias the WISE colour signature of the sources.

Figure~\ref{fig:cands} presents W1$-$W2 vs. $(G_{\rm RP}-$W1$)_0$ colour-colour diagrams for the resulting sample. The top row shows the bright regime at $G_{\rm RP}<16$, where we will aim for complete recovery of the quasar population. The bottom row shows the faintest magnitude bin of the sample, at $G_{\rm RP}=[17;18]$, which has the highest source counts as less luminous stars and quasars are intrinsically more common and sources are visible to larger distances covering a larger cosmic volume. The $G_{\rm RP}$ magnitude has been corrected for Galactic extinction using the GNILC map and an extinction coefficient of $R_{G_{\rm RP}}=2.035$ \citep{Casa18} in $A_{G_{\rm RP}}=0.86\times R_{G_{\rm RP}} E_{B-V}$, while the WISE bands had no correction applied due to the small coefficients of $R<0.15$ \citep{WangChen19}. The median reddening in the sample is $E_{B-V}=0.165$, and $\sim$90\% of the sample (given $|b|>10$) is below $E_{B-V}<0.5$.

Two clumps are visible in the distribution, a red clump with median colours of $\{(G_{\rm RP}-{\rm W1})_0,{\rm W1}-{\rm W2}\}\approx \{3.8,1.0\}$ and a blue clump centred on $\{(G_{\rm RP}-{\rm W1})_0,{\rm W1}-{\rm W2}\}\approx \{1.0,-0.1\}$. Object symbols are colour-coded by Galactic latitude $|b|$ in the left panel, which shows that the red clump is spread over moderate to high latitudes while the blue clump is dominated by the lowest latitudes. Milky Way stars are, of course, most abundant at the lowest latitudes, while quasars are expected to be uniformly spread over the sky apart from the depleting effect of low-latitude dust extinction in a flux-limited sample such as this one. 

Most convincing are the colours of the two clumps. While \citet{2010AJ....140.1868W} discuss the W1$-$W2 colours of a variety of astronomical objects as a function of W2$-$W3, we approach the question afresh in light of the different colour on the x-axis and the rejection of strongly extended objects. Quasars are characterised by (i) a blue accretion disc continuum in the UV-optical regime, (ii) an inflection point near a restframe wavelength of $\sim$1.5$\mu$m, and (iii) a red infrared continuum from thermal emission of a dust torus that is ubiquitous in quasars. Hence, they are expected to have red $(G_{\rm RP}-$W1$)_0$ and red W1$-$W2 colours.

Stars in contrast display a broad range of UV-optical colours corresponding to their photospheric temperatures, while the Rayleigh Jeans tails in their single-temperature spectra have all the same slopes. Thus they spread over a wide range in $(G_{\rm RP}-$W1$)_0$ but are tightly clustered around ${\rm W1}-{\rm W2}\approx 0$ (in Vega magnitudes). The reddest stars will have the lowest temperatures and come in two kinds: (i) cool dwarfs of type L or T will be of such low luminosity that they will be nearby and excluded from our parent sample by a large parallax, and (ii) red supergiants in nearby galaxies such as the Magellanic clouds are excluded with the LVG cut.

We thus adopt the working assumption that the red clump corresponds to quasars and the blue clump mostly to stars. We suggest that the grey dashed line, given by ${\rm W1}-{\rm W2}=-0.25(G_{\rm RP}-{\rm W1}-4)_0$, appears to be a balanced separator line between the clumps. As we will discuss later, we adopt the more inclusive black solid line, given by ${\rm W1}-{\rm W2}=-0.25(G_{\rm RP}-{\rm W1}-3.3)_0$, for our quasar selection.


Even at the $G_{\rm RP}=18$ faint end of the sample, photometric errors are negligible for $G_{\rm RP}$ (typically 0.01~mag), but they are noticeable in the WISE bands, where they increase towards the left of the diagram as the WISE bands get fainter for a given $G_{\rm RP}$ magnitude. At $G_{\rm RP}=18$, the mean error for the W1$-$W2 colour is $\sim$0.025~mag in the red quasar clump and $\sim$0.12~mag (with a tail up to 0.5~mag) in the blue star clump. It is clear that the extent of the red quasar clump is not driven by photometric errors but by the spread in appearance from both redshifting as well as a mixing sequence of quasar light dimming relative to host galaxy light as the W1$-$W2 colour gets bluer; the W1$-$W2 envelope of the blue star clump, in contrast, expands towards fainter magnitudes mostly due to photometric noise in WISE.

In the right panels of Figure~\ref{fig:cands}, objects are colour-coded by the BpRp Excess Factor. Most objects are consistent with point sources (indicated by values of $\sim$1.3), but a structure of extended-appearing objects at the red end of the $(G_{\rm RP}-{\rm W1})_0$ distribution forms a bridge between the quasar and star clumps. While the upper part in the distribution of extended objects is likely nearby quasars with discernible host galaxies, the lower part may include even galaxies without nuclear activity.

\begin{figure*}
    \centering
    \includegraphics[width=0.48\columnwidth]{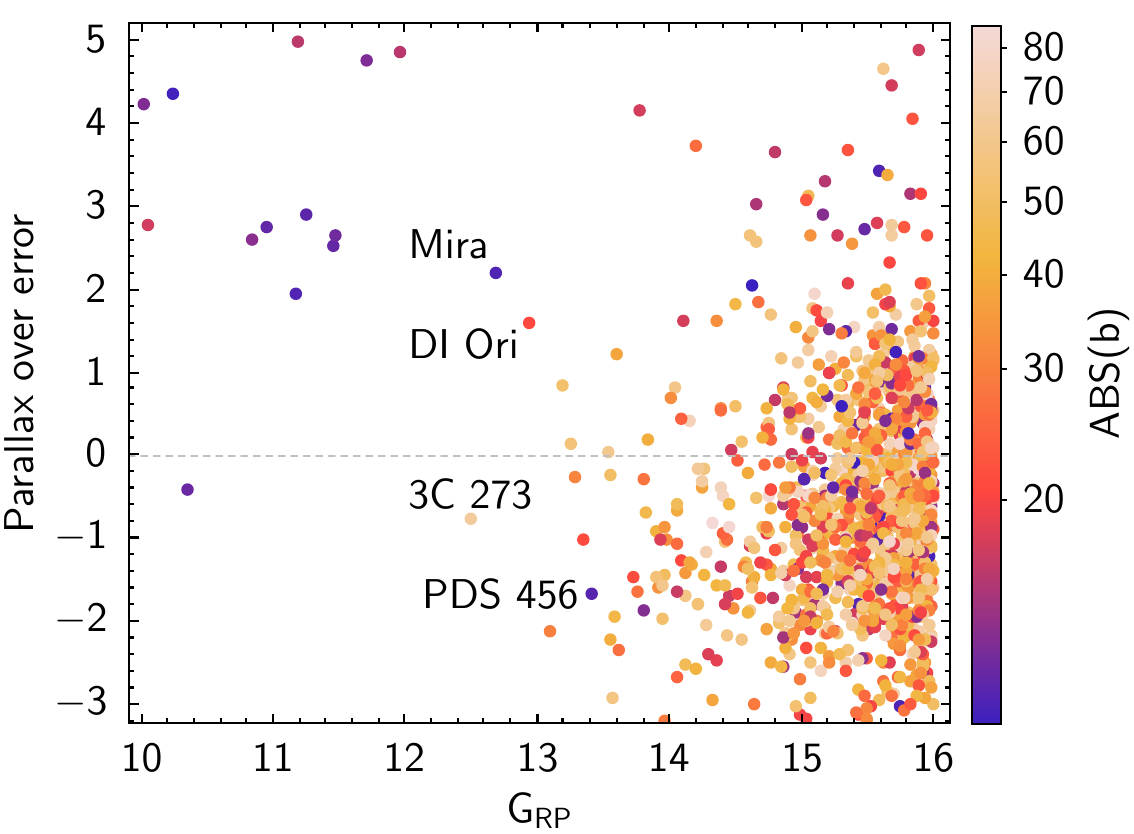} \hspace{5mm}
    \includegraphics[width=0.48\columnwidth]{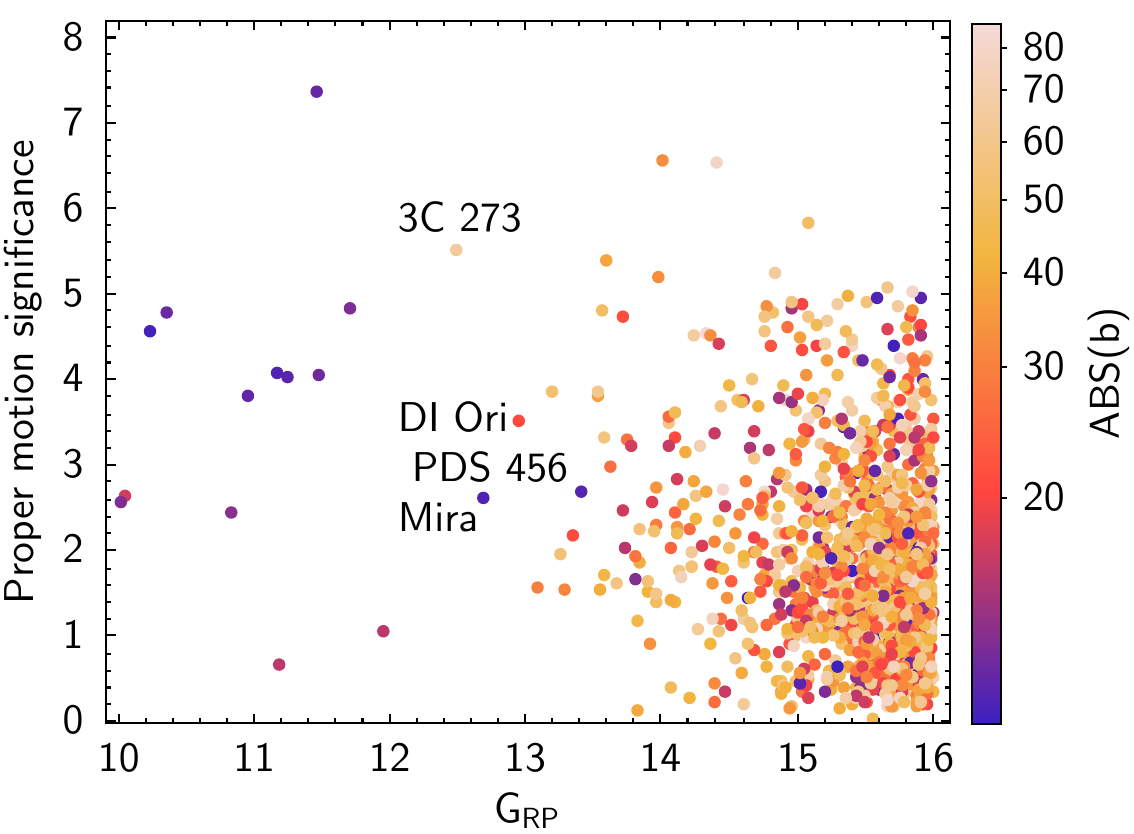} 
    \caption{Quasar candidates at $G_{\rm RP}<16$: significance of parallax signature (left) and proper motion signature (right) vs magnitude colour-coded by Galactic latitude. 3C 273 is the brightest known quasar; the 13 objects brighter than it have mostly more significant parallaxes and low Galactic latitude and are likely Galactic stars. Two objects just fainter than 3C~273 are Galactic stars at low latitude, DI~Orionis and a distant Mira-type variable star, IRAS~06214$-$1016 (labelled as Mira). The low-latitude candidate at $G_{\rm RP}\approx 13.4$ is a quasar (PDS~456) despite its position at $b=11.2$.}
    \label{fig:cands_RpPPMSN}
\end{figure*}

We now consider the $\sim$2\,000 brightest quasar candidates with $G_{\rm RP}<16$, i.e., all objects above the solid line selection line in the top row panels of Figure~\ref{fig:cands} but restricted to the sky at ${\rm Dec}<+20$, the search limits for the Southern AllBRICQS branch imposed by the needs of spectroscopic follow-up from Australia. Figure~\ref{fig:cands_RpPPMSN} shows the significance of their parallax and proper motion signature vs magnitude. Most of the candidates are fainter than $G_{\rm Rp}=13$, have a median parallax significance below 0 and a median proper motion significance of $\sim$1.7; they cover a range in Galactic latitude as shown in their colour coding, consistent with them being quasars. Brighter than $G_{\rm Rp}\approx 12$, there are 13 candidates with notably larger PPM significance and uniformly low Galactic latitude, which are presumably Galactic stars. Between these two groups, we note four particular objects: (i) closer in brightness to the presumed stars but more typical in latitude for the fainter quasars, is the brightest known quasar 3C~273; (ii) the brightest of the presumed quasars at $G_{\rm Rp}\approx 13$, both with elevated parallax significance, are low-latitude Galactic stars, the Mira-type variable star IRAS~06214$-$1016 and DI~Orionis; and (iii) the lowest-latitude object at $G_{\rm RP}\approx 13.5$ is the most luminous quasar at $z<0.2$, PDS~456, which is seen through Galactic extinction of $A_{\rm RP}>1$. Its location near the Galactic bulge at $(l,b)=(10.39,11.16)$ may explain why it was discovered as late as 35 years after 3C~273 by \citet{PDS456_97}. 

\begin{figure*}
    \centering
    \includegraphics[width=0.48\columnwidth]{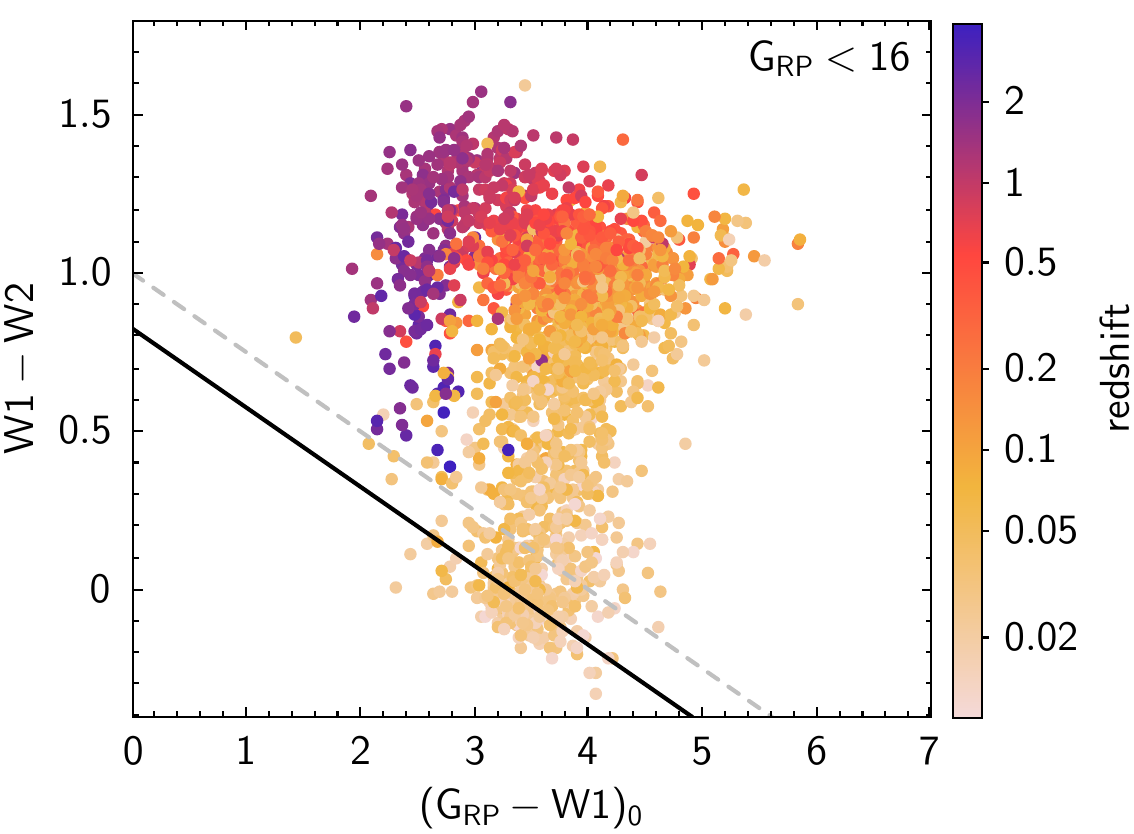} \hspace{5mm}
    \includegraphics[width=0.48\columnwidth]{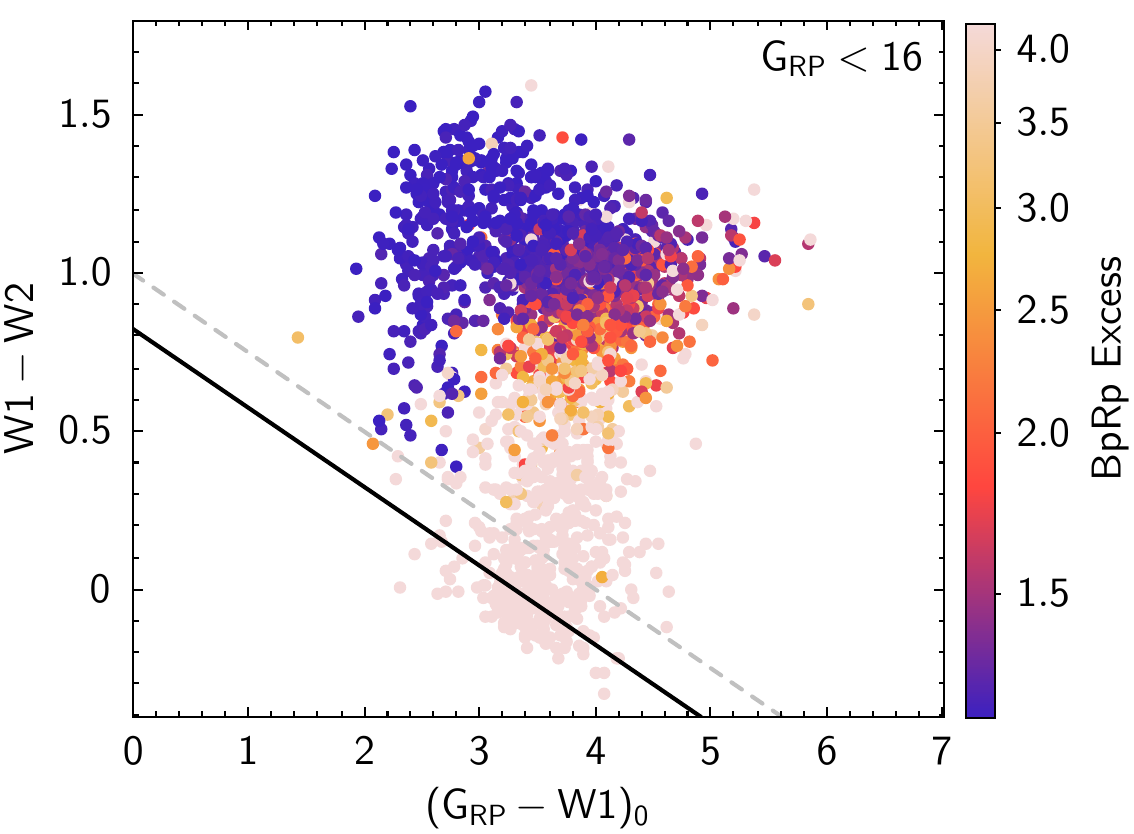} 
    \includegraphics[width=0.48\columnwidth]{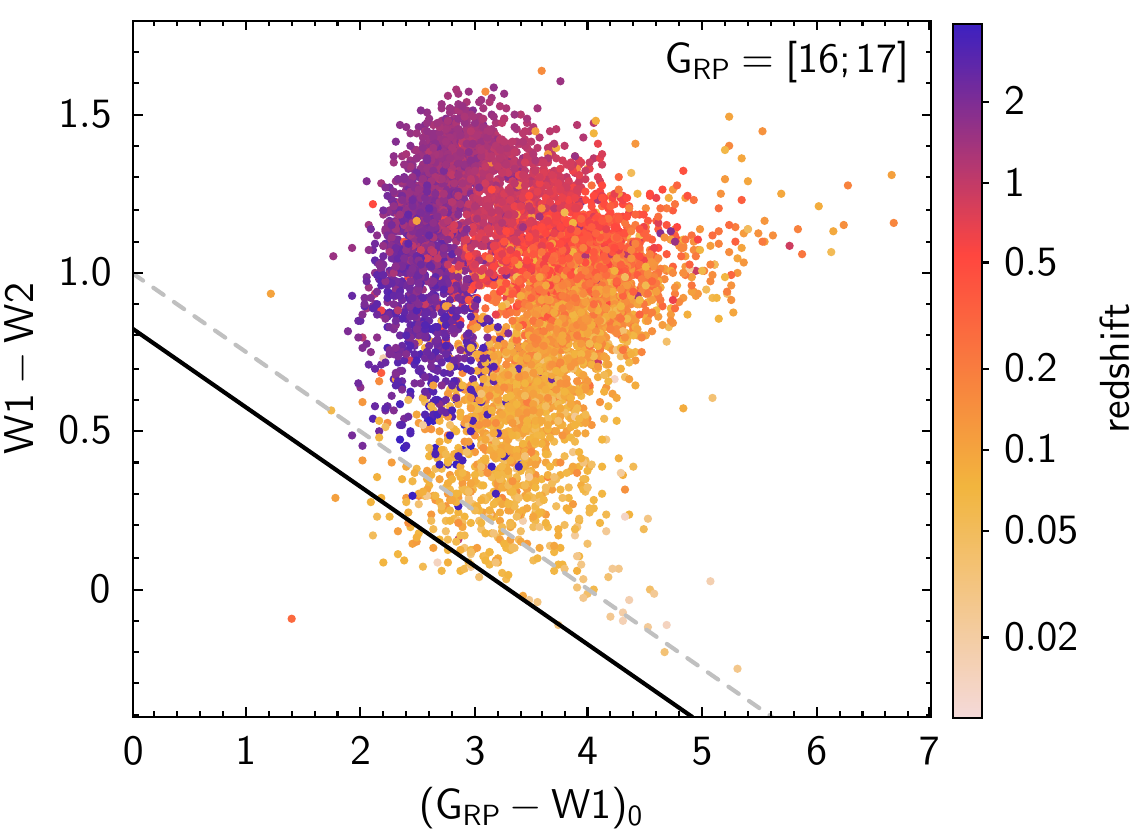} \hspace{5mm}
    \includegraphics[width=0.48\columnwidth]{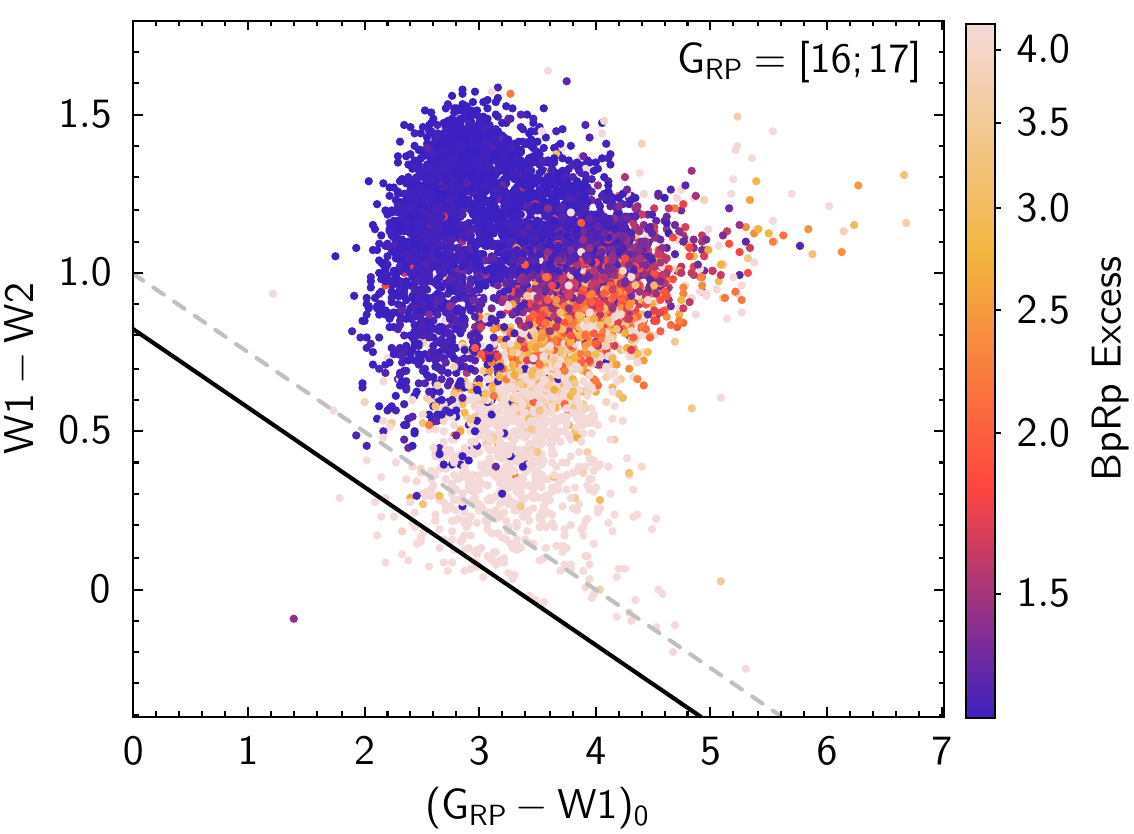} 
    \includegraphics[width=0.48\columnwidth]{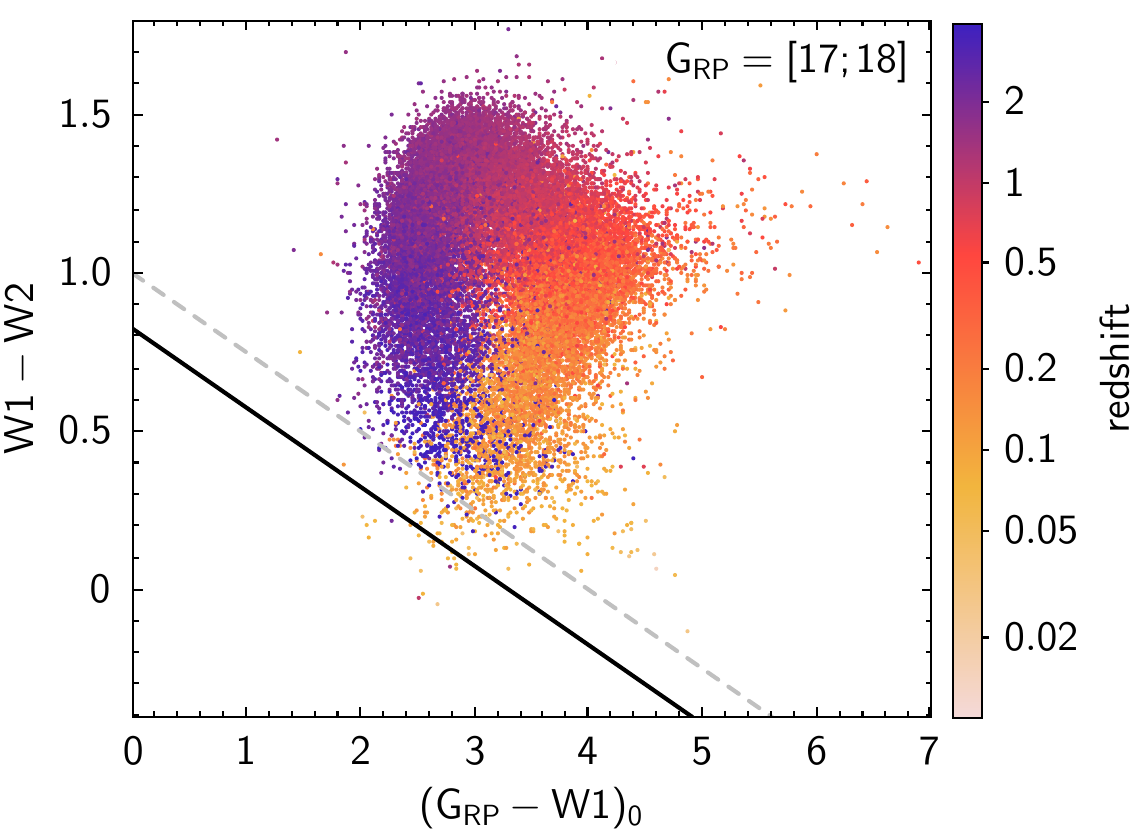} \hspace{5mm}
    \includegraphics[width=0.48\columnwidth]{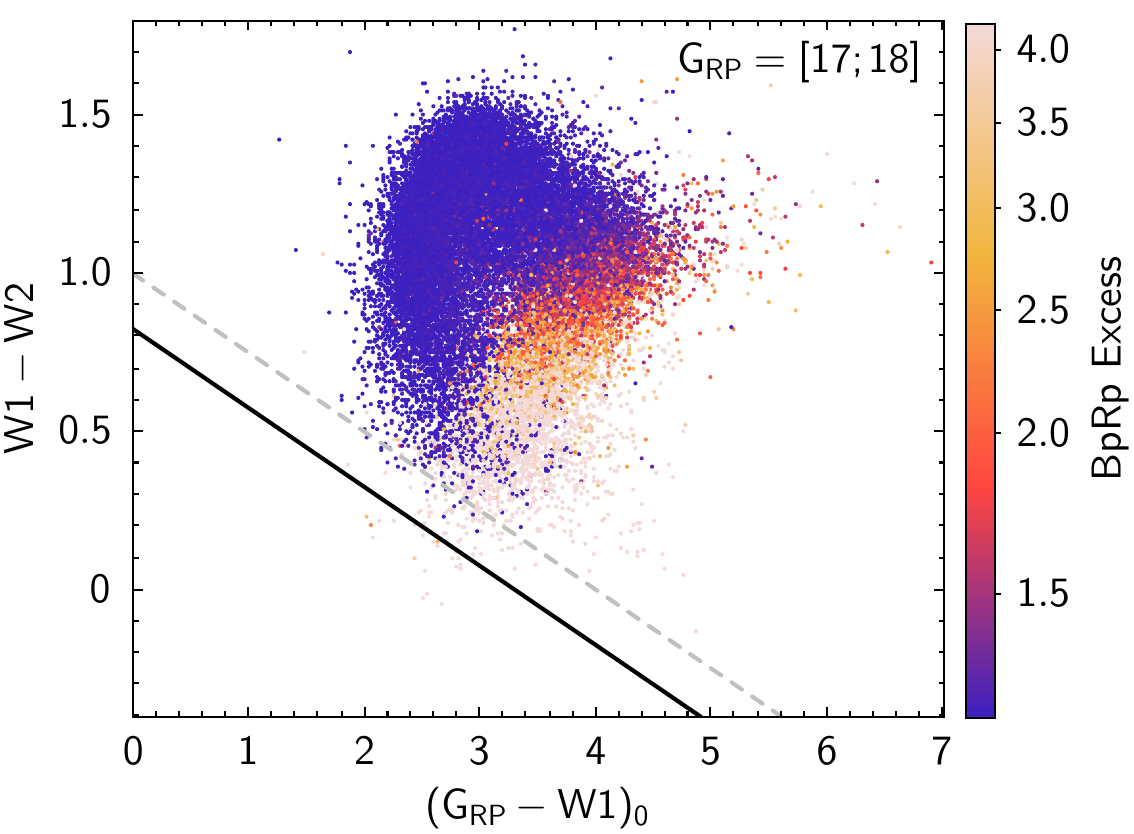} 
    \caption{Colour-colour diagram W1$-$W2 vs. $(G_{\rm RP}-$W1$)_0$ of known quasars from Milliquas v8 with redshift $z>0.01$ in the low-PPM signature parent sample of Section~\ref{sec:parent}, colour-coded by redshift (left) and BpRp Excess Factor (right). Rows step in magnitude from bright objects with $G_{\rm RP}<16$ (top) over $G_{\rm RP}=[16;17]$ (centre) to faint objects with $G_{\rm RP}=[17;18]$ (bottom). The reddest W1$-$W2 colour is seen at $z\sim 1.5$ where the NIR upturn in quasar SEDs due to thermal emission from the hot dust torus moves between the two WISE filters. Almost all known quasars reside above the solid selection line. Low-redshift Seyfert-1 galaxies with low nuclear luminosity drop below the line as the nuclear signature fades relative to the stellar photospheric light of the host galaxy. 
    }
    \label{fig:MQ_CCD}
\end{figure*}

\begin{table*}
\centering
{\begin{tabular}{
    rrlccrr
    S[table-format = 2.2(7)]
    crHHl}
\hline
RA & Dec & Popular name & $z$ & ref. & Sig(px) & Sig(PM) & {AEN} & {$G_{\rm RP}$} & Excess & {$(G_{\rm RP}-{\rm W1})_0$} & W1$-$W2 & Comment \\
\hline
\multicolumn{12}{l}{Literature objects that do not belong into the $G_{\rm RP}<16$ quasar sample} \\ 
\hline
172.1027 & $-$11.3717 & MRC 1125$-$110 & 0.077 & 2MAGN &   &   & 13.3 \pm 390.1 &15.26 & 31.25 &   & 0.07 & Not AGN \\
48.7149 & $-$41.1494 & DUKST 300$-$12 & 0.079 & 2MAGN &   &   & 13.1 \pm 103.3 & 15.23 & 29.19 &   & $$-$$0.07 & Not AGN \\
188.6844 & $-$6.2682 & NVSS J12347$-$0616 & 0.132 & 1225 & 11.77 & 122.50 & 0 \pm 0 & {\it 15.57} & 1.35 &   & 0.29 & faint, mismatch \\
16.5744 & 0.6219 & 
& 1.639 & DR16 &   &   & 0 \pm 0 & {\it 15.88} & 48.12 &   &   & faint, mismatch  \\
71.2147 & 27.3926 & 
& 1.845 & LAMQ1 &   &   & 10.9 \pm 22306.1 & 14.68 & 1.48 &   &   & Not AGN\\
\hline
\multicolumn{12}{l}{Literature objects that are not reported as gravitationally lensed quasars} \\ 
\hline
45.8151 & 8.8739 & NGC 7469 & 0.017 & 1315 & $-$0.67 & 10.19 & 0.3 \pm 19.4 & 13.08 & 2.57 & 4.477 & 0.87 & PM  \\
39.6881 & $-$30.8065 & ESO 416$-$5 & 0.063 & 0383 & $-$1.61 & 8.51 & 0.6 \pm 27.6 & 14.78 & 2.36 & 4.021 & 1.00 & PM  \\
229.8403 & 59.1399 & SBS 1518+593 & 0.078 & DR16Q & 7.51 & 9.94 & 0.3 \pm 7.6 & 15.16 & 1.59 &   & 1.03 & px, PM\\
151.1984 & 14.7793 & CGCG 93$-$37 & 0.084 & DR16Q & 1.44 & 5.67 & 0.4 \pm 6.1 & 15.32 & 2.60 & 3.781 & 0.69 & PM  \\
321.2596 & $-$4.6210 & IRAS 21224$-$0450 & 0.088 & 1225 & $-$2.62 & 9.5 & 0.2 \pm 9.4 & 14.85 & 1.54 & 4.515 & 0.95 & PM  \\
58.9863 & $-$51.3343 & PGC 457441 & 0.107 & 2086 & 2.44 & 8.27 & 1.8 \pm 13.6 & 15.77 & 7.46 & 3.667 & 0.49 & PM, BpRp Excess  \\
28.1129 & $-$23.3316 & RXS J01524$-$2319 & 0.113 & 0740 & 0.94 & 7.95 & 0.3 \pm 14.4 & 15.10 & 1.64 & 4.281 & 0.96 & PM  \\
118.0744 & 19.5951 & PGC 3095775 & 0.117 & DR16Q & $-$3.98 & 10.72 & 0.3 \pm 5.7 & 15.35 & 1.88 & 4.037 & 1.12 & PM  \\
232.7157 & $-$37.5826 & IGR J15311$-$3737 & 0.127 & 1147 & 5.77 & 1.68 & 0.2 \pm 1.2 & 16.00 & 1.38 &   & 0.93 & px\\
146.7191 & 13.3316 & MS 09441+1333 & 0.133 & DR16Q & $-$3.32 & 6.28 & 0.1 \pm 3.4 & 14.96 & 1.32 & 3.635 & 1.08 & PM  \\
119.5679 & 83.9434 & 
& 0.134 & BASS2 & $-$4.35 & 7.05 & 0.3 \pm 7.0 & 15.34 & 1.62 & 4.11 & 0.93 & PM  \\
29.9594 & 0.3947 & Mrk 1014 & 0.163 & DR16 & 2.78 & 5.92 & 0.6 \pm 14.1 & 15.42 & 2.01 & 4.61 & 0.97 & PM  \\
257.5559 & 33.7341 & RXS J17102+3344 & 0.207 & DR16Q &   &   & 8.0 \pm 4735.1 & 15.47 & 1.83 &   & 0.98 & No PPM data \\
79.1595 & $-$51.7804 & RXS J05166$-$5146 & 0.222 & 1677 & $-$0.24 & 6.27 & 0.1 \pm 1.0 & 15.66 & 1.42 & 4.305 & 0.99 & PM  \\
208.3589 & 36.3471 & CSO 1022 & 0.285 & DR16Q & $-$2.79 & 5.78 & 0.2 \pm 3.8 & 15.87 & 1.34 & 4.575 & 1.01 & PM  \\
186.2269 & 21.3796 & PKS 1222+21 & 0.433 & DR16Q & $-$0.82 & 7.45 & 0.1 \pm 0.8 & 14.50 & 1.21 & 3.853 & 1.05 & PM  \\ 
164.6234 & 1.5663 & PKS 1055+01 & 0.893 & DR16Q & 0.17 & 5.53 & 0.2 \pm 1.5 & 15.81 & 1.31 & 4.79 & 1.13 & PM  \\ 
\hline
\multicolumn{12}{l}{Literature objects that are gravitationally lensed quasars} \\ 
\hline
24.5914 & 48.6963 & PS J0138+4841 & 1.172 & LEMON4 & 1.33 & 5.74 & 2.7 \pm 594.9 & 15.58 & 1.68 & 3.47 & 1.40 & PM \\
123.3803 & 25.7509 & HS 0810+2554 & 1.512 & DR16Q & $-$1.82 & 8.37 & 3.1 \pm 712.8 & 15.17 & 2.05 & 3.416 & 1.32 & PM \\
235.8728 & 33.9857 & 
& 2.160 & DR16Q & 3.73 & 221.82 & 0.7 \pm 42.7 & 15.89 & 1.24 & 1.785 & 0.91 & PM  \\
202.5219 & $-$9.0844 & Persephone's Torch & 2.220 & Davies26 &  &  &  16.2 \pm 49613.6 & 14.38 & 1.85& No PPM data\\
127.9238 & 52.7549 & APM 08279+5255 & 3.911 & 0524 &   &   & 7.3 \pm 4687.8 & 14.70 & 2.14 &   & 0.61 & No PPM data\\
\hline
\end{tabular}}
\caption{Literature objects at $G_{\rm RP}<16$ that fail the selection criteria, ordered by redshift. These are either not AGN (see text for details) or truly fainter than $G_{\rm RP}=16$ (top section), or true quasars rejected by our selection cuts (middle section), with a subset of strongly gravitationally lensed objects in the bottom section. All of the true quasars pass the colour selection but fail the PPM criteria and show significant astrometric excess noise (AEN).
}
\label{tab:not_cands}
\end{table*}

\subsection{Properties of known quasars}

We now turn our attention to properties of spectroscopically confirmed quasars using the compilation MILLIQUAS v8 by \citet{Flesch23}; this catalogue is dominated by Northern-hemisphere quasars due to the Northern focus of past effort. Besides quasars identified by other groups, it also contains already those discovered by AllBRICQS that were reported in the first sample release by \citet{On23}.

We cross-match MILLIQUAS with our parent sample from Section~\ref{sec:parent} using the brightest Gaia source within a matching radius of 2"; Gaia's high spatial resolution allows it to record multiple sources inside such a matching radius if present, which occurs in $\sim$1\% of the quasars at $G_{\rm RP}<18$. Technically, it is unclear which of the sources in ambiguous cases is the quasar, given that ground-based spectroscopy uses slits or fibres with typical widths or diameters between 1" and 3". However, we ignore this problem in this section, where we are concerned with the properties of the vast majority of known quasars, which are well-isolated in the sky. 

Most objects from MILLIQUAS do not have counterparts in our parent sample, because they are too faint with $G_{\rm RP}>18$, and some are close to the Galactic plane with $|b|<10\deg$. Those which have brighter Gaia counterparts at $|b|>10\deg$ that are not in our parent sample will be discussed below in the context of our completeness discussion.

Figure~\ref{fig:MQ_CCD} shows $(G_{\rm RP}-$W1$)_0$ colour-colour diagrams for the matched quasars and Seyfert-1 galaxies (denoted with the descriptor 'Q' and 'A' in MILLIQUAS) at redshift $z>0.01$, colour-coded by redshift in the left column. Their colours are expected to change as the near-infrared inflection around $\sim 1.5\mu$m in their spectra is redshifted through the passbands and thus W1$-$W2 reaches its reddest level around $z\approx 1.5$. At the highest redshifts, they never go as blue as ${\rm W1}-{\rm W2}=0$ since the filter pair probes the restframe-optical spectral slopes of the mixed-temperature accretion disc continua that are always redder than the single-temperature Rayleigh-Jeans tail in a star. Towards the lowest redshifts, especially at $z<0.1$, the flux-limited sample of Seyfert-1 galaxies drawn from a shrinking nearby cosmic volume is dominated by galaxies with a nuclear luminosity that decreases with redshift so that their colours are increasingly dominated by photospheric emission from the host galaxy stars and thus reaches ${\rm W1}-{\rm W2}\approx 0$. The dilution effect of the host galaxies affects also the BpRp Excess Factor, which increases rapidly with the apparent angular size of the galaxies on the $z<0.1$ branch.

The rows of panels show the sample at different depths, where a deeper cut selects mostly many more high-redshift quasars, while fainter magnitudes at low redshift lead into the regime of dwarf galaxies, where nuclear activity is not yet well-catalogued.

\begin{figure*}
    \centering
    \includegraphics[width=0.48\columnwidth]{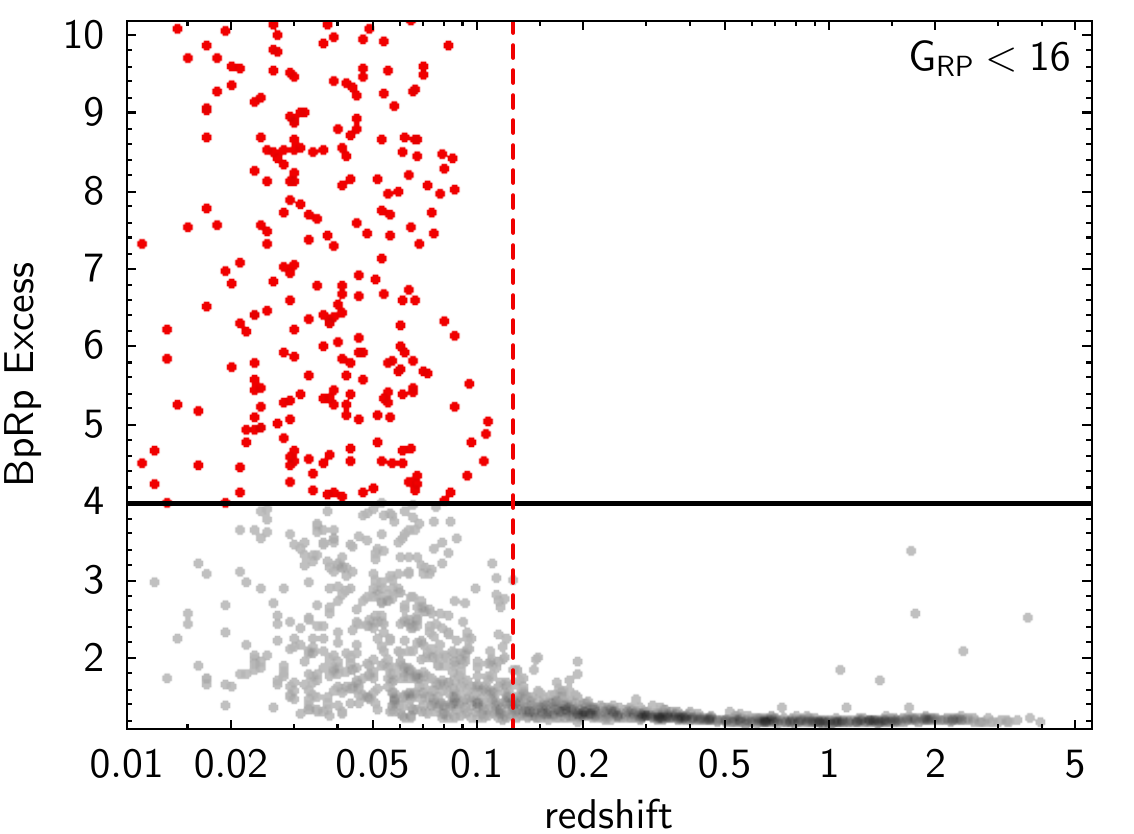} \hspace{5mm}
    \includegraphics[width=0.48\columnwidth]{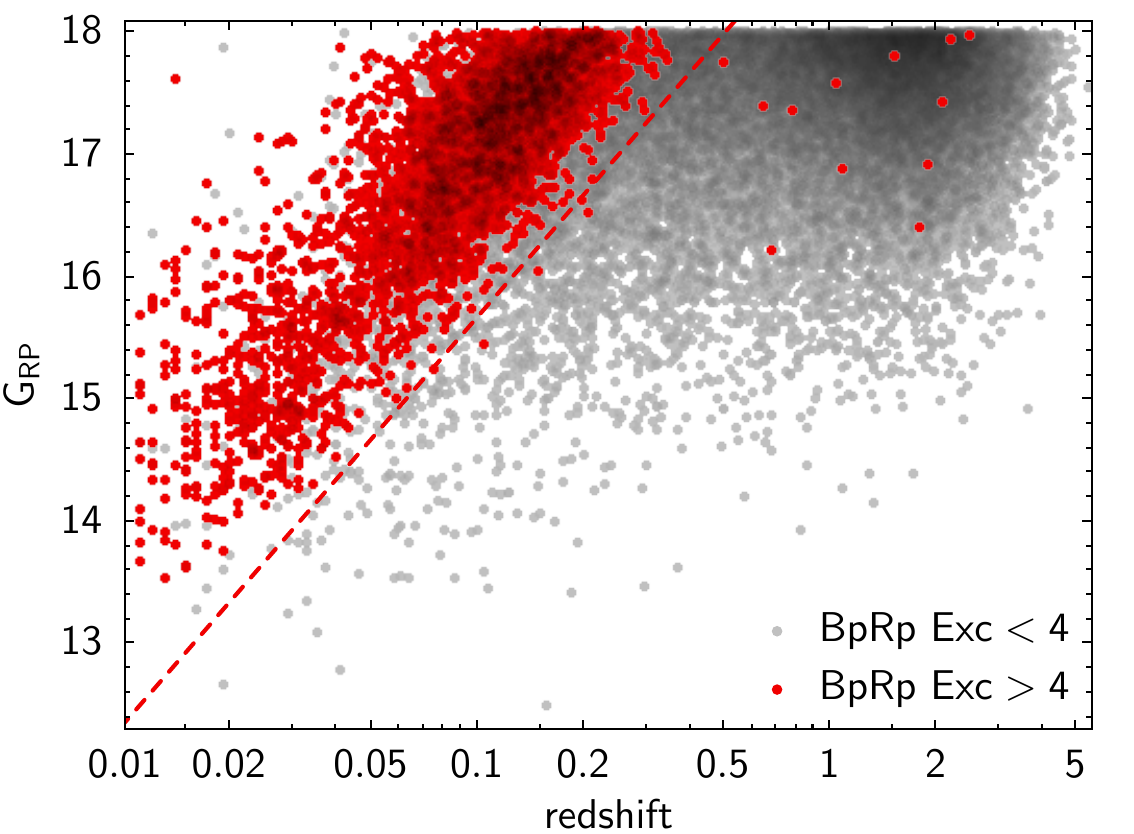} 
    \caption{Gaia BpRp Excess Factor vs. redshift and magnitude of known quasars (Milliquas v8) with Galactic latitude $|b|>10\deg$ and negligible PPM signatures. {\it Left:} BpRp Excess Factor at $G_{\rm RP}<16$. AllBRICQS limits its search to BpRp Excess less than 4 (solid line); nearly all known quasars at $z>0.1$ reside within the selection cut. Low-luminosity AGN at low redshift exceed the cut when the AGN signature fades relative to the extended light of the host galaxy. The few extended-looking quasars at high redshift have multiple images from gravitational lensing. {\it Right:} Objects rejected due to extended appearance follow a magnitude-dependent redshift trend produced by AGN-to-host contrast.}
    \label{fig:MQ_zExc}
\end{figure*}

\subsection{Quasar candidate selection}\label{sec:sel}

We select our quasar candidates by requiring a low parallax and proper motion signature, a (semi-)compact appearance, and a simple colour cut. While a cut aimed at purity for more efficient spectroscopic follow-up could use additional or tighter criteria, we remain more inclusive as we aim for completeness.

We now discuss our selection completeness by considering samples of known quasars; we will frequently refer to Table~\ref{tab:not_cands}, which lists the literature objects that fail any selection criteria (ignoring various objects at $z<0.125$ as explained in the following section). The table contains 26 objects, five of which should be discarded as discussed here:
\begin{itemize}
    \item Two of the objects, MRC 1125$-$110 and DUKST 300$-$12, are galaxies without any emission lines and dominated by old stellar populations, as seen in high-quality spectra of Amrutha et al.~(2026b), thus clearly not quasars. Broad emission lines are a common but also transient by-product of supernova explosions, and thus we speculate that past observations may have caught these two galaxies while a supernova provided a broad-line impression.
    \item Two further objects are fainter than our $G_{\rm RP}=16$ limit: NVSS J12347$-$0616 at $z=0.132$ is a blend of three sources: the listed properties belong to a star with strong PPM features, while the quasar is nearly 3" away from this star and has $G_{\rm RP}=17.58$. The object at $z=1.639$ has a $G_{\rm RP}$ magnitude with an error of $\sim$2 mag; with $G=21.00$ and $r_{\rm SDSS}=21.37$ it is truly much fainter and irrelevant here.
    \item The fifth object, at $z=1.845$, is referenced to the first data release of the LAMOST project, but the current LAMOST database has updated the object type to a Galactic F9 star, which is consistent with the public LAMOST spectrum. 
\end{itemize}
Where the following discussion refers to Table~\ref{tab:not_cands}, it will focus on the 21 leftover and likely true quasars.

\subsubsection{BpRp Excess Factor cut}

Our selection requires a Gaia BpRp Excess Factor below 4. Figure~\ref{fig:MQ_zExc} shows the BpRp Excess Factor against redshift for known quasars from the Milliquas compilation, limited to our primary search depth of $G_{\rm RP}<16$, avoiding the Galactic plane at $|b|<10 \deg$ because of its crowding, and restricting the sample to objects with a low PPM signature. Most quasars at $z>0.2$ have Excess Factor values of $\sim$1.2, which is typical for point sources. Some of the known quasars stray to larger values, especially those that are known to be gravitationally lensed into multiple images, which are sometimes resolved by Gaia and sometimes blended. The adopted threshold of 4 captures several lensed objects. At redshifts below 0.1 the known sample includes a rich tail shooting up to much larger values as extended host galaxies start to dominate the object appearance. This redshift range is further discussed by Amrutha et al.~(2026b), who publish a large Southern hemisphere catalogue and spectral atlas of type-1 AGN at $z<0.1$.

The right panel of Figure~\ref{fig:MQ_zExc} shows the distribution of selected and rejected objects in the magnitude-redshift plane, where we extend the magnitude cut to $G_{\rm RP}<18$. Here, we see that the bulk of rejected objects reaches higher redshifts at fainter magnitudes, which is a natural consequence of varying the contrast of the host galaxy when varying the luminosity of the nucleus. We suggest that our quasar selection will be complete to the right of the dashed line, which is given by 
\begin{equation}\label{zcomp}
    z > 2^{G_{\rm RP}-19}    ~.
\end{equation}
With a magnitude limit of $G_{\rm RP}<16$, the Milliquas sample includes 1\,025 quasars above this redshift limit, as well as four quasars in our LVG exclusion zones. Only one of the missed objects in Table~\ref{tab:not_cands} fails the BpRp Excess Factor selection, PGC 457441 at $z=0.107$.

\subsubsection{Parallax and proper motion (PPM) cut}

Our selection requires objects to have parallaxes and proper motion with less than 5$\sigma$ significance each; we allow higher significance for the proper motion when its absolute value is below 0.2~mas per year. In a Gaussian distribution, less than one in a million objects are expected outside a 5$\sigma$ cut, and thus we expect no quasar to be missed due to these cuts on statistical grounds. However, the selection rejects objects for which Gaia has not reported any PPM solution at all. 

All missed objects in Table~\ref{tab:not_cands} fail the PPM selection cut. Two of them have no PPM data at all, two further ones fail the parallax limit and 19 fail the proper motion limit. All 21 quasars have high astrometric excess noise (AEN), which implies that they comprise more than just a single point source. All five cases at $z>1$ are quasars that are multiply imaged by gravitational lensing. This includes, e.g., APM 08279+5255 with redshift $z=3.911$ and $G_{\rm RP}=14.70$, well-known to be the brightest lensed quasar at redshift $z>3$ \citep{Irwin98}, and the recently discovered Persephone's Torch at $z=2.22$ \citep{2026arXiv260413152D}. A few more objects at moderate redshift might be lensed or dual quasars that are not yet identified as such. The lowest-redshift cases are probably just affected by structure in their host galaxies that shift the astrometric centroids as the quasar nucleus varies in brightness.

\subsubsection{Colour cut} 

We select quasar candidates using a colour equation updated from the first observations published in \citet{On23}; there we used simply W1$-$W2$>0.2$, but we soon found that the increasing WISE errors towards objects that are blue in $(G_{\rm RP}-{\rm W1})_0$ inflates the stellar plume and produces candidates that are red in W1$-$W2. Our updated selection is: 
\begin{equation}
    {\rm W1}-{\rm W2}>-0.25(G_{\rm RP}-{\rm W1}-3.3)_0 
\end{equation}
(see Figures~\ref{fig:cands} and \ref{fig:MQ_CCD}). At $G_{\rm RP}<16$, this selection misses none of the confirmed broad-emission line AGN at $z>2^{G_{\rm RP}-19}$, i.e., at $z>(0.008,0.031,0.125)$, rounded to 0.001, for $G_{\rm RP}=(12,14,16)$. We have spot-checked several objects listed in Milliquas down to $G_{\rm RP}=18$ that are rejected by our colour cuts. These cases can be explained as source confusions or are objects where new and better spectra reveal that the quasar classification was not robust  (see appendix). 

\subsubsection{Local Volume Galaxy catalogue cut}

In our selection we avoid certain circular regions centred on the galaxies in the Local Volume Galaxies (LVG) catalogue by \citet{LVG13}. Among the 1\,029 quasars listed in Milliquas with $G_{\rm RP}<16$, $|b|>10\deg$, and $z>2^{G_{\rm RP}-10}$, four objects reside in such exclusion zones, one within the patch of the Andromeda Galaxy, one within the region of the Sgr Dwarf Galaxy, and two within the LMC's footprint. None of these four quasars violates the colour cuts, BpRp Excess cuts, or PPM cuts. The one seen behind the Sgr Dwarf Galaxy was indeed identified by AllBRICQS and published as part of the first batch of quasars in \citet{On23}.

\subsection{Overall selection completeness, dual and lensed quasars}

The global picture is then that none of the known quasars at redshifts larger than $z\sim 0.1$ fail the colour selection, but a fraction of sources with high AEN fail the PPM selection. At redshifts larger than $z\sim 0.1$, these tend to be gravitationally lensed or dual or blended quasars, while at the lowest redshifts these are Seyfert-1 galaxies with dominant host galaxy signatures. Lensed quasars, however, are not the focus of this work, and thus we tolerate incompleteness in discovering these. 

We note that lensed quasars with wide-separation images, such as HE 1104$-$1805 at $z=2.3$ with 3" separation, are of course resolved by Gaia and both image objects pass all the selection criteria individually, although with $G_{\rm RP}=(16.60,17.83)$ the two images of this object in particular are too faint for our main sample. Lensed or dual quasars with separations below 0.2" are marginally resolved by Gaia and likely included in our sample; \citet{2025ApJ...989..112G} and other works from the VODKA Collaboration, as well as the Cosmic Duets Program \citep{2026A&A...709A.236S} set examples for how to follow-up on these objects. It is the moderately resolved cases with large PPM signals that escape our present selection, but several other works are dedicated to discovering lensed quasars, including \citet{Le22}.

The blending of a quasar with another object near its line-of-sight tends to affect the catalogue colours of the resulting detection. While Gaia has a high spatial resolution, the WISE PSF in W1 and W2 is $\sim 6$" large and blending with a non-quasar reduces the quasar signature in W1$-$W2. This matters for faint quasars that are very close to bright stars. However, the sky surface density of quasars with $G_\mathrm{RP}<16$ is very low with 1 in every $\sim$17~deg$^2$ (from 2\,000 quasars in 34\,000~deg$^2$ at $|b|>10$). Thus, the probability of blending with a star that is at least one magnitude brighter is given by the density of stars with $G_\mathrm{RP}<15$, which is about 1 star per $\sim$60\,000~arcsec$^2$. 
We thus expect that we lost less than one quasar to blending with a brighter star in our search area. 

We further note, that technically the ubiquitous variability of quasars will affect the colours considered here as these are not observed simultaneously. While the magnitudes used here are mean values of multiple observations over several years (with survey mid-points in June 2014 and Dec 2015 for catWISE and Gaia DR3, respectively), the Gaia-WISE colour $G_\mathrm{RP}-$W1 has a mean time shift between the two passbands of $\sim$1.5~years.
Also, the variable brightness of quasars implies that some will cross the brightness threshold with time, and in the extreme case of changing-look AGN \citep[CLAGN;][]{RT23}, the spectral characteristics and colours change over time such that the quasar label would be used some of the time but not at other times. The completeness claim can thus only hold in terms of quasar states and magnitudes at the time of the Gaia and WISE observations used here.

Blending with fainter sources occurs preferentially in sky areas with a high density of faint stars such as in the direction of the Magellanic Clouds. While we have excluded these areas from our primary search, we are unsure about how complete we would recover the quasar population behind the Clouds.

\begin{figure*}
    \centering
    \includegraphics[width=0.75\columnwidth]{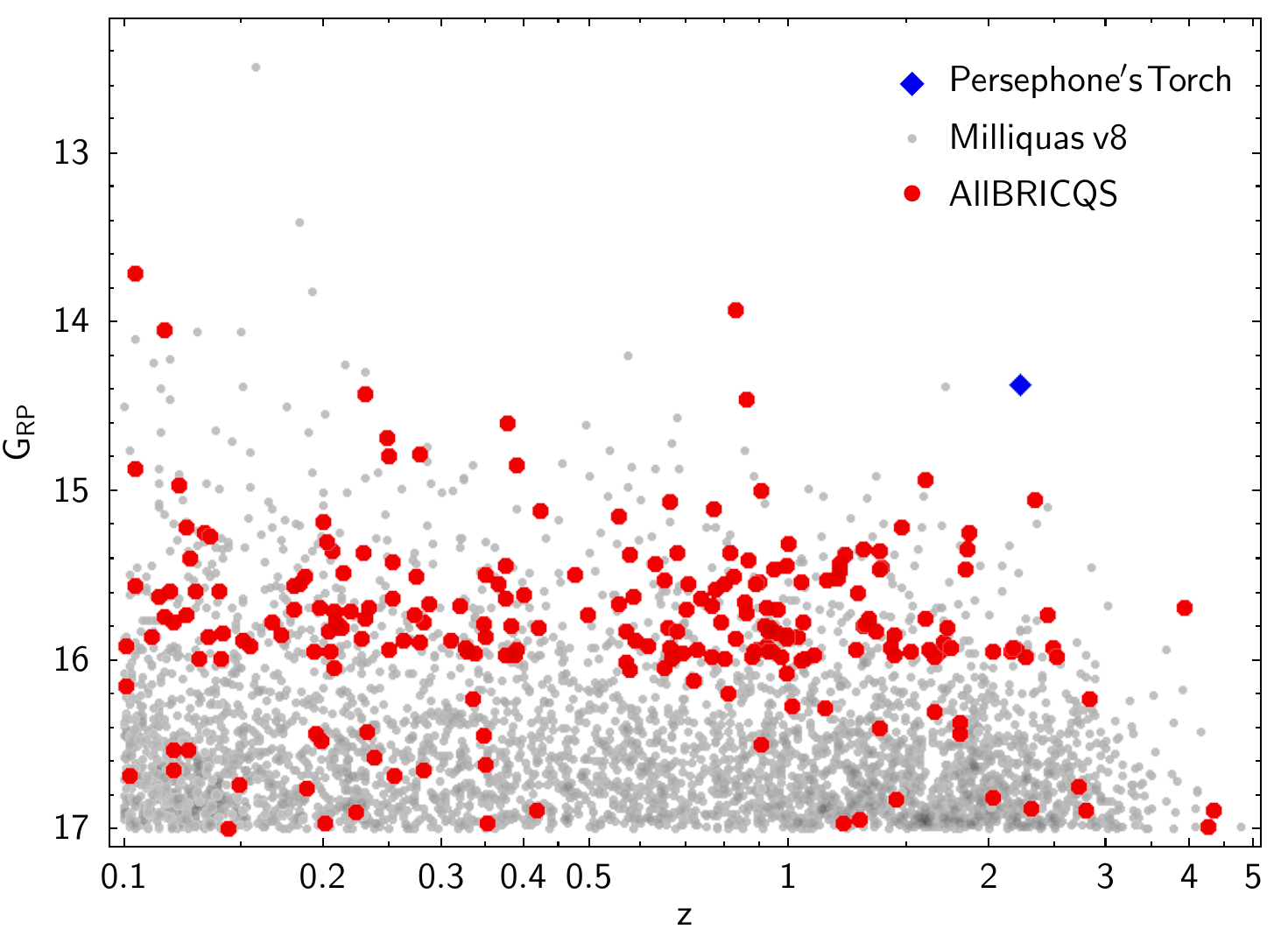} 
    \caption{Sample of quasars at declination $\delta<+20$, with $G_\mathrm{RP}<17$ and redshift $z>0.1$. Large round symbols are objects identified by AllBRICQS, including the first paper by \citet{On23}; while AllBRICQS aims to be complete to $G_\mathrm{RP}<16$, some fainter objects were used as filler targets. Some gravitationally lensed quasars are missed by AllBRICQS, e.g., the recently discovered Persephone's Torch. Small symbols are quasars in Milliquas v8 \citep{Flesch23} that are not AllBRICQS discoveries; 
    most of the previously known brightest Southern quasars are radio sources. }
    \label{fig:MQvsABQ_Rp17}
\end{figure*}

\section{Observations and data processing}

Optical spectra were obtained with the ANU 2.3m telescope \citep{Mathewson13} at Siding Spring Observatory (SSO) in Australia, using the Wide Field Spectrograph \cite[WiFeS;][]{Dop07,Dop10}. WiFeS is an image-slicing integral field spectrograph, with a field-of-view of $25\times38$~arcsec$^2$, typically with 1~arcsec$^2$ spaxels. With the $R\sim3000$ gratings B3000/R3000 and RT560 dichroic, the two-arm spectrograph covers the full optical range ($350-950$~nm). While the spatial resolution might provide information about the host galaxies for objects at low-luminosity and low-redshift, the main benefit of the integral field is avoiding slit losses, as the spectra can be extracted in large circular apertures and thus reflect correct spectral energy distributions (SEDs) at the time of observation. Brightness, colours or continuum slopes and emission-line properties are, of course, all known to vary in quasars.

Across the entire AllBRICQS project, observations were obtained between UT 2022-03-11 and 2026-08-14, during which time the 2.3m telescope operations were automated \citep{Price24}. Where possible, we used calibrations taken on the same night as the science data. However, when these were not available, the stability of the WiFeS instrument, located at one of the telescope's two Nasmyth foci, enables effective use of calibration data taken on nearby nights.


Data processing used version 2 of PyWiFeS\footnote{See \url{https://github.com/PyWiFeS/pywifes}.}, the WiFeS data pipeline \cite[][Onken et al., in prep.]{PyWiFeS}. PyWiFeS performs an automatic processing of calibration data that includes bad pixel masking, overscan subtraction, bias correction, internal lamp flatfields for pixel sensitivity correction, twilight flat fields for spatial sensitivity correction, arc lamps for wavelength correction, and standard star correction of throughput sensitivity. PyWiFeS v2 improves the wavelength and sensitivity calibration of the data cubes, and enables automated source detection and extraction, as well as splicing of the spectra between the blue and red arms of WiFeS.

Here, spectra were extracted using circular apertures with 2" to 3" radius depending on seeing conditions, while the PSF was on average $\sim2$~arcsec FWHM.
Some observations were obtained in nod-and-shuffle mode \citep{1994A&A...281..603C}, with 60-second sub-integrations at each offset position, which dramatically improves the sky subtraction of the data. The remainder of the data was taken in standard point-and-stare mode. For all data, sky subtraction was performed with either a circular source-free aperture near the quasar or an annulus centred on the quasar. Some objects in this new sample here have close neighbours, which we are trying to avoid extracting here, but in several cases the separation is less than 3" and even below 1", and in these cases the neighbour will clearly contaminate the target spectrum shown in the appendix. 

In a forthcoming paper, we will extract the spectra with wavelength-dependent PSF-weighted apertures to ensure best signal-to-noise and correct flux calibration across the full spectral range. These will then have spectral decomposition applied for tabulating emission-line properties.

\section{The Southern $G_{\rm RP}<16$ quasar sample}

Many bright quasars in the Southern sky known previously were originally discovered as radio sources, while the main large-area optical survey for bright quasars was the Hamburg-ESO Survey \citep{2000A&A...358...77W}. AllBRICQS has now added nearly all of the remaining, mostly radio-quiet, objects at $G_\mathrm{RP}<16$ and a few fainter filler targets. We list the properties of the newly confirmed AllBRICQS quasars in Table~\ref{tab:qsos_zgt01} for $z>0.1$ and Table~\ref{tab:qsos_zlt01} for $z<0.1$, both arranged in order of ascending RA. In the table comments, we indicate sources for which redshifts have been obtained in parallel by other efforts, including DESI DR1 \citep{DESIQSO23}. 
At $\delta<+20$, $|b|>10$, $G_\mathrm{RP}<16$ and $z>0.1$, there are now 745 quasars (or, more precisely, type-1 AGN) known, of which 185 are AllBRICQS discoveries. Fig.~\ref{fig:MQvsABQ_Rp17} shows the sample to a fainter limit of $G_\mathrm{RP}=17$, where 3\,855 quasars are known at $\delta<+20$, $|b|>10$, and $z>0.1$, including 230 AllBRICQS sources (of which 61 are also observed by SDSS DR20). 
%
%
Some noteworthy individual objects include:
\begin{itemize}
    \item The two brightest $z>0.1$ quasars known at Dec$<20^\circ$ (and indeed the whole sky), 3C~273 and PDS~456, are both at $z<0.2$ and have been known for over 60 years and $\sim$30 years, respectively.
    \item The two brightest $z>0.1$ quasars found by AllBRICQS are J0609$-$1735 at $z=0.104$ and J1144$-$4309 at $z=0.835$. The discovery of the latter, the most luminous quasar at $z<1.5$, was presented in \citet{On22} and triggered the AllBRICQS project.
    \item The highest-luminosity object discovered by AllBRICQS is J0529$-$4351, reported by \citet{On23} and independently discovered by QUBRICS \citep{Cristiani23}. Its VLTI/GRAVITY+ observations show the largest broad emission-line region in physical and angular size currently known \citep{Gravity26}.
    \item The highest redshift newly discovered object presented here is at $z=4.35$ (SMSS ID 308730116, $G_{\rm RP}=16.9$), which is is beyond our magnitude cut for complete follow-up. It is the second-highest redshift quasar at $G_{\rm RP}<17$ after the long known radio-loud object PMN~J1451-1512.
    \item The broadest emission line in the sample is observed in the quasar ESO 116-10 at $z=0.082$, whose double-peaked H$\alpha$ line has a peak separation of $\Delta(cz)\approx 12\,000$ km/s and FWHM $\approx 18\,000$ km/s \citep[see][]{Amrutha26}.
\end{itemize}

The quasar sample was split into two redshift ranges for further analysis: this paper presents the sample at $z>0.1$; the full spectral atlas with spectral decompositions in a public data repository will be presented in a forthcoming paper. Objects at $z<0.1$, newly discovered or previously known, were presented with their spectral decompositions in \citet{Amrutha26} and are available online already\footnote{See \url{https://www.mso.anu.edu.au/stromlo_agn/}}. 

Two of the $z<0.1$ candidates were found to be narrow-line AGN (see Table~\ref{tab:qsos_zlt01nl}). While it is entirely plausible that these are simple false positives in a selection aiming for completeness, there is also the possibility that these objects might have exhibited genuine type-1 spectra and colours at the time of the Gaia and WISE observations. Their Gaia XP spectra show strong emission signatures at the locations of H$\alpha$ and H$\beta$ suggesting a line equivalent width (EW) that is much higher than the narrow lines observed by us and in the case of J1830-4703 suggestive of EW$\approx 300\AA$, even larger than the median broad H$\alpha$ line in type-1 AGN. A caveat is that the low-resolution Gaia XP spectra might not be suitable for robustly measuring EWs. Our spectroscopy was obtained nearly a decade after the colours for selection were measured, and the high rate of changing-look AGN increases towards low luminosities; for the typical AGN population at $z<0.1$, \citet{Amrutha24} reported a turn-off fraction of 5\% per decade, which is certainly high enough to explain two objects.

Lastly, we consider candidates that were not proven to be AGN. First, we note that the candidate list is updated from the first release in \citet{On23} as some objects did not have the correct catWISE2020 data used in the process. This removes five of the six contaminating stars in the candidate list, which are now clearly outside the selection limits. We also change the selection rule for this final version, which has produced further candidates for follow-up. The latter change increases completeness at low luminosity and low redshift but deliberately accepts an increase in contamination from the stellar locus and non-AGN galaxies. Across the Southern AllBRICQS area, the candidate sample at $G_\mathrm{RP}<16$ has 1\,369 objects, of which 62 (or $<$5\%) do not have a quasar status; 39 of them are identified as stars on Simbad \citep{Simbad00} and 23 of them were followed up by us. Our follow-up includes six unknown sources with colours far from the stellar locus, which might be BL Lacs (one of them is identified as a BL Lac on Simbad), and close to the stellar locus we find five non-AGN low-redshift galaxies and twelve stars. Out of the 51 stars in total, 23 are in the sky area of the Orion-Eridanus superbubble \citep{OriEri09} and mostly kbnown as young stars, while the others are spread across the sky and are mostly long-period variables.



\section{Summary}

The All-sky BRIght Complete Quasar Survey (AllBRICQS) aimed at discovering the remaining unknown bright quasars in the sky at Galactic latitudes of $|b|>10\deg$ at nearly total completeness in the magnitude range of $G_{\rm RP}<16$ while adding fainter objects as filler targets. Here, we report 119 new quasars and conclude the Southern search at declination $<+20\deg$. The search area excludes a few small patches of sky amounting to less than 1\% of the original area, which are dominated by sources from nearby galaxies such as the Magellanic Clouds. Results of a separate Northern search were presented by \citet{Choi25}.

Given that the Southern sky was historically not well-covered by highly complete quasar surveys, the Southern search has been particularly fruitful with 270 new quasars identified, more than half of which were presented in the initial AllBRICQS paper by \citet{On23}. Given the Gaia-WISE colour cuts, the search focuses on redshifts of $z>0.1$, although $\sim 15$\% of the identified type-1 AGN are at lower redshifts of $z<0.1$. The median redshift of the Southern sample is $z\sim 0.5$, and the highest redshift quasar identified in this work is at $z=4.35$. 

This quasar sample is assumed to be complete in the search area with the exception of two types of objects, some of which are known to fail our selection criteria: (1) low-luminosity Seyfert galaxies at low redshift ($z<0.125$) may appear dominated by their host galaxies and thus not be evident AGN for our selection algorithm; and (2) some quasars that appear moderately extended in Gaia due to strong gravitational lensing or blending with another source at sub-arcsecond separation, including small-separation dual quasars. A small number of bright objects in our search area that were previously reported as quasars fail our colour selection but are here revealed as not being quasars after all.

Ultimately, the most complete and robust identification of quasars is provided by spectroscopic verification of quasar features together with their redshift measurement. The most desirable data resource for this purpose would be an all-sky spectroscopic resource. The Gaia XP spectra \citep{Montegriffo23}, e.g., have proven useful at identifying quasars \citep{Delchambre23}, but the spectral resolution is too low to disambiguate many objects. The new SphereX resource \citep{SphereX26} provides spectra constructed from imaging in 102 medium-band filters in the spirit of COMBO-17 \citep{W03}, MAHOROBA \citep{MAHOROBA04} and COSMOS \citep{Ilbert09} and takes this approach to a new level. SphereX data have already been used to find new quasars at $G_\mathrm{RP}>17.5$ and $z>4$ \citep{Davies26Sph}.

Some of the quasars, whether identified in this work or previously, are marginally resolved by Gaia and inside our selection limits; this implies that a fraction of them may be gravitationally lensed quasars, dual quasars or blended with another source near the line-of-sight. Careful inspection of their properties beyond data used in this paper reveals that some of these are radio-loud objects with known multi-wavelength jets. It is reasonable to assume that optical jets on sub-arcsecond scales may contribute to the extended appearance. We recommend that these objects are not used in BLR reverberation mapping studies, as their time lags and luminosities may be biased.
The forthcoming digital data release will be accompanied by a separate paper with spectral decomposition and estimates of continuum luminosities and emission-line properties.

\begin{acknowledgement}
We acknowledge the traditional owners of the land on which the ANU 2.3m telescope stands,
the Gamilaraay people, and pay our respects to their elders, past and present.
This paper is based on observations made with the Australian National University 2.3m Telescope at Siding Spring Observatory. 
We thank the WiFeS observers Katie Auchettl and Neelesh Amrutha for efforts in acquiring some spectra used in this paper.
The automation of the telescope was made possible through an initial grant provided by the Centre of Gravitational Astrophysics and the Research School of Astronomy and Astrophysics at the Australian National University and through a grant provided by the Australian Research Council through LE230100063. The Lens proposal system is maintained by the AAO Research Data \& Software team as part of the Data Central Science Platform. 
This research has made use of the SIMBAD database, operated at CDS, Strasbourg, France.
This work has made use of data from the European Space Agency (ESA) mission
{\it Gaia} (\url{https://www.cosmos.esa.int/gaia}), processed by the {\it Gaia}
Data Processing and Analysis Consortium (DPAC,
\url{https://www.cosmos.esa.int/web/gaia/dpac/consortium}). Funding for the DPAC
has been provided by national institutions, in particular the institutions
participating in the {\it Gaia} Multilateral Agreement. This work has made use of the {\sc Python} package \texttt{GaiaXPy}, developed and maintained by members of the Gaia Data Processing and Analysis Consortium (DPAC), and in particular, Coordination Unit 5 (CU5), and the Data Processing Centre located at the Institute of Astronomy, Cambridge, UK (DPCI).
This publication makes use of data products from the Wide-field Infrared Survey Explorer, which is a joint project of the University of California, Los Angeles, and the Jet Propulsion Laboratory/California Institute of Technology, and NEOWISE, which is a project of the Jet Propulsion Laboratory/California Institute of Technology. WISE and NEOWISE are funded by the National Aeronautics and Space Administration.

\end{acknowledgement}

\section*{Author Contribution Statement}

CW devised the AllBRICQS project with its selection rules and wrote the paper. CAO managed the observational program and the resulting data set and carried out the spectral data processing. All authors contributed to the observing proposals, the observations, and provided comments on the manuscript.

\clearpage

\onecolumn

\begin{longtable}{rrrccccl}
\caption{Newly identified bright quasars at $z>0.1$ ordered by RA. Some objects have been reported by DESI DR1 in the meantime; we also label objects previously observed by 6dFGS and update one redshift from \citet{Hon25}, and objects with close neighbours. The epoch is given in YYYYMMDD.} 
\label{tab:qsos_zgt01} \\
\hline
Gaia ID & RA & Dec & {$G_{\rm RP}$} & $b$ & Epoch & $z$ & Comment \\
\hline
4997246160302492800 &   5.5077 & $-$39.5159 & 16.476 & $-$76.18 & 20231130 & 0.198 & \\
2363775009866878848 &   6.5232 & $-$18.9152 & 15.839 & $-$79.93 & 20250720 & 0.141 & \\
2557794793048962816 &  10.4654 &  $+$8.2314 & 16.686 & $-$54.56 & 20231202 & 0.255 & DESI DR1 QSO (z=0.25456)\\
2352793912362096896 &  17.3088 & $-$20.7106 & 15.635 & $-$82.40 & 20260814 & 0.253 & \\
2354299727896266752 &  18.6082 & $-$20.1972 & 15.916 & $-$81.29 & 20250913 & 0.101 & \\
2455264101103968384 &  21.4515 & $-$14.3633 & 15.744 & $-$74.92 & 20250913 & 0.116 & DESI DR1 galaxy (z=0.116) at 4.7" offset\\
5008166788964165888 &  22.9331 & $-$40.6150 & 15.457 & $-$74.15 & 20260712 & 1.379 & \\
2511241921700772736 &  28.4575 &  $+$1.8799 & 15.599 & $-$57.33 & 20260712 & 0.128 & \\
5076554866023275520 &  43.8309 & $-$24.2667 & 15.859 & $-$62.01 & 20251001 & 0.110 & \\
4723537966228181376 &  46.1515 & $-$60.3482 & 15.105 & $-$49.84 & 20260814 & 0.770 & \\
4833899246842719616 &  51.8441 & $-$47.7192 & 15.221 & $-$53.39 & 20250715 & 0.124 & \\
3303572918621144320 &  62.9190 & $+$10.6699 & 15.704 & $-$28.49 & 20260823 & 0.180 & Star neighbour at 4" offset\\
4668888733634460032 &  63.8742 & $-$66.3291 & 15.562 & $-$39.97 & 20250715 & 0.104 & \\
3233227534771280128 &  74.1892 &  $+$3.7370 & 16.963 & $-$23.35 & 20241105 & 0.201 & Star neighbour at 3" offset\\
4783464888994718464 &  74.5651 & $-$52.0341 & 15.513 & $-$38.24 & 20230125 & 0.276 & \\
3401651962865377664 &  79.4489 & $+$19.2119 & 15.733 & $-$10.59 & 20230125 & 0.497 & \\
4772554057155329280 &  81.1671 & $-$51.0787 & 15.780 & $-$34.15 & 20230125 & 0.283 & Double-peaked [OIII] emission line\\
3223978756491532544 &  82.6201 &  $+$3.2334 & 15.790 & $-$16.35 & 20230125 & 0.348 & \\
4821279499056544128 &  82.6541 & $-$37.5029 & 16.228 & $-$31.43 & 20231203 & 2.825 & See also \citet{Scialpi26} \\ 
4660112947260694528 &  83.6841 & $-$67.6306 & 15.310 & $-$32.20 & 20220523 & 1.000 & SAGE0534AGN in \citet{Pennock22} \\ 
2972012094891415168 &  85.0961 & $-$15.7192 & 15.934 & $-$22.63 & 20230125 & 2.490 & \\
2963965834441633536 &  85.7600 & $-$22.7421 & 15.949 & $-$24.68 & 20230218 & 0.193 & \\
2914195787773870976 &  86.2417 & $-$25.9180 & 15.944 & $-$25.35 & 20230218 & 2.200 & \\
4756282247150094592 &  86.2739 & $-$64.1887 & 15.861 & $-$31.40 & 20230219 & 1.030 & \\
2891221801548638464 &  90.5821 & $-$30.5874 & 15.545 & $-$23.26 & 20230219 & 0.900 & \\
2943054669468180736 &  92.2645 & $-$17.5858 & 13.719 & $-$17.08 & 20260803 & 0.104 & See also \citet{OriEri09} \\
5568922701292781568 &  97.7837 & $-$43.4490 & 15.989 & $-$21.76 & 20241229 & 0.130 & \\
5569767847777314432 &  99.7300 & $-$41.0445 & 15.953 & $-$19.63 & 20230218 & 0.204 & \\
5579276493054921088 & 102.5784 & $-$35.3333 & 15.920 & $-$15.50 & 20230219 & 0.924 & \\
5505166316966035584 & 107.3712 & $-$50.5300 & 15.945 & $-$17.90 & 20230218 & 2.020 & \\
3166755670255692160 & 111.7075 & $+$14.1805 & 15.884 & $+$14.14 & 20230219 & 0.310 & \\
3168691016880413568 & 111.7237 & $+$16.0566 & 15.866 & $+$14.94 & 20230314 & 0.349 & Star neighbour at 3" offset\\
3169243624552708864 & 113.4660 & $+$17.2305 & 15.804 & $+$16.94 & 20230220 & 1.290 & \\
3039081956293667840 & 120.1745 &  $-$9.1183 & 15.891 & $+$10.91 & 20230218 & 1.705 & \\
5210440227979225216 & 123.4418 & $-$76.7601 & 15.986 & $-$21.85 & 20230218 & 2.270 &\\ 
5753659621333635584 & 128.7924 &  $-$8.5186 & 15.949 & $+$18.50 & 20230219 & 2.160 & \\
5733690256790128768 & 134.1787 & $-$15.1530 & 15.460 & $+$19.10 & 20241224 & 1.839 & 6dFGS AGN, redshift updated from $z=0.57$ \\
5248127642587863936 & 136.4086 & $-$66.7662 & 16.826 & $-$13.00 & 20231218 & 1.450 & \\
5218645337799040512 & 138.1331 & $-$72.9904 & 15.943 & $-$16.60 & 20230218 & 0.898 & \\
5243626826099437824 & 143.3145 & $-$69.6076 & 16.399 & $-$13.07 & 20231218 & 1.370 & Close neighbour at 0.5" offset\\
5664735415846236288 & 144.6925 & $-$21.9299 & 14.690 & $+$22.35 & 20241225 & 0.249 & 6dFGS AGN\\
5657302785962730880 & 148.0679 & $-$27.3476 & 15.955 & $+$20.58 & 20230218 & 1.680 & \\
5392438062695693056 & 158.7130 & $-$41.0565 & 15.806 & $+$14.82 & 20241227 & 0.658 & 6dFGS AGN\\
3477562009423846528 & 176.3351 & $-$32.9191 & 15.781 & $+$27.93 & 20231221 & 0.790 & Star neighbour at 1.3" offset\\
5380347832834307328 & 178.6570 & $-$41.7528 & 15.483 & $+$19.87 & 20250125 & 0.214 & 6dFGS AGN\\
3571109455146781440 & 182.4321 & $-$13.4874 & 16.374 & $+$48.14 & 20250125 & 1.810 & \\
3713158076758219136 & 202.3253 &  $+$4.1386 & 15.865 & $+$65.29 & 20250130 & 0.134 & DESI DR1 QSO (z=0.1342)\\
3620423307566344960 & 207.6557 &  $-$6.5376 & 15.365 & $+$53.41 & 20230218 & 0.229 & \\
6097617473029986048 & 212.4350 & $-$43.7051 & 16.751 & $+$16.92 & 20240410 & 2.720 & Star neighbour at 0.7" offset\\
6098678875410849664 & 220.8103 & $-$44.8604 & 16.886 & $+$13.62 & 20240403 & 4.350 & \\
6339227284852156032 & 224.0478 &  $-$3.2574 & 15.359 & $+$47.24 & 20260621 & 0.206 & \\
6255479923711772288 & 233.5082 & $-$17.2200 & 16.992 & $+$30.66 & 20240318 & 0.144 & \\
6008184365469774336 & 235.2632 & $-$39.9645 & 16.947 & $+$12.16 & 20240318 & 1.280 & \\
6009890395197284992 & 235.7628 & $-$36.7636 & 16.879 & $+$14.43 & 20240317 & 2.310 & Star neighbour at 2.6" offset\\
6016290033903149568 & 236.6031 & $-$32.1111 & 16.151 & $+$17.59 & 20240318 & 0.101 & \\
4488321539645489280 & 267.6126 &  $+$9.0600 & 16.740 & $+$17.63 & 20240408 & 0.149 & \\
4471290821790202368 & 270.8058 &  $+$4.8543 & 15.632 & $+$12.93 & 20250604 & 0.113 & \\
6649380995152426368 & 274.5538 & $-$55.7786 & 16.960 & $-$17.74 & 20240409 & 1.205 & Close pair of sources\\
6703279055149055616 & 275.1377 & $-$49.7102 & 16.889 & $-$15.69 & 20240418 & 0.418 & \\
6631168925066469248 & 278.3833 & $-$62.1643 & 16.620 & $-$21.76 & 20240413 & 0.350 & \\
6723453787853466240 & 279.1403 & $-$40.8884 & 16.437 & $-$14.76 & 20240516 & 1.800 & \\
6711767864732502656 & 284.4584 & $-$44.3957 & 16.434 & $-$19.67 & 20240417 & 0.195 & \\
6446290786497271296 & 290.0615 & $-$59.6736 & 16.888 & $-$26.73 & 20240515 & 2.800 & Star neighbour at 2" offset\\
6366763041740526848 & 298.3570 & $-$75.4514 & 16.764 & $-$30.07 & 20240507 & 0.189 & \\
4241989187948111616 & 298.7870 &  $+$3.2831 & 16.963 & $-$12.57 & 20240506 & 0.352 & \\
4236979091414475264 & 299.8961 &  $-$0.1234 & 16.899 & $-$15.17 & 20240512 & 0.223 & \\
6424163973984288640 & 301.8268 & $-$67.8611 & 15.782 & $-$32.09 & 20250617 & 0.119 & \\
6853627305918774400 & 303.0645 & $-$21.5761 & 16.278 & $-$26.97 & 20240517 & 1.010 & Star neighbour at 1" offset\\
4243318634937262848 & 304.3265 &  $+$2.4287 & 16.683 & $-$17.82 & 20240513 & 0.102 & \\
1809005457324343680 & 305.9086 & $+$17.0434 & 15.249 & $-$11.50 & 20250826 & 0.132 & \\
1804559204095620992 & 306.9724 & $+$14.1474 & 16.536 & $-$13.95 & 20240513 & 0.125 & \\
6374428974250236416 & 309.3359 & $-$71.4537 & 16.424 & $-$33.95 & 20240516 & 0.232 & \\
6454240874042761984 & 312.3595 & $-$61.6068 & 15.736 & $-$37.55 & 20250618 & 0.124 & \\
4227633070771957632 & 313.4180 &  $+$0.2700 & 16.816 & $-$26.76 & 20240516 & 2.020 & DESI DR1 QSO (z=2.01889)\\
6916153680330103552 & 315.5970 &  $-$2.5480 & 16.575 & $-$30.07 & 20240518 & 0.238 & \\
6776822123749931648 & 316.4526 & $-$36.5620 & 16.655 & $-$41.84 & 20240519 & 0.283 & \\
6892240569441090048 & 321.1449 & $-$11.5232 & 15.591 & $-$39.15 & 20250814 & 0.118 & \\
6783515499567624960 & 322.0791 & $-$31.8904 & 16.446 & $-$45.94 & 20240519 & 0.348 & \\
6613812167652033408 & 335.4524 & $-$30.8806 & 16.657 & $-$57.22 & 20240519 & 0.119 & \\
6379142718036753664 & 344.9484 & $-$74.0836 & 16.228 & $-$40.85 & 20231203 & 0.334 & \\
6557165642121205376 & 349.1586 & $-$30.4916 & 15.420 & $-$69.00 & 20260621 & 0.253 & \\
2821215827487699584 & 355.0920 & $+$18.6109 & 15.989 & $-$41.12 & 20250629 & 0.140 & \\
2769998720439761152 & 356.5646 & $+$12.7855 & 16.538 & $-$47.06 & 20240704 & 0.119 & DESI DR1 QSO (z=0.12023)\\
\hline
\end{longtable}

\begin{longtable}{rrrccccH}
\caption{Newly identified bright quasars at $z<0.1$ ordered by RA. Some objects have been reported by DESI DR1 in the meantime. The epoch is given in YYYYMMDD. These objects are analysed in detail in the spectral atlas by \citet{Amrutha26}.} 
\label{tab:qsos_zlt01} \\
\hline
Gaia ID & RA & Dec & {$G_{\rm RP}$} & $b$ & Epoch & $z$ & Comment \\
\hline
2562194974159162368 &  26.3897 &  $+$6.1904 & 15.785 & $-$54.20 & 20251001 & 0.0885 &  \\
2574811182892898432 &  29.3327 & $+$12.2507 & 15.369 & $-$47.49 & 20251001 & 0.0522 &  \\
  32630657549817216 &  43.6515 & $+$15.4289 & 15.555 & $-$38.00 & 20251001 & 0.0980 &  \\
  35735987624247936 &  45.4495 & $+$18.5682 & 15.639 & $-$34.44 & 20251001 & 0.0444 &  \\
4727308947514100736 &  47.5442 & $-$57.5116 & 15.446 & $-$51.05 & 20250716 & 0.0826 &  \\
3178283328120135296 &  68.8856 & $-$11.7405 & 15.698 & $-$35.43 & 20250715 & 0.0590 &  \\
4663228001099609984 &  70.1484 & $-$65.1302 & 15.892 & $-$38.02 & 20250715 & 0.0811 &  \\
4624111912867529344 &  76.4164 & $-$79.2003 & 15.716 & $-$31.47 & 20230125 & 0.0830 &  \\
3210353024284579840 &  79.2157 &  $-$4.4252 & 15.574 & $-$23.01 & 20251111 & 0.0942 &  \\
4791972738172468352 &  89.7974 & $-$53.4273 & 15.894 & $-$29.08 & 20250101 & 0.0790 &  \\
5590460656614152576 & 107.3835 & $-$35.6296 & 15.827 & $-$12.03 & 20241228 & 0.0307 &  \\
3155588858367280640 & 111.7287 &  $+$9.8346 & 15.409 & $+$12.30 & 20241229 & 0.0880 &  \\
3067175994508522752 & 121.3432 &  $-$6.0082 & 14.967 & $+$13.46 & 20241212 & 0.0990 &  \\
5740917243639422336 & 144.1945 &  $-$8.9503 & 15.971 & $+$30.64 & 20241227 & 0.0860 &  \\
5242563770154481408 & 145.9809 & $-$70.8281 & 15.604 & $-$13.35 & 20230218 & 0.0720 &  \\
5383331804314573056 & 169.6023 & $-$41.2681 & 15.767 & $+$18.28 & 20241231 & 0.0900 &  \\
5377665986541771776 & 177.4450 & $-$46.4643 & 15.996 & $+$15.09 & 20250505 & 0.0950 &  \\
6144691036149038208 & 183.2729 & $-$44.1354 & 15.427 & $+$18.19 & 20250223 & 0.0690 &  \\
6078410413651757824 & 192.7494 & $-$51.1269 & 15.717 & $+$11.74 & 20250505 & 0.0950 &  \\
6296194358245802752 & 214.3295 & $-$18.5001 & 15.895 & $+$39.84 & 20250503 & 0.0835 &  \\
6118645804714832896 & 216.4691 & $-$36.3584 & 15.894 & $+$22.73 & 20250913 & 0.0860 &  \\
6106685626462934016 & 218.5163 & $-$37.1103 & 15.704 & $+$21.38 & 20250913 & 0.0790 &  \\
6199896346751788160 & 228.9971 & $-$36.0427 & 15.902 & $+$18.23 & 20250617 & 0.0730 &  \\
4127689216182079360 & 255.9000 & $-$20.7889 & 15.157 & $+$12.43 & 20250616 & 0.0271 &  \\
4386899735239115904 & 258.0307 &  $+$1.3627 & 15.823 & $+$22.60 & 20260814 & 0.0594 &  \\
5954488134593658752 & 269.2794 & $-$45.2652 & 15.897 & $-$10.18 & 20250604 & 0.0701 &  \\
4467985174438222976 & 269.3139 &  $+$1.5769 & 16.483 & $+$12.76 & 20240408 & 0.0735 &  \\
4474366297548005120 & 270.2555 &  $+$5.2124 & 15.987 & $+$13.58 & 20250616 & 0.0875 &  \\
6706071161850699264 & 271.7447 & $-$50.0474 & 15.008 & $-$13.84 & 20250630 & 0.0500 &  \\
6714297501686651776 & 288.3627 & $-$40.5057 & 15.999 & $-$21.16 & 20250407 & 0.0590 &  \\
4236646008108570496 & 302.1287 &  $+$0.3593 & 15.637 & $-$16.90 & 20250506 & 0.0600 &  \\
1754118146098637952 & 306.5419 & $+$11.4911 & 15.462 & $-$15.03 & 20250704 & 0.0860 &  \\
6904608636440724096 & 308.0105 & $-$10.3921 & 15.915 & $-$27.01 & 20250716 & 0.0810 &  \\
6801636829880596352 & 313.2312 & $-$28.1277 & 15.740 & $-$37.72 & 20250716 & 0.0770 &  \\
6773969440832096896 & 314.9206 & $-$38.9785 & 15.558 & $-$40.85 & 20250716 & 0.0455 &  \\
2666892872242400384 & 328.4807 &  $-$7.1582 & 15.940 & $-$43.46 & 20250530 & 0.0592 &  \\
2596616800574068864 & 339.7580 & $-$16.0930 & 15.857 & $-$57.42 & 20250831 & 0.0726 &  \\
\hline
\end{longtable}

\begin{longtable}{rrrccccH}
\caption{Newly identified narrow-line AGN at $z<0.1$ that showed colour characteristics of broad-line AGN at the time of the Gaia and WISE observations, which are centred in time around 10 years prior to the optical spectra analysed here. They can be either CLAGN or simply contaminants in our selection.} 
\label{tab:qsos_zlt01nl} \\
\hline
Gaia ID & RA & Dec & {$G_{\rm RP}$} & $b$ & Epoch & $z$ & Comment \\
\hline
5551654179686639872 & 100.4075 & $-$47.9169 & 15.814 & $-$21.34 & 20241229 & 0.057 &  \\
6708181330821888384 & 277.6944 & $-$47.0513 & 16.116 & $-$16.21 & 20240412 & 0.084 &  \\
\hline
\end{longtable}

\clearpage
\twocolumn

%
%

\bibliography{bib}

@ARTICLE{OriEri09,
       author = {{Lee}, Hsu-Tai and {Chen}, W.~P.},
        title = "{Triggered Star Formation on the Border of the Orion-Eridanus Superbubble}",
      journal = {\apj},
         year = 2009,
        month = apr,
       volume = {694},
       number = {2},
        pages = {1423-1434},
          doi = {10.1088/0004-637X/694/2/1423},
archivePrefix = {arXiv},
       eprint = {astro-ph/0608216},
 primaryClass = {astro-ph},
       adsurl = {https://ui.adsabs.harvard.edu/abs/2009ApJ...694.1423L}
}

@ARTICLE{Scialpi26,
       author = {{Scialpi}, M. and {Mannucci}, F. and {D'Amato}, Q. and {Marconcini}, C. and {Cresci}, G. and {Marconi}, A. and {Ulivi}, L. and {Fumagalli}, M. and {Rosati}, P. and {Tozzi}, G. and {Zanchettin}, M.~V. and {Battistini}, L. and {Bertola}, E. and {Bracci}, C. and {Carniani}, S. and {Cataldi}, E. and {Ceci}, M. and {Chakraborty}, A. and {Cicone}, C. and {Ciurlo}, A. and {De Rosa}, A. and {Di Rosa}, G. and {Feltre}, A. and {Ginolfi}, M. and {Lamperti}, I. and {Lusso}, E. and {Moreschini}, B. and {Nardini}, E. and {Parvatikar}, M. and {Perna}, M. and {Rubinur}, K. and {Severgnini}, P. and {Singh}, J. and {Spingola}, C. and {Venturi}, G. and {Vignali}, C. and {Volonteri}, M.},
        title = "{Cosmic Duets: I. High-spatial resolution spectroscopy of dual and lensed active galactic nuclei with MUSE-NFM}",
      journal = {\aap},
         year = 2026,
        month = may,
       volume = {709},
          eid = {A236},
        pages = {A236},
          doi = {10.1051/0004-6361/202558548},
archivePrefix = {arXiv},
       eprint = {2512.11960},
 primaryClass = {astro-ph.GA},
       adsurl = {https://ui.adsabs.harvard.edu/abs/2026A&A...709A.236S}
}

@INPROCEEDINGS{MAHOROBA04,
       author = {{Taniguchi}, Yoshiaki},
        title = "{The Mahoroba Project --- Deep Survey with an Optical Intermediate-Band Filter System on the Subaru Telescope}",
    booktitle = {Studies of Galaxies in the Young Universe with New Generation Telescope},
         year = 2004,
       editor = {{Arimoto}, Nobuo and {Duschl}, Wolfgang J.},
        month = feb,
        pages = {107-111},
          doi = {10.48550/arXiv.astro-ph/0301097},
archivePrefix = {arXiv},
       eprint = {astro-ph/0301097},
 primaryClass = {astro-ph},
       adsurl = {https://ui.adsabs.harvard.edu/abs/2004sgyu.conf..107T}
}

@ARTICLE{Ilbert09,
       author = {{Ilbert}, O. and {Capak}, P. and {Salvato}, M. and {Aussel}, H. and {McCracken}, H.~J. and {Sanders}, D.~B. and {Scoville}, N. and {Kartaltepe}, J. and {Arnouts}, S. and {Le Floc'h}, E. and {Mobasher}, B. and {Taniguchi}, Y. and {Lamareille}, F. and {Leauthaud}, A. and {Sasaki}, S. and {Thompson}, D. and {Zamojski}, M. and {Zamorani}, G. and {Bardelli}, S. and {Bolzonella}, M. and {Bongiorno}, A. and {Brusa}, M. and {Caputi}, K.~I. and {Carollo}, C.~M. and {Contini}, T. and {Cook}, R. and {Coppa}, G. and {Cucciati}, O. and {de la Torre}, S. and {de Ravel}, L. and {Franzetti}, P. and {Garilli}, B. and {Hasinger}, G. and {Iovino}, A. and {Kampczyk}, P. and {Kneib}, J.-P. and {Knobel}, C. and {Kovac}, K. and {Le Borgne}, J.~F. and {Le Brun}, V. and {Le F{\`e}vre}, O. and {Lilly}, S. and {Looper}, D. and {Maier}, C. and {Mainieri}, V. and {Mellier}, Y. and {Mignoli}, M. and {Murayama}, T. and {Pell{\`o}}, R. and {Peng}, Y. and {P{\'e}rez-Montero}, E. and {Renzini}, A. and {Ricciardelli}, E. and {Schiminovich}, D. and {Scodeggio}, M. and {Shioya}, Y. and {Silverman}, J. and {Surace}, J. and {Tanaka}, M. and {Tasca}, L. and {Tresse}, L. and {Vergani}, D. and {Zucca}, E.},
        title = "{Cosmos Photometric Redshifts with 30-Bands for 2-deg$^{2}$}",
      journal = {\apj},
         year = 2009,
        month = jan,
       volume = {690},
       number = {2},
        pages = {1236-1249},
          doi = {10.1088/0004-637X/690/2/1236},
archivePrefix = {arXiv},
       eprint = {0809.2101},
 primaryClass = {astro-ph},
       adsurl = {https://ui.adsabs.harvard.edu/abs/2009ApJ...690.1236I}
}

@ARTICLE{SphereX26,
       author = {{Bock}, James J. and {Aboobaker}, Asad M. and {Adamo}, Joseph and {Akeson}, Rachel and {Alred}, John M. and {Alibay}, Farah and {Ashby}, Matthew L.~N. and {Bach}, Yoonsoo P. and {Bleem}, Lindsey E. and {Bolton}, Douglas and {Braun}, David F. and {Bruton}, Sean and {Bryan}, Sean A. and {Chang}, Tzu-Ching and {Chen}, Shuang-Shuang and {Cheng}, Yun-Ting and {Cheshire}, IV, James R. and {Chiang}, Yi-Kuan and {Choppin de Janvry}, Jean and {Condon}, Samuel and {Cook}, Walter R. and {Cooray}, Asantha and {Crill}, Brendan P. and {Cukierman}, Ari J. and {Dor{\'e}}, Olivier and {Dowell}, C. Darren and {Dubois-Felsmann}, Gregory P. and {Eifler}, Tim and {Everett}, Spencer and {Fabinsky}, Beth E. and {Faisst}, Andreas L. and {Fanson}, James L. and {Farrington}, Allen H. and {Fatahi}, Tamim and {Fazar}, Candice M. and {Feder}, Richard M. and {Frater}, Eric H. and {Grasshorn Gebhardt}, Henry S. and {Giri}, Utkarsh and {Goldina}, Tatiana and {Gorjian}, Varoujan and {Habib}, Salman and {Hart}, William G. and {Heinrich}, Chen and {Hora}, Joseph L. and {Huai}, Zhaoyu and {Hui}, Howard and {Jo}, Young-Soo and {Jeong}, Woong-Seob and {Kang}, Jae Hwan and {Kang}, Miju and {Kecman}, Branislav and {Kim}, Chul-Hwan and {Kim}, Jaeyeong and {Kim}, Minjin and {Kim}, Young-Jun and {Kim}, Yongjung and {Kirkpatrick}, J. Davy and {Kobayashi}, Yosuke and {Korngut}, Phil M. and {Krause}, Elisabeth and {Lee}, Bomee and {Lee}, Ho-Gyu and {Lee}, Jae-Joon and {Lee}, Jeong-Eun and {Lisse}, Carey M. and {Mariani}, Giacomo and {Masters}, Daniel C. and {Mauskopf}, Philip D. and {Melnick}, Gary J. and {Minasyan}, Mary H. and {Mirocha}, Jordan and {Miyasaka}, Hiromasa and {Moore}, Anne and {Moore}, Bradley D. and {Murgia}, Giulia and {Naylor}, Bret J. and {Nelson}, Christina and {Nguyen}, Chi H. and {Nguyen}, Hien T. and {Noh}, Jinyoung K. and {Padin}, Stephen and {Paladini}, Roberta and {Park}, Sung-Joon and {Penanen}, Konstantin I. and {Putnam}, Dustin S. and {Pyo}, Jeonghyun and {Ramachandra}, Nesar and {Ramanathan}, Keshav and {Rustamkulov}, Zafar and {Reiley}, Daniel J. and {Rice}, Eric B. and {Rocca}, Jennifer M. and {Seok}, Ji Yeon and {Smith}, Roger and {Stober}, Jeremy and {Susca}, Sara and {Teplitz}, Harry I. and {Thelen}, Michael P. and {Tolls}, Volker and {Torrini}, Gabriela and {Trangsrud}, Amy R. and {Unwin}, Stephen and {Velicheti}, Phani and {Wang}, Pao-Yu and {Wen}, Robin Y. and {Werner}, Michael W. and {Williams}, Abby E. and {Williamson}, Ross and {Wincentsen}, James and {Windhorst}, Rogier A. and {Yang}, Soung-Chul and {Yang}, Yujin and {Zemcov}, Michael},
        title = "{The SPHEREx Satellite Mission}",
      journal = {\apj},
         year = 2026,
        month = mar,
       volume = {999},
       number = {1},
          eid = {139},
        pages = {139},
          doi = {10.3847/1538-4357/ae2be2},
archivePrefix = {arXiv},
       eprint = {2511.02985},
 primaryClass = {astro-ph.IM},
       adsurl = {https://ui.adsabs.harvard.edu/abs/2026ApJ...999..139B}
}

@ARTICLE{Davies26Sph,
       author = {{Davies}, Frederick B. and {Bosman}, Sarah E.~I. and {Ganguly}, Arpita and {Ba{\~n}ados}, Eduardo and {Belladitta}, Silvia and {Stern}, Daniel and {Acevedo Barroso}, Javier A. and {Yang}, Daming and {Hennawi}, Joseph F. and {Wang}, Feige and {Yang}, Jinyi and {Fan}, Xiaohui},
        title = "{Three Hundred Quasars from the Couch: A first look at high-redshift quasar discovery with SPHEREx}",
      journal = {arXiv e-prints},
         year = 2026,
        month = mar,
          eid = {arXiv:2603.10135},
        pages = {arXiv:2603.10135},
          doi = {10.48550/arXiv.2603.10135},
archivePrefix = {arXiv},
       eprint = {2603.10135},
 primaryClass = {astro-ph.GA},
       adsurl = {https://ui.adsabs.harvard.edu/abs/2026arXiv260310135D}
}

@ARTICLE{Hon25,
       author = {{Hon}, Wei Jeat and {Webster}, Rachel L. and {Wolf}, Christian},
        title = "{Broad-line active galactic nuclei in the 6dF Galaxy Survey}",
      journal = {\mnras},
         year = 2025,
        month = feb,
       volume = {536},
       number = {4},
        pages = {3611-3630},
          doi = {10.1093/mnras/stae2815},
archivePrefix = {arXiv},
       eprint = {2410.08536},
 primaryClass = {astro-ph.GA},
       adsurl = {https://ui.adsabs.harvard.edu/abs/2025MNRAS.536.3611H}
}

@ARTICLE{Amrutha24,
       author = {{Amrutha}, Neelesh and {Wolf}, Christian and {Onken}, Christopher A. and {Hon}, Wei Jeat and {Lai}, Samuel and {Tonry}, John L. and {Webster}, Rachel},
        title = "{Discovering changing-look AGN in the 6dF Galaxy Survey using ATLAS light curves}",
      journal = {\mnras},
         year = 2024,
        month = dec,
       volume = {535},
       number = {3},
        pages = {2322-2340},
          doi = {10.1093/mnras/stae2470},
archivePrefix = {arXiv},
       eprint = {2410.22671},
 primaryClass = {astro-ph.GA},
       adsurl = {https://ui.adsabs.harvard.edu/abs/2024MNRAS.535.2322A}
}

@ARTICLE{Amrutha26,
       author = {{Amrutha}, Neelesh and {Wolf}, Christian and {Onken}, Christopher A. and {Jeat Hon}, Wei and {Lai}, Samuel and {Webster}, Rachel},
        title = "{OzSSy1: The Australian Southern Seyfert-1 Spectroscopic Atlas and Catalogue at z < 0.1}",
      journal = {arXiv e-prints},
         year = 2026,
        month = jun,
          eid = {arXiv:2606.27675},
        pages = {arXiv:2606.27675},
          doi = {10.48550/arXiv.2606.27675},
archivePrefix = {arXiv},
       eprint = {2606.27675},
 primaryClass = {astro-ph.GA},
       adsurl = {https://ui.adsabs.harvard.edu/abs/2026arXiv260627675A}
}

@ARTICLE{Cristiani23,
       author = {{Cristiani}, Stefano and {Porru}, Matteo and {Guarneri}, Francesco and {Calderone}, Giorgio and {Boutsia}, Konstantina and {Grazian}, Andrea and {Cupani}, Guido and {D'Odorico}, Valentina and {Fontanot}, Fabio and {Martins}, Carlos J.~A.~P. and {Marques}, Catarina M.~J. and {Maitra}, Soumak and {Trost}, Andrea},
        title = "{Spectroscopy of QUBRICS quasar candidates: 1672 new redshifts and a golden sample for the Sandage test of the redshift drift}",
      journal = {\mnras},
         year = 2023,
        month = jun,
       volume = {522},
       number = {2},
        pages = {2019-2028},
          doi = {10.1093/mnras/stad1007},
archivePrefix = {arXiv},
       eprint = {2304.00362},
 primaryClass = {astro-ph.CO},
       adsurl = {https://ui.adsabs.harvard.edu/abs/2023MNRAS.522.2019C}
}

@ARTICLE{Pennock22,
       author = {{Pennock}, Clara M. and {van Loon}, Jacco Th and {Anih}, Joy O. and {Maitra}, Chandreyee and {Haberl}, Frank and {Sansom}, Anne E. and {Ivanov}, Valentin D. and {Cowley}, Michael J. and {Afonso}, Jos{\'e} and {Ant{\'o}n}, Sonia and {Cioni}, Maria-Rosa L. and {Craig}, Jessica E.~M. and {Filipovi{\'c}}, Miroslav D. and {Hopkins}, Andrew M. and {Nanni}, Ambra and {Prandoni}, Isabella and {Vardoulaki}, Eleni},
        title = "{The VMC survey - XLIX. Discovery of a population of quasars dominated by nuclear dust emission behind the Magellanic Clouds}",
      journal = {\mnras},
         year = 2022,
        month = oct,
       volume = {515},
       number = {4},
        pages = {6046-6065},
          doi = {10.1093/mnras/stac2096},
archivePrefix = {arXiv},
       eprint = {2207.12301},
 primaryClass = {astro-ph.GA},
       adsurl = {https://ui.adsabs.harvard.edu/abs/2022MNRAS.515.6046P}
}

@ARTICLE{2026arXiv260413152D,
       author = {{Davies}, Frederick B. and {Ba{\~n}ados}, Eduardo and {Bosman}, Sarah E.~I. and {Ganguly}, Arpita and {Belladitta}, Silvia and {Power}, Jennifer and {Rees}, Jon},
        title = "{Persephone's Torch: A 15th Magnitude Quadruply-Lensed Quasar From the Couch Discovered with SPHEREx and the LBT}",
      journal = {arXiv e-prints},
         year = 2026,
        month = apr,
          eid = {arXiv:2604.13152},
        pages = {arXiv:2604.13152},
          doi = {10.48550/arXiv.2604.13152},
archivePrefix = {arXiv},
       eprint = {2604.13152},
 primaryClass = {astro-ph.GA},
       adsurl = {https://ui.adsabs.harvard.edu/abs/2026arXiv260413152D}
}

@ARTICLE{2025ApJ...989..112G,
       author = {{Gross}, Arran C. and {Chen}, Yu-Ching and {Oguri}, Masamune and {Nolan}, Liam and {Liu}, Xin and {Shen}, Yue and {Zhuang}, Ming-Yang and {Li}, Junyao and {Zakamska}, Nadia L. and {Hwang}, Hsiang-Chih and {Ishikawa}, Yuzo},
        title = "{Varstrometry for Off-nucleus and Dual Sub-Kpc AGN (VODKA): A Mix of Singles, Lenses, and True Duals at Cosmic Noon}",
      journal = {\apj},
         year = 2025,
        month = aug,
       volume = {989},
       number = {1},
          eid = {112},
        pages = {112},
          doi = {10.3847/1538-4357/ade671},
archivePrefix = {arXiv},
       eprint = {2409.16363},
 primaryClass = {astro-ph.GA},
       adsurl = {https://ui.adsabs.harvard.edu/abs/2025ApJ...989..112G}
}

@ARTICLE{sdss_dr20_2026,
       author = {{SDSS Collaboration} and others},
        title = "{The Twentieth Data Release of the Sloan Digital Sky Survey: First All-Sky BOSS Spectra, eROSITA-SDSS-V Mapper Coordinated Observations, and a Preview of the Local Volume Mapper}",
      journal = {arXiv e-prints},
         year = 2026,
        month = jul,
          eid = {arXiv:2607.26149},
        pages = {arXiv:2607.26149},
          doi = {10.48550/arXiv.2607.26149},
archivePrefix = {arXiv},
       eprint = {2607.26149},
 primaryClass = {astro-ph.GA},
       adsurl = {https://harvard.edu}
}

@ARTICLE{Mathewson13,
       author = {{Mathewson}, Don S. and {Hart}, John and {Wehner}, Hermann P. and {Hovey}, Gary R. and {van Harmelen}, Jan},
        title = "{The Australian National University's 2.3m New Generation Telescope at Siding Spring Observatory}",
      journal = {Journal of Astronomical History and Heritage},
         year = 2013,
        month = mar,
       volume = {16},
       number = {1},
        pages = {2-28},
       adsurl = {https://ui.adsabs.harvard.edu/abs/2013JAHH...16....2M}
}

@ARTICLE{Dop07,
       author = {{Dopita}, Michael and {Hart}, John and {McGregor}, Peter and {Oates}, Patrick and {Bloxham}, Gabe and {Jones}, Damien},
        title = "{The Wide Field Spectrograph (WiFeS)}",
      journal = {\apss},
         year = 2007,
        month = aug,
       volume = {310},
       number = {3-4},
        pages = {255-268},
          doi = {10.1007/s10509-007-9510-z},
archivePrefix = {arXiv},
       eprint = {0705.0287},
 primaryClass = {astro-ph},
       adsurl = {https://ui.adsabs.harvard.edu/abs/2007Ap&SS.310..255D}
}

@ARTICLE{Dop10,
       author = {{Dopita}, Michael and {Rhee}, Jonghwan and {Farage}, Catherine and {McGregor}, Peter and {Bloxham}, Gabe and {Green}, Anthony and {Roberts}, Bill and {Neilson}, Jon and {Wilson}, Greg and {Young}, Peter and {Firth}, Peter and {Busarello}, Gianni and {Merluzzi}, Paola},
        title = "{The Wide Field Spectrograph (WiFeS): performance and data reduction}",
      journal = {\apss},
         year = 2010,
        month = jun,
       volume = {327},
       number = {2},
        pages = {245-257},
          doi = {10.1007/s10509-010-0335-9},
archivePrefix = {arXiv},
       eprint = {1002.4472},
 primaryClass = {astro-ph.IM},
       adsurl = {https://ui.adsabs.harvard.edu/abs/2010Ap&SS.327..245D}
}

@ARTICLE{Price24,
       author = {{Price}, Ian and {Nielsen}, Jon and {Lidman}, Chris and {Soon}, Jamie and {Travouillon}, Tony and {Sharp}, Rob},
        title = "{Converting the ANU 2.3 telescope to fully automated operation}",
      journal = {\pasa},
         year = 2024,
        month = sep,
       volume = {41},
          eid = {e057},
        pages = {e057},
          doi = {10.1017/pasa.2024.75},
archivePrefix = {arXiv},
       eprint = {2409.19842},
 primaryClass = {astro-ph.IM},
       adsurl = {https://ui.adsabs.harvard.edu/abs/2024PASA...41...57P}
}

@ARTICLE{PyWiFeS,
       author = {{Childress}, Michael J. and {Vogt}, Fr{\'e}d{\'e}ric P.~A. and {Nielsen}, Jon and {Sharp}, Robert G.},
        title = "{PyWiFeS: a rapid data reduction pipeline for the Wide Field Spectrograph (WiFeS)}",
      journal = {\apss},
         year = 2014,
        month = feb,
       volume = {349},
       number = {2},
        pages = {617-636},
          doi = {10.1007/s10509-013-1682-0},
archivePrefix = {arXiv},
       eprint = {1311.2666},
 primaryClass = {astro-ph.IM},
       adsurl = {https://ui.adsabs.harvard.edu/abs/2014Ap&SS.349..617C}
}

@ARTICLE{Gravity26,
       author = {{GRAVITY+ Collaboration} and {Abd El Dayem}, K. and {Aimar}, N. and {Berdeu}, A. and {Berger}, J.-P. and {Bourdarot}, G. and {Bourget}, P. and {Brandner}, W. and {Cao}, Y. and {Correia}, C. and {Cuevas Cardona}, S. and {Davies}, R. and {Defr{\`e}re}, D. and {Drescher}, A. and {Eckart}, A. and {Eisenhauer}, F. and {Fabricius}, M. and {Farah}, A. and {Feuchtgruber}, H. and {F{\"o}rster Schreiber}, N.~M. and {Foschi}, A. and {Garcia}, P. and {Garcia Lopez}, R. and {Genzel}, R. and {Gillessen}, S. and {Gomes}, T. and {Gont{\'e}}, F. and {Gopinath}, V. and {Graf}, J. and {Hartl}, M. and {Haubois}, X. and {Hau{\ss}mann}, F. and {Ho}, L.~C. and {H{\"o}nig}, S. and {Houll{\'e}}, M. and {Joharle}, S. and {Keiman}, C. and {Kervella}, P. and {Kolb}, J. and {Kreidberg}, L. and {Labdon}, A. and {Lacour}, S. and {Lai}, O. and {Lai}, S. and {Laugier}, R. and {Le Bouquin}, J.-B. and {Leftley}, J. and {Li}, R. and {Lopez}, B. and {Lutz}, D. and {Mang}, F. and {M{\'e}rand}, A. and {Millour}, F. and {Montarg{\`e}s}, M. and {More}, N. and {Moruj{\~a}o}, N. and {Nowacki}, H. and {Nowak}, M. and {Oberti}, S. and {Onken}, C. and {Osorno}, J. and {Ott}, T. and {Paumard}, T. and {Perraut}, K. and {Perrin}, G. and {Petrov}, R. and {Petrucci}, P.-O. and {Pourr{\'e}}, N. and {Rabien}, S. and {Rau}, C. and {Ribeiro}, D.~C. and {Robbe-Dubois}, S. and {Sadun Bordoni}, M. and {Salman}, M. and {Sanchez-Bermudez}, J. and {Santos}, D. and {Sauter}, J. and {Scialpi}, M. and {Scigliuto}, J. and {Shangguan}, J. and {Shchekaturov}, P. and {Shimizu}, T. and {Soulez}, F. and {Straubmeier}, C. and {Sturm}, E. and {Subroweit}, M. and {Sykes}, C. and {Tacconi}, L.~J. and {{\"U}bler}, H. and {Ulbricht}, G. and {Vincent}, F. and {Webster}, R. and {Wieprecht}, E. and {Woillez}, J. and {Wolf}, C.},
        title = "{Spatially resolved broad-line region in a quasar at z = 4: Dynamical black hole mass and prominent outflow}",
      journal = {\aap},
         year = 2026,
        month = feb,
       volume = {706},
          eid = {A99},
        pages = {A99},
          doi = {10.1051/0004-6361/202557285},
archivePrefix = {arXiv},
       eprint = {2509.13911},
 primaryClass = {astro-ph.GA},
       adsurl = {https://ui.adsabs.harvard.edu/abs/2026A&A...706A..99G}
}

@ARTICLE{Gravity17,
       author = {{GRAVITY Collaboration} and {Abuter}, R. and {Accardo}, M. and {Amorim}, A. and {Anugu}, N. and {{\'A}vila}, G. and {Azouaoui}, N. and {Benisty}, M. and {Berger}, J.~P. and {Blind}, N. and {Bonnet}, H. and {Bourget}, P. and {Brandner}, W. and {Brast}, R. and {Buron}, A. and {Burtscher}, L. and {Cassaing}, F. and {Chapron}, F. and {Choquet}, {\'E}. and {Cl{\'e}net}, Y. and {Collin}, C. and {Coud{\'e} Du Foresto}, V. and {de Wit}, W. and {de Zeeuw}, P.~T. and {Deen}, C. and {Delplancke-Str{\"o}bele}, F. and {Dembet}, R. and {Derie}, F. and {Dexter}, J. and {Duvert}, G. and {Ebert}, M. and {Eckart}, A. and {Eisenhauer}, F. and {Esselborn}, M. and {F{\'e}dou}, P. and {Finger}, G. and {Garcia}, P. and {Garcia Dabo}, C.~E. and {Garcia Lopez}, R. and {Gendron}, E. and {Genzel}, R. and {Gillessen}, S. and {Gonte}, F. and {Gordo}, P. and {Grould}, M. and {Gr{\"o}zinger}, U. and {Guieu}, S. and {Haguenauer}, P. and {Hans}, O. and {Haubois}, X. and {Haug}, M. and {Haussmann}, F. and {Henning}, Th. and {Hippler}, S. and {Horrobin}, M. and {Huber}, A. and {Hubert}, Z. and {Hubin}, N. and {Hummel}, C.~A. and {Jakob}, G. and {Janssen}, A. and {Jochum}, L. and {Jocou}, L. and {Kaufer}, A. and {Kellner}, S. and {Kendrew}, S. and {Kern}, L. and {Kervella}, P. and {Kiekebusch}, M. and {Klein}, R. and {Kok}, Y. and {Kolb}, J. and {Kulas}, M. and {Lacour}, S. and {Lapeyr{\`e}re}, V. and {Lazareff}, B. and {Le Bouquin}, J.-B. and {L{\`e}na}, P. and {Lenzen}, R. and {L{\'e}v{\^e}que}, S. and {Lippa}, M. and {Magnard}, Y. and {Mehrgan}, L. and {Mellein}, M. and {M{\'e}rand}, A. and {Moreno-Ventas}, J. and {Moulin}, T. and {M{\"u}ller}, E. and {M{\"u}ller}, F. and {Neumann}, U. and {Oberti}, S. and {Ott}, T. and {Pallanca}, L. and {Panduro}, J. and {Pasquini}, L. and {Paumard}, T. and {Percheron}, I. and {Perraut}, K. and {Perrin}, G. and {Pfl{\"u}ger}, A. and {Pfuhl}, O. and {Phan Duc}, T. and {Plewa}, P.~M. and {Popovic}, D. and {Rabien}, S. and {Ram{\'\i}rez}, A. and {Ramos}, J. and {Rau}, C. and {Riquelme}, M. and {Rohloff}, R.-R. and {Rousset}, G. and {Sanchez-Bermudez}, J. and {Scheithauer}, S. and {Sch{\"o}ller}, M. and {Schuhler}, N. and {Spyromilio}, J. and {Straubmeier}, C. and {Sturm}, E. and {Suarez}, M. and {Tristram}, K.~R.~W. and {Ventura}, N. and {Vincent}, F. and {Waisberg}, I. and {Wank}, I. and {Weber}, J. and {Wieprecht}, E. and {Wiest}, M. and {Wiezorrek}, E. and {Wittkowski}, M. and {Woillez}, J. and {Wolff}, B. and {Yazici}, S. and {Ziegler}, D. and {Zins}, G.},
        title = "{First light for GRAVITY: Phase referencing optical interferometry for the Very Large Telescope Interferometer}",
      journal = {\aap},
         year = 2017,
        month = jun,
       volume = {602},
          eid = {A94},
        pages = {A94},
          doi = {10.1051/0004-6361/201730838},
archivePrefix = {arXiv},
       eprint = {1705.02345},
 primaryClass = {astro-ph.IM},
       adsurl = {https://ui.adsabs.harvard.edu/abs/2017A&A...602A..94G}
}

@ARTICLE{Gravity26LGSAO,
       author = {{Gravity+ Collaboration} and {Abuter}, R. and {Allouche}, F. and {Bailet}, C. and {Benisty}, M. and {Berdeu}, A. and {Berger}, J.-P. and {Berio}, P. and {Bigioli}, A. and {Blanchard}, C. and {Boebion}, O. and {Bonnet}, H. and {Bourdarot}, G. and {Bourget}, P. and {Brandner}, W. and {Brul{\'e}}, J. and {Burgos}, P. and {Carbillet}, M. and {Correia}, C. and {Courtney-Barrer}, B. and {Curaba}, S. and {Davies}, R. and {Defr{\`e}re}, D. and {Delboulb{\'e}}, A. and {Delplancke}, F. and {Dembet}, R. and {Drescher}, A. and {Dubost}, N. and {Eckart}, A. and {{\'E}douard}, C. and {Eisenhauer}, F. and {Esteras Otal}, L. and {Fabricius}, M. and {Feuchtgruber}, H. and {F{\'e}dou}, P. and {Finger}, G. and {Schreiber}, N.~M. F{\"o}rster and {Frahm}, R. and {Garcia}, E. and {Garcia}, P. and {Lopez}, R. Garcia and {Genzel}, R. and {Gil}, J.~P. and {Gillessen}, S. and {Gomes}, T. and {Gont{\'e}}, F. and {Gopinath}, V. and {Gouvret}, C. and {Graf}, J. and {Guajardo}, P. and {Guieu}, S. and {Hackenberg}, W. and {Hartl}, M. and {Haubois}, X. and {Hau{\ss}mann}, F. and {Henning}, T. and {Hibon}, P. and {H{\"o}nig}, S. and {Horrobin}, M. and {Houll{\'e}}, M. and {Hubin}, N. and {Taieb}, I. Ibn and {Jochum}, L. and {Jocou}, L. and {Jost}, A. and {Kammerer}, J. and {Karl}, L. and {Kaufer}, A. and {Kern}, P. and {Kervella}, P. and {Kolb}, J. and {Korhonen}, H. and {Kreidberg}, L. and {Krempl}, P. and {Lacour}, S. and {Lagarde}, S. and {Lai}, O. and {Lapeyr{\`e}re}, V. and {Laugier}, R. and {Leal}, V. and {Le Bouquin}, J.-B. and {Leftley}, J. and {L{\'e}na}, P. and {Lopez}, B. and {Lutz}, D. and {Magnard}, Y. and {Mang}, F. and {Marcotto}, A. and {Maurel}, D. and {M{\'e}rand}, A. and {Millour}, F. and {Montarges}, M. and {More}, N. and {Moruj{\~a}o}, N. and {Moulin}, T. and {Nowacki}, H. and {Nowak}, M. and {Oberti}, S. and {Ott}, T. and {Pallanca}, L. and {Patru}, F. and {Paumard}, T. and {Perraut}, K. and {Perrin}, G. and {Petrucci}, P.~O. and {Petrov}, R. and {Pfuhl}, O. and {Pourr{\'e}}, N. and {Rabien}, S. and {Rau}, C. and {Riquelme}, M. and {Robbe-Dubois}, S. and {Rochat}, S. and {Salman}, M. and {S{\'a}nchez-Berm{\'u}dez}, J. and {Schubert}, J. and {Scigliuto}, J. and {Shchekaturov}, P. and {Schuhler}, N. and {Shangguan}, J. and {Shimizu}, T. and {Scheithauer}, S. and {Soenke}, C. and {Soulez}, F. and {Stadler}, E. and {Stadler}, J. and {Straubmeier}, C. and {Sturm}, E. and {Subroweit}, M. and {Sykes}, C. and {Tacconi}, L.~J. and {Tristram}, K.~R.~W. and {Uysal}, S. and {von Fellenberg}, S. and {Widmann}, F. and {Wieprecht}, E. and {Wiezorrek}, E. and {Woillez}, J. and {Yazici}, S. and {Zins}, G.},
        title = "{First light for the GRAVITY+ Adaptive Optics: Extreme adaptive optics for the Very Large Telescope Interferometer}",
      journal = {\aap},
         year = 2026,
        month = mar,
       volume = {707},
          eid = {A115},
        pages = {A115},
          doi = {10.1051/0004-6361/202555666},
archivePrefix = {arXiv},
       eprint = {2509.21431},
 primaryClass = {astro-ph.IM},
       adsurl = {https://ui.adsabs.harvard.edu/abs/2026A&A...707A.115G}
}

@ARTICLE{Wolf24,
       author = {{Wolf}, Christian and {Lai}, Samuel and {Onken}, Christopher A. and {Amrutha}, Neelesh and {Bian}, Fuyan and {Hon}, Wei Jeat and {Tisserand}, Patrick and {Webster}, Rachel L.},
        title = "{The accretion of a solar mass per day by a 17-billion solar mass black hole}",
      journal = {Nature Astronomy},
         year = 2024,
        month = apr,
       volume = {8},
        pages = {520-529},
          doi = {10.1038/s41550-024-02195-x},
archivePrefix = {arXiv},
       eprint = {2402.15101},
 primaryClass = {astro-ph.CO},
       adsurl = {https://ui.adsabs.harvard.edu/abs/2024NatAs...8..520W}
}

@ARTICLE{On22,
       author = {{Onken}, Christopher A. and {Lai}, Samuel and {Wolf}, Christian and {Lucy}, Adrian B. and {Hon}, Wei Jeat and {Tisserand}, Patrick and {Sokoloski}, Jennifer L. and {Luna}, Gerardo J.~M. and {Manick}, Rajeev and {Fan}, Xiaohui and {Bian}, Fuyan},
        title = "{Discovery of the most luminous quasar of the last 9 Gyr}",
      journal = {\pasa},
         year = 2022,
        month = jan,
       volume = {39},
          eid = {e037},
        pages = {e037},
          doi = {10.1017/pasa.2022.36},
archivePrefix = {arXiv},
       eprint = {2206.04204},
 primaryClass = {astro-ph.GA},
       adsurl = {https://ui.adsabs.harvard.edu/abs/2022PASA...39...37O}
}

@ARTICLE{On23,
       author = {{Onken}, Christopher A. and {Wolf}, Christian and {Hon}, Wei Jeat and {Lai}, Samuel and {Tisserand}, Patrick and {Webster}, Rachel},
        title = "{AllBRICQS: The All-sky BRIght, Complete Quasar Survey}",
      journal = {\pasa},
         year = 2023,
        month = mar,
       volume = {40},
          eid = {e010},
        pages = {e010},
          doi = {10.1017/pasa.2023.7},
archivePrefix = {arXiv},
       eprint = {2209.09342},
 primaryClass = {astro-ph.GA},
       adsurl = {https://ui.adsabs.harvard.edu/abs/2023PASA...40...10O}
}

@ARTICLE{PDS456_97,
       author = {{Torres}, Carlos A.~O. and {Quast}, Germano R. and {Coziol}, Roger and {Jablonski}, Francisco and {de la Reza}, Ramiro and {L{\'e}pine}, J.~R.~D. and {Greg{\'o}rio-Hetem}, J.},
        title = "{Discovery of a Luminous Quasar in the Nearby Universe}",
      journal = {\apjl},
         year = 1997,
        month = oct,
       volume = {488},
       number = {1},
        pages = {L19-L22},
          doi = {10.1086/310913},
archivePrefix = {arXiv},
       eprint = {astro-ph/9707258},
 primaryClass = {astro-ph},
       adsurl = {https://ui.adsabs.harvard.edu/abs/1997ApJ...488L..19T}
}

@ARTICLE{1994A&A...281..603C,
       author = {{Cuillandre}, J.~C. and {Fort}, B. and {Picat}, J.~P. and {Soucail}, J.~P. and {Altieri}, B. and {Beigbeder}, F. and {Duplin}, J.~P. and {Pourthie}, T. and {Ratier}, G.},
        title = "{'Va-et-Vient' spectroscopy: A new mode for faint object CCD spectroscopy with very large telescopes}",
      journal = {\aap},
         year = 1994,
        month = jan,
       volume = {281},
       number = {2},
        pages = {603-612},
       adsurl = {https://ui.adsabs.harvard.edu/abs/1994A&A...281..603C}
}

@ARTICLE{Ina20,
       author = {{Inayoshi}, Kohei and {Visbal}, Eli and {Haiman}, Zolt{\'a}n},
        title = "{The Assembly of the First Massive Black Holes}",
      journal = {\araa},
         year = 2020,
        month = aug,
       volume = {58},
        pages = {27-97},
          doi = {10.1146/annurev-astro-120419-014455},
archivePrefix = {arXiv},
       eprint = {1911.05791},
 primaryClass = {astro-ph.GA},
       adsurl = {https://ui.adsabs.harvard.edu/abs/2020ARA&A..58...27I}
}

@INPROCEEDINGS{Smith96,
       author = {{Smith}, L.~J. and {Pettini}, M. and {King}, D.~L. and {Hunstead}, R.~W.},
        title = "{The Chemical Evolution of Galaxies at High Redshift}",
    booktitle = {From Stars to Galaxies: the Impact of Stellar Physics on Galaxy Evolution},
         year = 1996,
       editor = {{Leitherer}, C. and {Fritze-von-Alvensleben}, U. and {Huchra}, J.},
       series = {Astronomical Society of the Pacific Conference Series},
       volume = {98},
        month = jan,
        pages = {559},
          doi = {10.48550/arXiv.astro-ph/9601153},
archivePrefix = {arXiv},
       eprint = {astro-ph/9601153},
 primaryClass = {astro-ph},
       adsurl = {https://ui.adsabs.harvard.edu/abs/1996ASPC...98..559S}
}

@ARTICLE{Rauch98,
       author = {{Rauch}, Michael},
        title = "{The Lyman Alpha Forest in the Spectra of QSOs}",
      journal = {\araa},
         year = 1998,
        month = jan,
       volume = {36},
        pages = {267-316},
          doi = {10.1146/annurev.astro.36.1.267},
archivePrefix = {arXiv},
       eprint = {astro-ph/9806286},
 primaryClass = {astro-ph},
       adsurl = {https://ui.adsabs.harvard.edu/abs/1998ARA&A..36..267R}
}

@ARTICLE{Fabian12,
       author = {{Fabian}, A.~C.},
        title = "{Observational Evidence of Active Galactic Nuclei Feedback}",
      journal = {\araa},
         year = 2012,
        month = sep,
       volume = {50},
        pages = {455-489},
          doi = {10.1146/annurev-astro-081811-125521},
archivePrefix = {arXiv},
       eprint = {1204.4114},
 primaryClass = {astro-ph.CO},
       adsurl = {https://ui.adsabs.harvard.edu/abs/2012ARA&A..50..455F}
}

@ARTICLE{W03,
       author = {{Wolf}, C. and {Wisotzki}, L. and {Borch}, A. and {Dye}, S. and {Kleinheinrich}, M. and {Meisenheimer}, K.},
        title = "{The evolution of faint AGN between z  =\raisebox{-0.5ex}\textasciitilde  1 and z =\raisebox{-0.5ex}\textasciitilde 5 from the COMBO-17 survey}",
      journal = {\aap},
         year = 2003,
        month = sep,
       volume = {408},
        pages = {499-514},
          doi = {10.1051/0004-6361:20030990},
archivePrefix = {arXiv},
       eprint = {astro-ph/0304072},
 primaryClass = {astro-ph},
       adsurl = {https://ui.adsabs.harvard.edu/abs/2003A&A...408..499W}
}

@ARTICLE{Wu12,
       author = {{Wu}, Xue-Bing and {Hao}, Guoqiang and {Jia}, Zhendong and {Zhang}, Yanxia and {Peng}, Nanbo},
        title = "{SDSS Quasars in the WISE Preliminary Data Release and Quasar Candidate Selection with Optical/Infrared Colors}",
      journal = {\aj},
         year = 2012,
        month = aug,
       volume = {144},
       number = {2},
          eid = {49},
        pages = {49},
          doi = {10.1088/0004-6256/144/2/49},
archivePrefix = {arXiv},
       eprint = {1204.6197},
 primaryClass = {astro-ph.CO},
       adsurl = {https://ui.adsabs.harvard.edu/abs/2012AJ....144...49W}
}

@ARTICLE{Quaia24,
       author = {{Storey-Fisher}, Kate and {Hogg}, David W. and {Rix}, Hans-Walter and {Eilers}, Anna-Christina and {Fabbian}, Giulio and {Blanton}, Michael R. and {Alonso}, David},
        title = "{Quaia, the Gaia-unWISE Quasar Catalog: An All-sky Spectroscopic Quasar Sample}",
      journal = {\apj},
         year = 2024,
        month = mar,
       volume = {964},
       number = {1},
          eid = {69},
        pages = {69},
          doi = {10.3847/1538-4357/ad1328},
archivePrefix = {arXiv},
       eprint = {2306.17749},
 primaryClass = {astro-ph.GA},
       adsurl = {https://ui.adsabs.harvard.edu/abs/2024ApJ...964...69S}
}

@ARTICLE{Dipole24,
       author = {{Mittal}, Vasudev and {Oayda}, Oliver T. and {Lewis}, Geraint F.},
        title = "{The cosmic dipole in the Quaia sample of quasars: a Bayesian analysis}",
      journal = {\mnras},
         year = 2024,
        month = jan,
       volume = {527},
       number = {3},
        pages = {8497-8510},
          doi = {10.1093/mnras/stad3706},
archivePrefix = {arXiv},
       eprint = {2311.14938},
 primaryClass = {astro-ph.CO},
       adsurl = {https://ui.adsabs.harvard.edu/abs/2024MNRAS.527.8497M}
}

@ARTICLE{Choi25,
       author = {{Choi}, Yunyi and {Fu}, Yuming and {Im}, Myungshin and {Wu}, Xue-Bing and {Onken}, Christopher A. and {Wolf}, Christian and {Chang}, Seo-Won and {Choi}, Hyeonho and {Jeong}, Mankeun and {Kim}, Yongjung and {Lim}, Gu and {Pang}, Yuxuan and {Kim}, Taewan and {Sohn}, Jubee and {Kim}, Dohyeong and {Kim}, Ji Hoon and {Ko}, Eunhee and {Paek}, Gregory S.~H. and {Jung}, Sungho},
        title = "{AllBRICQS: The Discovery of Luminous Quasars in the Northern Hemisphere}",
      journal = {\apjs},
         year = 2025,
        month = oct,
       volume = {280},
       number = {2},
          eid = {73},
        pages = {73},
          doi = {10.3847/1538-4365/adf8ed},
archivePrefix = {arXiv},
       eprint = {2508.06028},
 primaryClass = {astro-ph.GA},
       adsurl = {https://ui.adsabs.harvard.edu/abs/2025ApJS..280...73C}
}

@ARTICLE{Lucy25,
       author = {{Lucy}, A.~B. and {Sokoloski}, J.~L. and {Luna}, G.~J.~M. and {Mukai}, K. and {Nu{\~n}ez}, N.~E. and {Buckley}, D.~A.~H. and {Breytenbach}, H. and {Paul}, B. and {Potter}, S.~B. and {Manick}, R. and {Howell}, D.~A. and {Wolf}, C. and {Onken}, C.~A.},
        title = "{A new way to find symbiotic stars: accretion disc detection with continuum survey photometry}",
      journal = {\mnras},
         year = 2025,
        month = nov,
       volume = {543},
       number = {3},
        pages = {2292-2325},
          doi = {10.1093/mnras/staf1351},
archivePrefix = {arXiv},
       eprint = {2412.00855},
 primaryClass = {astro-ph.SR},
       adsurl = {https://ui.adsabs.harvard.edu/abs/2025MNRAS.543.2292L}
}

@ARTICLE{catWISE,
       author = {{Marocco}, Federico and {Eisenhardt}, Peter R.~M. and {Fowler}, John W. and {Kirkpatrick}, J. Davy and {Meisner}, Aaron M. and {Schlafly}, Edward F. and {Stanford}, S.~A. and {Garcia}, Nelson and {Caselden}, Dan and {Cushing}, Michael C. and {Cutri}, Roc M. and {Faherty}, Jacqueline K. and {Gelino}, Christopher R. and {Gonzalez}, Anthony H. and {Jarrett}, Thomas H. and {Koontz}, Renata and {Mainzer}, Amanda and {Marchese}, Elijah J. and {Mobasher}, Bahram and {Schlegel}, David J. and {Stern}, Daniel and {Teplitz}, Harry I. and {Wright}, Edward L.},
        title = "{The CatWISE2020 Catalog}",
      journal = {\apjs},
         year = 2021,
        month = mar,
       volume = {253},
       number = {1},
          eid = {8},
        pages = {8},
          doi = {10.3847/1538-4365/abd805},
archivePrefix = {arXiv},
       eprint = {2012.13084},
 primaryClass = {astro-ph.IM},
       adsurl = {https://ui.adsabs.harvard.edu/abs/2021ApJS..253....8M}
}

@ARTICLE{WHO91,
       author = {{Warren}, Stephen J. and {Hewett}, Paul C. and {Irwin}, Michael J. and {Osmer}, Patrick S.},
        title = "{A Wide-Field Multicolor Survey for High-Redshift Quasars, Z >= 2.2. I. Photometric Catalog and Survey Selection Function}",
      journal = {\apjs},
         year = 1991,
        month = may,
       volume = {76},
        pages = {1},
          doi = {10.1086/191563},
       adsurl = {https://ui.adsabs.harvard.edu/abs/1991ApJS...76....1W}
}

@ARTICLE{Flesch23,
       author = {{Flesch}, Eric Wim},
        title = "{The Million Quasars (Milliquas) Catalogue, v8}",
      journal = {The Open Journal of Astrophysics},
         year = 2023,
        month = dec,
       volume = {6},
          eid = {49},
        pages = {49},
          doi = {10.21105/astro.2308.01505},
archivePrefix = {arXiv},
       eprint = {2308.01505},
 primaryClass = {astro-ph.GA},
       adsurl = {https://ui.adsabs.harvard.edu/abs/2023OJAp....6E..49F}
}

@ARTICLE{Jones09,
       author = {{Jones}, D. Heath and {Read}, Mike A. and {Saunders}, Will and {Colless}, Matthew and {Jarrett}, Tom and {Parker}, Quentin A. and {Fairall}, Anthony P. and {Mauch}, Thomas and {Sadler}, Elaine M. and {Watson}, Fred G. and {Burton}, Donna and {Campbell}, Lachlan A. and {Cass}, Paul and {Croom}, Scott M. and {Dawe}, John and {Fiegert}, Kristin and {Frankcombe}, Leela and {Hartley}, Malcolm and {Huchra}, John and {James}, Dionne and {Kirby}, Emma and {Lahav}, Ofer and {Lucey}, John and {Mamon}, Gary A. and {Moore}, Lesa and {Peterson}, Bruce A. and {Prior}, Sayuri and {Proust}, Dominique and {Russell}, Ken and {Safouris}, Vicky and {Wakamatsu}, Ken-Ichi and {Westra}, Eduard and {Williams}, Mary},
        title = "{The 6dF Galaxy Survey: final redshift release (DR3) and southern large-scale structures}",
      journal = {\mnras},
         year = 2009,
        month = oct,
       volume = {399},
       number = {2},
        pages = {683-698},
          doi = {10.1111/j.1365-2966.2009.15338.x},
archivePrefix = {arXiv},
       eprint = {0903.5451},
 primaryClass = {astro-ph.CO},
       adsurl = {https://ui.adsabs.harvard.edu/abs/2009MNRAS.399..683J}
}

@ARTICLE{Irwin98,
       author = {{Irwin}, Michael J. and {Ibata}, Rodrigo A. and {Lewis}, Geraint F. and {Totten}, Edward J.},
        title = "{APM 08279+5255: an Ultraluminous Broad Absorption Line Quasar at a Redshift Z = 3.87}",
      journal = {\apj},
         year = 1998,
        month = oct,
       volume = {505},
       number = {2},
        pages = {529-535},
          doi = {10.1086/306213},
archivePrefix = {arXiv},
       eprint = {astro-ph/9806171},
 primaryClass = {astro-ph},
       adsurl = {https://ui.adsabs.harvard.edu/abs/1998ApJ...505..529I}
}

@ARTICLE{2020ApJS..250...26B,
       author = {{Boutsia}, Konstantina and {Grazian}, Andrea and {Calderone}, Giorgio and {Cristiani}, Stefano and {Cupani}, Guido and {Guarneri}, Francesco and {Fontanot}, Fabio and {Amorin}, Ricardo and {D'Odorico}, Valentina and {Giallongo}, Emanuele and {Salvato}, Mara and {Omizzolo}, Alessandro and {Romano}, Michael and {Menci}, Nicola},
        title = "{The Spectroscopic Follow-up of the QUBRICS Bright Quasar Survey}",
      journal = {\apjs},
         year = 2020,
        month = oct,
       volume = {250},
       number = {2},
          eid = {26},
        pages = {26},
          doi = {10.3847/1538-4365/abafc1},
archivePrefix = {arXiv},
       eprint = {2008.03865},
 primaryClass = {astro-ph.GA},
       adsurl = {https://ui.adsabs.harvard.edu/abs/2020ApJS..250...26B}
}

@ARTICLE{Delchambre23,
       author = {{Delchambre}, L. and {Bailer-Jones}, C.~A.~L. and {Bellas-Velidis}, I. and {Drimmel}, R. and {Garabato}, D. and {Carballo}, R. and {Hatzidimitriou}, D. and {Marshall}, D.~J. and {Andrae}, R. and {Dafonte}, C. and {Livanou}, E. and {Fouesneau}, M. and {Licata}, E.~L. and {Lindstr{\o}m}, H.~E.~P. and {Manteiga}, M. and {Robin}, C. and {Silvelo}, A. and {Abreu Aramburu}, A. and {{\'A}lvarez}, M.~A. and {Bakker}, J. and {Bijaoui}, A. and {Brouillet}, N. and {Brugaletta}, E. and {Burlacu}, A. and {Casamiquela}, L. and {Chaoul}, L. and {Chiavassa}, A. and {Contursi}, G. and {Cooper}, W.~J. and {Creevey}, O.~L. and {Dapergolas}, A. and {de Laverny}, P. and {Demouchy}, C. and {Dharmawardena}, T.~E. and {Edvardsson}, B. and {Fr{\'e}mat}, Y. and {Garc{\'\i}a-Lario}, P. and {Garc{\'\i}a-Torres}, M. and {Gavel}, A. and {Gomez}, A. and {Gonz{\'a}lez-Santamar{\'\i}a}, I. and {Heiter}, U. and {Jean-Antoine Piccolo}, A. and {Kontizas}, M. and {Kordopatis}, G. and {Korn}, A.~J. and {Lanzafame}, A.~C. and {Lebreton}, Y. and {Lobel}, A. and {Lorca}, A. and {Magdaleno Romeo}, A. and {Marocco}, F. and {Mary}, N. and {Nicolas}, C. and {Ordenovic}, C. and {Pailler}, F. and {Palicio}, P.~A. and {Pallas-Quintela}, L. and {Panem}, C. and {Pichon}, B. and {Poggio}, E. and {Recio-Blanco}, A. and {Riclet}, F. and {Rybizki}, J. and {Santove{\~n}a}, R. and {Sarro}, L.~M. and {Schultheis}, M.~S. and {Segol}, M. and {Slezak}, I. and {Smart}, R.~L. and {Sordo}, R. and {Soubiran}, C. and {S{\"u}veges}, M. and {Th{\'e}venin}, F. and {Torralba Elipe}, G. and {Ulla}, A. and {Utrilla}, E. and {Vallenari}, A. and {van Dillen}, E. and {Zhao}, H. and {Zorec}, J.},
        title = "{Gaia Data Release 3. Apsis. III. Non-stellar content and source classification}",
      journal = {\aap},
         year = 2023,
        month = jun,
       volume = {674},
          eid = {A31},
        pages = {A31},
          doi = {10.1051/0004-6361/202243423},
archivePrefix = {arXiv},
       eprint = {2206.06710},
 primaryClass = {astro-ph.GA},
       adsurl = {https://ui.adsabs.harvard.edu/abs/2023A&A...674A..31D}
}

@ARTICLE{WangChen19,
       author = {{Wang}, Shu and {Chen}, Xiaodian},
        title = "{The Optical to Mid-infrared Extinction Law Based on the APOGEE, Gaia DR2, Pan-STARRS1, SDSS, APASS, 2MASS, and WISE Surveys}",
      journal = {\apj},
         year = 2019,
        month = jun,
       volume = {877},
       number = {2},
          eid = {116},
        pages = {116},
          doi = {10.3847/1538-4357/ab1c61},
archivePrefix = {arXiv},
       eprint = {1904.04575},
 primaryClass = {astro-ph.GA},
       adsurl = {https://ui.adsabs.harvard.edu/abs/2019ApJ...877..116W}
}

@Article{Casa18,
  author        = {Casagrande, L. and VandenBerg, Don A.},
  journal       = {\mnras},
  title         = {On the use of Gaia magnitudes and new tables of bolometric corrections},
  year          = {2018},
  month         = sep,
  number        = {1},
  pages         = {L102-L107},
  volume        = {479},
  archiveprefix = {arXiv},
  doi           = {10.1093/mnrasl/sly104},
  eprint        = {1806.01953},
  primaryclass  = {astro-ph.SR},
  url           = {https://ui.adsabs.harvard.edu/abs/2018MNRAS.479L.102C},
}

@ARTICLE{LVG13,
       author = {{Karachentsev}, Igor D. and {Makarov}, Dmitry I. and {Kaisina}, Elena I.},
        title = "{Updated Nearby Galaxy Catalog}",
      journal = {\aj},
         year = 2013,
        month = apr,
       volume = {145},
       number = {4},
          eid = {101},
        pages = {101},
          doi = {10.1088/0004-6256/145/4/101},
archivePrefix = {arXiv},
       eprint = {1303.5328},
 primaryClass = {astro-ph.CO},
       adsurl = {https://ui.adsabs.harvard.edu/abs/2013AJ....145..101K}
}

@ARTICLE{Malkan83,
       author = {{Malkan}, M.~A.},
        title = "{The ultraviolet excess of luminous quasars. II. Evidence for massive accretion disks.}",
      journal = {\apj},
         year = 1983,
        month = may,
       volume = {268},
        pages = {582-590},
          doi = {10.1086/160981},
       adsurl = {https://ui.adsabs.harvard.edu/abs/1983ApJ...268..582M}
}

@ARTICLE{An93,
       author = {{Antonucci}, Robert},
        title = "{Unified models for active galactic nuclei and quasars.}",
      journal = {\araa},
         year = 1993,
        month = jan,
       volume = {31},
        pages = {473-521},
          doi = {10.1146/annurev.aa.31.090193.002353},
       adsurl = {https://ui.adsabs.harvard.edu/abs/1993ARA&A..31..473A}
}

@ARTICLE{2018A&A...616A...1G,
       author = {{Gaia Collaboration}},
        title = "{Gaia Data Release 2. Summary of the contents and survey properties}",
      journal = {\aap},
         year = 2018,
        month = aug,
       volume = {616},
          eid = {A1},
        pages = {A1},
          doi = {10.1051/0004-6361/201833051},
archivePrefix = {arXiv},
       eprint = {1804.09365},
 primaryClass = {astro-ph.GA},
       adsurl = {https://ui.adsabs.harvard.edu/abs/2018A&A...616A...1G}
}

@ARTICLE{2021A&A...649A...1G,
       author = {{Gaia Collaboration}},
        title = "{Gaia Early Data Release 3. Summary of the contents and survey properties}",
      journal = {\aap},
         year = 2021,
        month = may,
       volume = {649},
          eid = {A1},
        pages = {A1},
          doi = {10.1051/0004-6361/202039657},
archivePrefix = {arXiv},
       eprint = {2012.01533},
 primaryClass = {astro-ph.GA},
       adsurl = {https://ui.adsabs.harvard.edu/abs/2021A&A...649A...1G}
}

@ARTICLE{2018Natur.563..657G,
       author = {{Gravity Collaboration} and {Sturm}, E. and {Dexter}, J. and {Pfuhl}, O. and {Stock}, M.~R. and {Davies}, R.~I. and {Lutz}, D. and {Cl{\'e}net}, Y. and {Eckart}, A. and {Eisenhauer}, F. and {Genzel}, R. and {Gratadour}, D. and {H{\"o}nig}, S.~F. and {Kishimoto}, M. and {Lacour}, S. and {Millour}, F. and {Netzer}, H. and {Perrin}, G. and {Peterson}, B.~M. and {Petrucci}, P.~O. and {Rouan}, D. and {Waisberg}, I. and {Woillez}, J. and {Amorim}, A. and {Brandner}, W. and {F{\"o}rster Schreiber}, N.~M. and {Garcia}, P.~J.~V. and {Gillessen}, S. and {Ott}, T. and {Paumard}, T. and {Perraut}, K. and {Scheithauer}, S. and {Straubmeier}, C. and {Tacconi}, L.~J. and {Widmann}, F.},
        title = "{Spatially resolved rotation of the broad-line region of a quasar at sub-parsec scale}",
      journal = {\nat},
         year = 2018,
        month = nov,
       volume = {563},
       number = {7733},
        pages = {657-660},
          doi = {10.1038/s41586-018-0731-9},
archivePrefix = {arXiv},
       eprint = {1811.11195},
 primaryClass = {astro-ph.GA},
       adsurl = {https://ui.adsabs.harvard.edu/abs/2018Natur.563..657G}
}

@ARTICLE{Le22,
       author = {{Lemon}, C. and {Anguita}, T. and {Auger-Williams}, M.~W. and {Courbin}, F. and {Galan}, A. and {McMahon}, R. and {Neira}, F. and {Oguri}, M. and {Schechter}, P. and {Shajib}, A. and {Treu}, T. and {Agnello}, A. and {Spiniello}, C.},
        title = "{Gravitationally lensed quasars in Gaia - IV. 150 new lenses, quasar pairs, and projected quasars}",
      journal = {\mnras},
         year = 2023,
        month = apr,
       volume = {520},
       number = {3},
        pages = {3305-3328},
          doi = {10.1093/mnras/stac3721},
archivePrefix = {arXiv},
       eprint = {2206.07714},
 primaryClass = {astro-ph.GA},
       adsurl = {https://ui.adsabs.harvard.edu/abs/2023MNRAS.520.3305L}
}

@ARTICLE{2008MNRAS.386.1192L,
       author = {{Liske}, J. and {Grazian}, A. and {Vanzella}, E. and {Dessauges}, M. and {Viel}, M. and {Pasquini}, L. and {Haehnelt}, M. and {Cristiani}, S. and {Pepe}, F. and {Avila}, G. and {Bonifacio}, P. and {Bouchy}, F. and {Dekker}, H. and {Delabre}, B. and {D'Odorico}, S. and {D'Odorico}, V. and {Levshakov}, S. and {Lovis}, C. and {Mayor}, M. and {Molaro}, P. and {Moscardini}, L. and {Murphy}, M.~T. and {Queloz}, D. and {Shaver}, P. and {Udry}, S. and {Wiklind}, T. and {Zucker}, S.},
        title = "{Cosmic dynamics in the era of Extremely Large Telescopes}",
      journal = {\mnras},
         year = 2008,
        month = may,
       volume = {386},
       number = {3},
        pages = {1192-1218},
          doi = {10.1111/j.1365-2966.2008.13090.x},
archivePrefix = {arXiv},
       eprint = {0802.1532},
 primaryClass = {astro-ph},
       adsurl = {https://ui.adsabs.harvard.edu/abs/2008MNRAS.386.1192L}
}

@ARTICLE{DESIQSO23,
       author = {{Chaussidon}, Edmond and {Y{\`e}che}, Christophe and {Palanque-Delabrouille}, Nathalie and {Alexander}, David M. and {Yang}, Jinyi and {Ahlen}, Steven and {Bailey}, Stephen and {Brooks}, David and {Cai}, Zheng and {Chabanier}, Sol{\`e}ne and {Davis}, Tamara M. and {Dawson}, Kyle and {de laMacorra}, Axel and {Dey}, Arjun and {Dey}, Biprateep and {Eftekharzadeh}, Sarah and {Eisenstein}, Daniel J. and {Fanning}, Kevin and {Font-Ribera}, Andreu and {Gazta{\~n}aga}, Enrique and {A Gontcho}, Satya Gontcho and {Gonzalez-Morales}, Alma X. and {Guy}, Julien and {Herrera-Alcantar}, Hiram K. and {Honscheid}, Klaus and {Ishak}, Mustapha and {Jiang}, Linhua and {Juneau}, Stephanie and {Kehoe}, Robert and {Kisner}, Theodore and {Kov{\'a}cs}, Andras and {Kremin}, Anthony and {Lan}, Ting-Wen and {Landriau}, Martin and {Le Guillou}, Laurent and {Levi}, Michael E. and {Magneville}, Christophe and {Martini}, Paul and {Meisner}, Aaron M. and {Moustakas}, John and {Mu{\~n}oz-Guti{\'e}rrez}, Andrea and {Myers}, Adam D. and {Newman}, Jeffrey A. and {Nie}, Jundan and {Percival}, Will J. and {Poppett}, Claire and {Prada}, Francisco and {Raichoor}, Anand and {Ravoux}, Corentin and {Ross}, Ashley J. and {Schlafly}, Edward and {Schlegel}, David and {Tan}, Ting and {Tarl{\'e}}, Gregory and {Zhou}, Rongpu and {Zhou}, Zhimin and {Zou}, Hu},
        title = "{Target Selection and Validation of DESI Quasars}",
      journal = {\apj},
         year = 2023,
        month = feb,
       volume = {944},
       number = {1},
          eid = {107},
        pages = {107},
          doi = {10.3847/1538-4357/acb3c2},
archivePrefix = {arXiv},
       eprint = {2208.08511},
 primaryClass = {astro-ph.CO},
       adsurl = {https://ui.adsabs.harvard.edu/abs/2023ApJ...944..107C}
}

@ARTICLE{PDvar16,
       author = {{Palanque-Delabrouille}, N. and {Magneville}, Ch. and {Y{\`e}che}, Ch. and {P{\^a}ris}, I. and {Petitjean}, P. and {Burtin}, E. and {Dawson}, K. and {McGreer}, I. and {Myers}, A.~D. and {Rossi}, G. and {Schlegel}, D. and {Schneider}, D. and {Streblyanska}, A. and {Tinker}, J.},
        title = "{The extended Baryon Oscillation Spectroscopic Survey: Variability selection and quasar luminosity function}",
      journal = {\aap},
         year = 2016,
        month = mar,
       volume = {587},
          eid = {A41},
        pages = {A41},
          doi = {10.1051/0004-6361/201527392},
archivePrefix = {arXiv},
       eprint = {1509.05607},
 primaryClass = {astro-ph.CO},
       adsurl = {https://ui.adsabs.harvard.edu/abs/2016A&A...587A..41P}
}

@ARTICLE{Ross20eBOSS,
       author = {{Ross}, Ashley J. and {Bautista}, Julian and {Tojeiro}, Rita and {Alam}, Shadab and {Bailey}, Stephen and {Burtin}, Etienne and {Comparat}, Johan and {Dawson}, Kyle S. and {de Mattia}, Arnaud and {du Mas des Bourboux}, H{\'e}lion and {Gil-Mar{\'\i}n}, H{\'e}ctor and {Hou}, Jiamin and {Kong}, Hui and {Lyke}, Brad W. and {Mohammad}, Faizan G. and {Moustakas}, John and {Mueller}, Eva-Maria and {Myers}, Adam D. and {Percival}, Will J. and {Raichoor}, Anand and {Rezaie}, Mehdi and {Seo}, Hee-Jong and {Smith}, Alex and {Tinker}, Jeremy L. and {Zarrouk}, Pauline and {Zhao}, Cheng and {Zhao}, Gong-Bo and {Bizyaev}, Dmitry and {Brinkmann}, Jonathan and {Brownstein}, Joel R. and {Rosell}, Aurelio Carnero and {Chabanier}, Sol{\`e}ne and {Choi}, Peter D. and {Chuang}, Chia-Hsun and {Cruz-Gonzalez}, Irene and {de la Macorra}, Axel and {de la Torre}, Sylvain and {Escoffier}, Stephanie and {Fromenteau}, Sebastien and {Higley}, Alexandra and {Jullo}, Eric and {Kneib}, Jean-Paul and {McLane}, Jacob N. and {Mu{\~n}oz-Guti{\'e}rrez}, Andrea and {Neveux}, Richard and {Newman}, Jeffrey A. and {Nitschelm}, Christian and {Palanque-Delabrouille}, Nathalie and {Paviot}, Romain and {Pullen}, Anthony R. and {Rossi}, Graziano and {Ruhlmann-Kleider}, Vanina and {Schneider}, Donald P. and {Maga{\~n}a}, Mariana Vargas and {Vivek}, M. and {Zhang}, Yucheng},
        title = "{The Completed SDSS-IV extended Baryon Oscillation Spectroscopic Survey: Large-scale structure catalogues for cosmological analysis}",
      journal = {\mnras},
         year = 2020,
        month = oct,
       volume = {498},
       number = {2},
        pages = {2354-2371},
          doi = {10.1093/mnras/staa2416},
archivePrefix = {arXiv},
       eprint = {2007.09000},
 primaryClass = {astro-ph.CO},
       adsurl = {https://ui.adsabs.harvard.edu/abs/2020MNRAS.498.2354R}
}

@ARTICLE{RT23,
       author = {{Ricci}, Claudio and {Trakhtenbrot}, Benny},
        title = "{Changing-look active galactic nuclei}",
      journal = {Nature Astronomy},
         year = 2023,
        month = nov,
       volume = {7},
        pages = {1282-1294},
          doi = {10.1038/s41550-023-02108-4},
archivePrefix = {arXiv},
       eprint = {2211.05132},
 primaryClass = {astro-ph.GA},
       adsurl = {https://ui.adsabs.harvard.edu/abs/2023NatAs...7.1282R}
}

@ARTICLE{Montegriffo23,
       author = {{Montegriffo}, P. and {De Angeli}, F. and {Andrae}, R. and {Riello}, M. and {Pancino}, E. and {Sanna}, N. and {Bellazzini}, M. and {Evans}, D.~W. and {Carrasco}, J.~M. and {Sordo}, R. and {Busso}, G. and {Cacciari}, C. and {Jordi}, C. and {van Leeuwen}, F. and {Vallenari}, A. and {Altavilla}, G. and {Barstow}, M.~A. and {Brown}, A.~G.~A. and {Burgess}, P.~W. and {Castellani}, M. and {Cowell}, S. and {Davidson}, M. and {De Luise}, F. and {Delchambre}, L. and {Diener}, C. and {Fabricius}, C. and {Fr{\'e}mat}, Y. and {Fouesneau}, M. and {Gilmore}, G. and {Giuffrida}, G. and {Hambly}, N.~C. and {Harrison}, D.~L. and {Hidalgo}, S. and {Hodgkin}, S.~T. and {Holland}, G. and {Marinoni}, S. and {Osborne}, P.~J. and {Pagani}, C. and {Palaversa}, L. and {Piersimoni}, A.~M. and {Pulone}, L. and {Ragaini}, S. and {Rainer}, M. and {Richards}, P.~J. and {Rowell}, N. and {Ruz-Mieres}, D. and {Sarro}, L.~M. and {Walton}, N.~A. and {Yoldas}, A.},
        title = "{Gaia Data Release 3. External calibration of BP/RP low-resolution spectroscopic data}",
      journal = {\aap},
         year = 2023,
        month = jun,
       volume = {674},
          eid = {A3},
        pages = {A3},
          doi = {10.1051/0004-6361/202243880},
archivePrefix = {arXiv},
       eprint = {2206.06205},
 primaryClass = {astro-ph.IM},
       adsurl = {https://ui.adsabs.harvard.edu/abs/2023A&A...674A...3M}
}

@ARTICLE{2022MNRAS.511..572O,
       author = {{Onken}, Christopher A. and {Wolf}, Christian and {Bian}, Fuyan and {Fan}, Xiaohui and {Hon}, Wei Jeat and {Raithel}, David and {Tisserand}, Patrick and {Lai}, Samuel},
        title = "{Ultraluminous high-redshift quasars from SkyMapper - II. New quasars and the bright end of the luminosity function}",
      journal = {\mnras},
         year = 2022,
        month = mar,
       volume = {511},
       number = {1},
        pages = {572-594},
          doi = {10.1093/mnras/stac051},
archivePrefix = {arXiv},
       eprint = {2105.12215},
 primaryClass = {astro-ph.CO},
       adsurl = {https://ui.adsabs.harvard.edu/abs/2022MNRAS.511..572O}
}

@ARTICLE{2021A&A...647A...1P,
       author = {{Predehl}, P. and {Andritschke}, R. and {Arefiev}, V. and {Babyshkin}, V. and {Batanov}, O. and {Becker}, W. and {B{\"o}hringer}, H. and {Bogomolov}, A. and {Boller}, T. and {Borm}, K. and {Bornemann}, W. and {Br{\"a}uninger}, H. and {Br{\"u}ggen}, M. and {Brunner}, H. and {Brusa}, M. and {Bulbul}, E. and {Buntov}, M. and {Burwitz}, V. and {Burkert}, W. and {Clerc}, N. and {Churazov}, E. and {Coutinho}, D. and {Dauser}, T. and {Dennerl}, K. and {Doroshenko}, V. and {Eder}, J. and {Emberger}, V. and {Eraerds}, T. and {Finoguenov}, A. and {Freyberg}, M. and {Friedrich}, P. and {Friedrich}, S. and {F{\"u}rmetz}, M. and {Georgakakis}, A. and {Gilfanov}, M. and {Granato}, S. and {Grossberger}, C. and {Gueguen}, A. and {Gureev}, P. and {Haberl}, F. and {H{\"a}lker}, O. and {Hartner}, G. and {Hasinger}, G. and {Huber}, H. and {Ji}, L. and {Kienlin}, A. v. and {Kink}, W. and {Korotkov}, F. and {Kreykenbohm}, I. and {Lamer}, G. and {Lomakin}, I. and {Lapshov}, I. and {Liu}, T. and {Maitra}, C. and {Meidinger}, N. and {Menz}, B. and {Merloni}, A. and {Mernik}, T. and {Mican}, B. and {Mohr}, J. and {M{\"u}ller}, S. and {Nandra}, K. and {Nazarov}, V. and {Pacaud}, F. and {Pavlinsky}, M. and {Perinati}, E. and {Pfeffermann}, E. and {Pietschner}, D. and {Ramos-Ceja}, M.~E. and {Rau}, A. and {Reiffers}, J. and {Reiprich}, T.~H. and {Robrade}, J. and {Salvato}, M. and {Sanders}, J. and {Santangelo}, A. and {Sasaki}, M. and {Scheuerle}, H. and {Schmid}, C. and {Schmitt}, J. and {Schwope}, A. and {Shirshakov}, A. and {Steinmetz}, M. and {Stewart}, I. and {Str{\"u}der}, L. and {Sunyaev}, R. and {Tenzer}, C. and {Tiedemann}, L. and {Tr{\"u}mper}, J. and {Voron}, V. and {Weber}, P. and {Wilms}, J. and {Yaroshenko}, V.},
        title = "{The eROSITA X-ray telescope on SRG}",
      journal = {\aap},
         year = 2021,
        month = mar,
       volume = {647},
          eid = {A1},
        pages = {A1},
          doi = {10.1051/0004-6361/202039313},
archivePrefix = {arXiv},
       eprint = {2010.03477},
 primaryClass = {astro-ph.HE},
       adsurl = {https://ui.adsabs.harvard.edu/abs/2021A&A...647A...1P}
}

@ARTICLE{2020ApJS..249...17R,
       author = {{Rakshit}, Suvendu and {Stalin}, C.~S. and {Kotilainen}, Jari},
        title = "{Spectral Properties of Quasars from Sloan Digital Sky Survey Data Release 14: The Catalog}",
      journal = {\apjs},
         year = 2020,
        month = jul,
       volume = {249},
       number = {1},
          eid = {17},
        pages = {17},
          doi = {10.3847/1538-4365/ab99c5},
archivePrefix = {arXiv},
       eprint = {1910.10395},
 primaryClass = {astro-ph.GA},
       adsurl = {https://ui.adsabs.harvard.edu/abs/2020ApJS..249...17R}
}

@ARTICLE{2005MNRAS.360..839R,
       author = {{Richards}, Gordon T. and {Croom}, Scott M. and {Anderson}, Scott F. and {Bland-Hawthorn}, Joss and {Boyle}, Brian J. and {De Propris}, Roberto and {Drinkwater}, Michael J. and {Fan}, Xiaohui and {Gunn}, James E. and {Ivezi{\'c}}, {\v{Z}}eljko and {Jester}, Sebastian and {Loveday}, Jon and {Meiksin}, Avery and {Miller}, Lance and {Myers}, Adam and {Nichol}, Robert C. and {Outram}, Phil J. and {Pimbblet}, Kevin A. and {Roseboom}, Isaac G. and {Ross}, Nic and {Schneider}, Donald P. and {Shanks}, Tom and {Sharp}, Robert G. and {Stoughton}, Chris and {Strauss}, Michael A. and {Szalay}, Alexander S. and {Vanden Berk}, Daniel E. and {York}, Donald G.},
        title = "{The 2dF-SDSS LRG and QSO (2SLAQ) Survey: the z < 2.1 quasar luminosity function from 5645 quasars to g= 21.85}",
      journal = {\mnras},
         year = 2005,
        month = jul,
       volume = {360},
       number = {3},
        pages = {839-852},
          doi = {10.1111/j.1365-2966.2005.09096.x},
archivePrefix = {arXiv},
       eprint = {astro-ph/0504300},
 primaryClass = {astro-ph},
       adsurl = {https://ui.adsabs.harvard.edu/abs/2005MNRAS.360..839R}
}

@ARTICLE{1963Natur.197.1040S,
       author = {{Schmidt}, M.},
        title = "{3C 273 : A Star-Like Object with Large Red-Shift}",
      journal = {\nat},
         year = 1963,
        month = mar,
       volume = {197},
       number = {4872},
        pages = {1040},
          doi = {10.1038/1971040a0},
       adsurl = {https://ui.adsabs.harvard.edu/abs/1963Natur.197.1040S}
}

@ARTICLE{2026A&A...709A.236S,
       author = {{Scialpi}, M. and {Mannucci}, F. and {D'Amato}, Q. and {Marconcini}, C. and {Cresci}, G. and {Marconi}, A. and {Ulivi}, L. and {Fumagalli}, M. and {Rosati}, P. and {Tozzi}, G. and {Zanchettin}, M.~V. and {Battistini}, L. and {Bertola}, E. and {Bracci}, C. and {Carniani}, S. and {Cataldi}, E. and {Ceci}, M. and {Chakraborty}, A. and {Cicone}, C. and {Ciurlo}, A. and {De Rosa}, A. and {Di Rosa}, G. and {Feltre}, A. and {Ginolfi}, M. and {Lamperti}, I. and {Lusso}, E. and {Moreschini}, B. and {Nardini}, E. and {Parvatikar}, M. and {Perna}, M. and {Rubinur}, K. and {Severgnini}, P. and {Singh}, J. and {Spingola}, C. and {Venturi}, G. and {Vignali}, C. and {Volonteri}, M.},
        title = "{Cosmic Duets: I. High-spatial resolution spectroscopy of dual and lensed active galactic nuclei with MUSE-NFM}",
      journal = {\aap},
         year = 2026,
        month = may,
       volume = {709},
          eid = {A236},
        pages = {A236},
          doi = {10.1051/0004-6361/202558548},
archivePrefix = {arXiv},
       eprint = {2512.11960},
 primaryClass = {astro-ph.GA},
       adsurl = {https://ui.adsabs.harvard.edu/abs/2026A&A...709A.236S}
}

@Article{UP95,
  author        = {Urry, C. Megan and Padovani, Paolo},
  journal       = {\pasp},
  title         = {Unified Schemes for Radio-Loud Active Galactic Nuclei},
  year          = {1995},
  month         = sep,
  pages         = {803},
  volume        = {107},
  archiveprefix = {arXiv},
  doi           = {10.1086/133630},
  eprint        = {astro-ph/9506063},
  primaryclass  = {astro-ph},
  url           = {https://ui.adsabs.harvard.edu/abs/1995PASP..107..803U},
}

@ARTICLE{Simbad00,
       author = {{Wenger}, M. and {Ochsenbein}, F. and {Egret}, D. and {Dubois}, P. and {Bonnarel}, F. and {Borde}, S. and {Genova}, F. and {Jasniewicz}, G. and {Lalo{\"e}}, S. and {Lesteven}, S. and {Monier}, R.},
        title = "{The SIMBAD astronomical database. The CDS reference database for astronomical objects}",
      journal = {\aaps},
         year = 2000,
        month = apr,
       volume = {143},
        pages = {9-22},
          doi = {10.1051/aas:2000332},
archivePrefix = {arXiv},
       eprint = {astro-ph/0002110},
 primaryClass = {astro-ph},
       adsurl = {https://ui.adsabs.harvard.edu/abs/2000A&AS..143....9W}
}

@ARTICLE{2000A&A...358...77W,
       author = {{Wisotzki}, L. and {Christlieb}, N. and {Bade}, N. and {Beckmann}, V. and {K{\"o}hler}, T. and {Vanelle}, C. and {Reimers}, D.},
        title = "{The Hamburg/ESO survey for bright QSOs. III. A large flux-limited sample of QSOs}",
      journal = {\aap},
         year = 2000,
        month = jun,
       volume = {358},
        pages = {77-87},
archivePrefix = {arXiv},
       eprint = {astro-ph/0004162},
 primaryClass = {astro-ph},
       adsurl = {https://ui.adsabs.harvard.edu/abs/2000A&A...358...77W}
}

@ARTICLE{2018PASA...35...10W,
       author = {{Wolf}, Christian and {Onken}, Christopher A. and {Luvaul}, Lance C. and {Schmidt}, Brian P. and {Bessell}, Michael S. and {Chang}, Seo-Won and {Da Costa}, Gary S. and {Mackey}, Dougal and {Martin-Jones}, Tony and {Murphy}, Simon J. and {Preston}, Tim and {Scalzo}, Richard A. and {Shao}, Li and {Smillie}, Jon and {Tisserand}, Patrick and {White}, Marc C. and {Yuan}, Fang},
        title = "{SkyMapper Southern Survey: First Data Release (DR1)}",
      journal = {\pasa},
         year = 2018,
        month = feb,
       volume = {35},
          eid = {e010},
        pages = {e010},
          doi = {10.1017/pasa.2018.5},
archivePrefix = {arXiv},
       eprint = {1801.07834},
 primaryClass = {astro-ph.IM},
       adsurl = {https://ui.adsabs.harvard.edu/abs/2018PASA...35...10W}
}

@ARTICLE{2018PASA...35...24W,
       author = {{Wolf}, Christian and {Bian}, Fuyan and {Onken}, Christopher A. and {Schmidt}, Brian P. and {Tisserand}, Patrick and {Alonzi}, Noura and {Hon}, Wei Jeat and {Tonry}, John L.},
        title = "{Discovery of the Most Ultra-Luminous QSO Using GAIA, SkyMapper, and WISE}",
      journal = {\pasa},
         year = 2018,
        month = jun,
       volume = {35},
          eid = {e024},
        pages = {e024},
          doi = {10.1017/pasa.2018.22},
archivePrefix = {arXiv},
       eprint = {1805.04317},
 primaryClass = {astro-ph.GA},
       adsurl = {https://ui.adsabs.harvard.edu/abs/2018PASA...35...24W}
}

@ARTICLE{2020MNRAS.491.1970W,
       author = {{Wolf}, Christian and {Hon}, Wei Jeat and {Bian}, Fuyan and {Onken}, Christopher A. and {Alonzi}, Noura and {Bessell}, Michael A. and {Li}, Zefeng and {Schmidt}, Brian P. and {Tisserand}, Patrick},
        title = "{Ultra-luminous quasars at redshift z > 4.5 from SkyMapper}",
      journal = {\mnras},
         year = 2020,
        month = jan,
       volume = {491},
       number = {2},
        pages = {1970-1979},
          doi = {10.1093/mnras/stz2955},
archivePrefix = {arXiv},
       eprint = {1910.05856},
 primaryClass = {astro-ph.GA},
       adsurl = {https://ui.adsabs.harvard.edu/abs/2020MNRAS.491.1970W}
}

@ARTICLE{Stern12,
       author = {{Stern}, Daniel and {Assef}, Roberto J. and {Benford}, Dominic J. and {Blain}, Andrew and {Cutri}, Roc and {Dey}, Arjun and {Eisenhardt}, Peter and {Griffith}, Roger L. and {Jarrett}, T.~H. and {Lake}, Sean and {Masci}, Frank and {Petty}, Sara and {Stanford}, S.~A. and {Tsai}, Chao-Wei and {Wright}, E.~L. and {Yan}, Lin and {Harrison}, Fiona and {Madsen}, Kristin},
        title = "{Mid-infrared Selection of Active Galactic Nuclei with the Wide-Field Infrared Survey Explorer. I. Characterizing WISE-selected Active Galactic Nuclei in COSMOS}",
      journal = {\apj},
         year = 2012,
        month = jul,
       volume = {753},
       number = {1},
          eid = {30},
        pages = {30},
          doi = {10.1088/0004-637X/753/1/30},
archivePrefix = {arXiv},
       eprint = {1205.0811},
 primaryClass = {astro-ph.CO},
       adsurl = {https://ui.adsabs.harvard.edu/abs/2012ApJ...753...30S}
}

@ARTICLE{HB93,
       author = {{Hewitt}, A. and {Burbidge}, G.},
        title = "{A Revised and Updated Catalog of Quasi-stellar Objects}",
      journal = {\apjs},
         year = 1993,
        month = aug,
       volume = {87},
        pages = {451},
          doi = {10.1086/191811},
       adsurl = {https://ui.adsabs.harvard.edu/abs/1993ApJS...87..451H}
}

@ARTICLE{2010AJ....140.1868W,
       author = {{Wright}, Edward L. and {Eisenhardt}, Peter R.~M. and {Mainzer}, Amy K. and {Ressler}, Michael E. and {Cutri}, Roc M. and {Jarrett}, Thomas and {Kirkpatrick}, J. Davy and {Padgett}, Deborah and {McMillan}, Robert S. and {Skrutskie}, Michael and {Stanford}, S.~A. and {Cohen}, Martin and {Walker}, Russell G. and {Mather}, John C. and {Leisawitz}, David and {Gautier}, Thomas N., III and {McLean}, Ian and {Benford}, Dominic and {Lonsdale}, Carol J. and {Blain}, Andrew and {Mendez}, Bryan and {Irace}, William R. and {Duval}, Valerie and {Liu}, Fengchuan and {Royer}, Don and {Heinrichsen}, Ingolf and {Howard}, Joan and {Shannon}, Mark and {Kendall}, Martha and {Walsh}, Amy L. and {Larsen}, Mark and {Cardon}, Joel G. and {Schick}, Scott and {Schwalm}, Mark and {Abid}, Mohamed and {Fabinsky}, Beth and {Naes}, Larry and {Tsai}, Chao-Wei},
        title = "{The Wide-field Infrared Survey Explorer (WISE): Mission Description and Initial On-orbit Performance}",
      journal = {\aj},
         year = 2010,
        month = dec,
       volume = {140},
       number = {6},
        pages = {1868-1881},
          doi = {10.1088/0004-6256/140/6/1868},
archivePrefix = {arXiv},
       eprint = {1008.0031},
 primaryClass = {astro-ph.IM},
       adsurl = {https://ui.adsabs.harvard.edu/abs/2010AJ....140.1868W}
}

@ARTICLE{2000AJ....120.1579Y,
       author = {{York}, Donald G. and {Adelman}, J. and {Anderson}, John E., Jr. and {Anderson}, Scott F. and {Annis}, James and {Bahcall}, Neta A. and {Bakken}, J.~A. and {Barkhouser}, Robert and {Bastian}, Steven and {Berman}, Eileen and {Boroski}, William N. and {Bracker}, Steve and {Briegel}, Charlie and {Briggs}, John W. and {Brinkmann}, J. and {Brunner}, Robert and {Burles}, Scott and {Carey}, Larry and {Carr}, Michael A. and {Castander}, Francisco J. and {Chen}, Bing and {Colestock}, Patrick L. and {Connolly}, A.~J. and {Crocker}, J.~H. and {Csabai}, Istv{\'a}n and {Czarapata}, Paul C. and {Davis}, John Eric and {Doi}, Mamoru and {Dombeck}, Tom and {Eisenstein}, Daniel and {Ellman}, Nancy and {Elms}, Brian R. and {Evans}, Michael L. and {Fan}, Xiaohui and {Federwitz}, Glenn R. and {Fiscelli}, Larry and {Friedman}, Scott and {Frieman}, Joshua A. and {Fukugita}, Masataka and {Gillespie}, Bruce and {Gunn}, James E. and {Gurbani}, Vijay K. and {de Haas}, Ernst and {Haldeman}, Merle and {Harris}, Frederick H. and {Hayes}, J. and {Heckman}, Timothy M. and {Hennessy}, G.~S. and {Hindsley}, Robert B. and {Holm}, Scott and {Holmgren}, Donald J. and {Huang}, Chi-hao and {Hull}, Charles and {Husby}, Don and {Ichikawa}, Shin-Ichi and {Ichikawa}, Takashi and {Ivezi{\'c}}, {\v{Z}}eljko and {Kent}, Stephen and {Kim}, Rita S.~J. and {Kinney}, E. and {Klaene}, Mark and {Kleinman}, A.~N. and {Kleinman}, S. and {Knapp}, G.~R. and {Korienek}, John and {Kron}, Richard G. and {Kunszt}, Peter Z. and {Lamb}, D.~Q. and {Lee}, B. and {Leger}, R. French and {Limmongkol}, Siriluk and {Lindenmeyer}, Carl and {Long}, Daniel C. and {Loomis}, Craig and {Loveday}, Jon and {Lucinio}, Rich and {Lupton}, Robert H. and {MacKinnon}, Bryan and {Mannery}, Edward J. and {Mantsch}, P.~M. and {Margon}, Bruce and {McGehee}, Peregrine and {McKay}, Timothy A. and {Meiksin}, Avery and {Merelli}, Aronne and {Monet}, David G. and {Munn}, Jeffrey A. and {Narayanan}, Vijay K. and {Nash}, Thomas and {Neilsen}, Eric and {Neswold}, Rich and {Newberg}, Heidi Jo and {Nichol}, R.~C. and {Nicinski}, Tom and {Nonino}, Mario and {Okada}, Norio and {Okamura}, Sadanori and {Ostriker}, Jeremiah P. and {Owen}, Russell and {Pauls}, A. George and {Peoples}, John and {Peterson}, R.~L. and {Petravick}, Donald and {Pier}, Jeffrey R. and {Pope}, Adrian and {Pordes}, Ruth and {Prosapio}, Angela and {Rechenmacher}, Ron and {Quinn}, Thomas R. and {Richards}, Gordon T. and {Richmond}, Michael W. and {Rivetta}, Claudio H. and {Rockosi}, Constance M. and {Ruthmansdorfer}, Kurt and {Sandford}, Dale and {Schlegel}, David J. and {Schneider}, Donald P. and {Sekiguchi}, Maki and {Sergey}, Gary and {Shimasaku}, Kazuhiro and {Siegmund}, Walter A. and {Smee}, Stephen and {Smith}, J. Allyn and {Snedden}, S. and {Stone}, R. and {Stoughton}, Chris and {Strauss}, Michael A. and {Stubbs}, Christopher and {SubbaRao}, Mark and {Szalay}, Alexander S. and {Szapudi}, Istvan and {Szokoly}, Gyula P. and {Thakar}, Anirudda R. and {Tremonti}, Christy and {Tucker}, Douglas L. and {Uomoto}, Alan and {Vanden Berk}, Dan and {Vogeley}, Michael S. and {Waddell}, Patrick and {Wang}, Shu-i. and {Watanabe}, Masaru and {Weinberg}, David H. and {Yanny}, Brian and {Yasuda}, Naoki and {SDSS Collaboration}},
        title = "{The Sloan Digital Sky Survey: Technical Summary}",
      journal = {\aj},
         year = 2000,
        month = sep,
       volume = {120},
       number = {3},
        pages = {1579-1587},
          doi = {10.1086/301513},
archivePrefix = {arXiv},
       eprint = {astro-ph/0006396},
 primaryClass = {astro-ph},
       adsurl = {https://ui.adsabs.harvard.edu/abs/2000AJ....120.1579Y}
}

\appendix


\clearpage

\section{Notes on miscellaneous claimed literature quasars with $G_\mathrm{RP}>16$ that are mostly not quasars}

Here, we list the results of our inquiries into the physical nature of several objects that were reported as quasars in MILLIQUAS v8 but did not meet our selection criteria. These objects are  fainter than $G_\mathrm{RP}=16$, while brighter objects of this kind are discussed in Section~\ref{sec:sel} and listed in Table~\ref{tab:not_cands}. The following twelve objects discussed in this section are not an exhaustive list, but are those that we explored for a start. As the details show, eleven of these objects have no evidence at present for being type-1 AGN, while one is a likely lensed quasar with a high Gaia parallax and AEN. We did not pursue a similarly detailed review of the complete faint sample: 

\begin{itemize}
\item Q~2211-183, redshift $z=1.18$, SMSS ID 5857733, Gaia \texttt{source\_id} 6825778256931732864 -- the MILLIQUAS position is coincident with a Galactic star of $G_{\rm RP}=16.29$ at $\sim$300~pc distance; there is no second Gaia source within 15" to explain a small coordinate confusion. The object is listed in a compilation of quasars by \citet{HB93} with a reference to M.~Drinkwater's PhD thesis. Private communication with Michael Drinkwater has clarified that there is no position confusion but that the spectral signature was marginal and the object's type uncertain at the time. It is thus established now as a Galactic star. 

\item Q 0123-527, redshift $z=2.32$, SMSS ID 503652057, Gaia \texttt{source\_id} 4915326088698873088 -- the MILLIQUAS position is coincident with a Galactic star of $G_{\rm RP}=16.19$ at $\sim$600~pc distance, which has a $\sim$3 mag brighter neighbour with 7" offset that is a star at $\sim$300~pc distance; this is another entry from the Hewett catalogue going back to an uncertain object in Michael Drinkwater's list that is established now as a Galactic star.

\item PGC~1889865, redshift $z=0.77$ -- a Northern object, whose MILLIQUAS position coincides with a Galactic star of $G_{\rm RP}=16.53$, while the Simbad position for this object identifier is 12" different and coincides with a Gaia source of $G_{\rm RP}=20.1$, which is clearly too faint for the scope of the Gaia-WISE study in this paper.

\item LAMOST J081244.17+245840.8, redshift $z=0.286$ -- the MILLIQUAS position is coincident with a Galactic star of $G_{\rm RP}=16.66$. A fainter Gaia source of $G_{\rm RP}=18.96$ is 4.5" away and has no parallax; while the latter object might be a quasar, it is too faint for the scope of this paper.

\item HE 0541$-$4135, redshift $z=0.219$, 
SMSS ID 18767126 -- the MILLIQUAS position is coincident with a Galactic star of $G_{\rm RP}=16.33$. However, 5" to the East is another source with $G_{\rm RP}=17.08$, which has no parallax and the right colours for a quasar. That other source has a spectrum from the 6dF Galaxy Survey \citep[6dFGS;][]{Jones09}, where it was recorded as a catalogue entry with $z=0.047$. The 6dF spectrum appears visually inconclusive, and the redshift value is not supported by it. The nature of this object is thus unclear at present. 

\item SDSS J075136.64+303942.4, redshift $z=0.143$, $G_{\rm RP}=16.78$ -- inspection with the SDSS DR16 Navigate Tool shows a regular old galaxy spectrum with weak narrow emission lines suggesting a LINER galaxy but not a broad emission-line Seyfert. The weak narrow emission lines argue also against a CLAGN.
    
\item LAMOSTJ042052.94+161248.1, redshift $z=0.291$, $G_{\rm RP}=16.95$; the online resource for LAMOST~DR5~v3 now classifies this object as a star or galaxy, and the spectra show no evidence for a quasar at the quoted redshift.
    
\item LAMOSTJ081844.94+005934.3, redshift $z=1.042$, $G_{\rm RP}=17.57$ -- this object also has a BpRp Excess factor of 5.94 and is thus rejected as a candidate on two grounds. The LAMOST interpretation is based on an apparent weak broad MgII emission line in a noisy spectrum; a newer, higher-quality spectrum from DESI DR1 shows it clearly as a normal galaxy at $z=0.0521$.
    
\item LAMOSTJ015141.38+030353.5, redshift $z=1.527$, $G_{\rm RP}=17.81$ -- this object also has a BpRp Excess factor of 6.56 and is thus rejected as a candidate on two grounds. LAMOST DR5 now rates it as a galaxy at $z=0.103$.

\item Q 2217-4208, redshift $z=2.21$, $G_{\rm RP}=17.94$ -- this object also has a BpRp Excess factor of 5.94 and is thus rejected as a candidate on two grounds. This object was originally listed in \citet{HB93}, who cite M.~Drinkwater's PhD thesis. It may have a similar interpretation as the two objects above, pending spectroscopic confirmation.

\item DESI 39633411337359156, redshift $z=0.983$ -- a Northern object, whose MILLIQUAS position is 1.2" from the nearest Gaia source (\texttt{source\_id} 2159258214943609856), which is a Galactic star of $G_{\rm RP}=16.01$. In the DESI DR1 catalogue, the star is listed as a K star, while the object at the position of the quasar is faint with $r\approx 19$ and thus beyond the scope of this work.

\item DESJ023004.58-070445.9, redshift $z=2.005$, SMSS ID 15384292 -- the MILLIQUAS position is coincident with two Gaia sources separated by $\sim$0.5", the brighter of which has $G_{\rm RP}=16.87$. Both have DESI DR1 spectra that appear virtually identical, with $z=2.007$, suggesting a lensed quasar. The parallax has a significance of 5.2$\sigma$ but comes with a large Gaia AEN of $5.3\pm 427.6$~mas. Large AEN values invalidate the parallax significance and commonly occur in blended sources such as lensed or dual quasars.

\end{itemize}

\vspace{1cm}

\section{Gallery of AllBRICQS Quasar Spectra}\label{app:gallery}
\renewcommand\thefigure{A.\arabic{figure}}    
\setcounter{figure}{0}    

We present the WiFeS spectra for the AllBRICQS objects from Tables~\ref{tab:qsos_zgt01}, \ref{tab:qsos_zlt01} and \ref{tab:qsos_zlt01nl} (120 broad-line and two narrow-line AGN) in Figure~\ref{fig:spec1}, arranged in order of ascending redshift.

\begin{figure*}[p]
    \centering
    \includegraphics[width=\columnwidth]{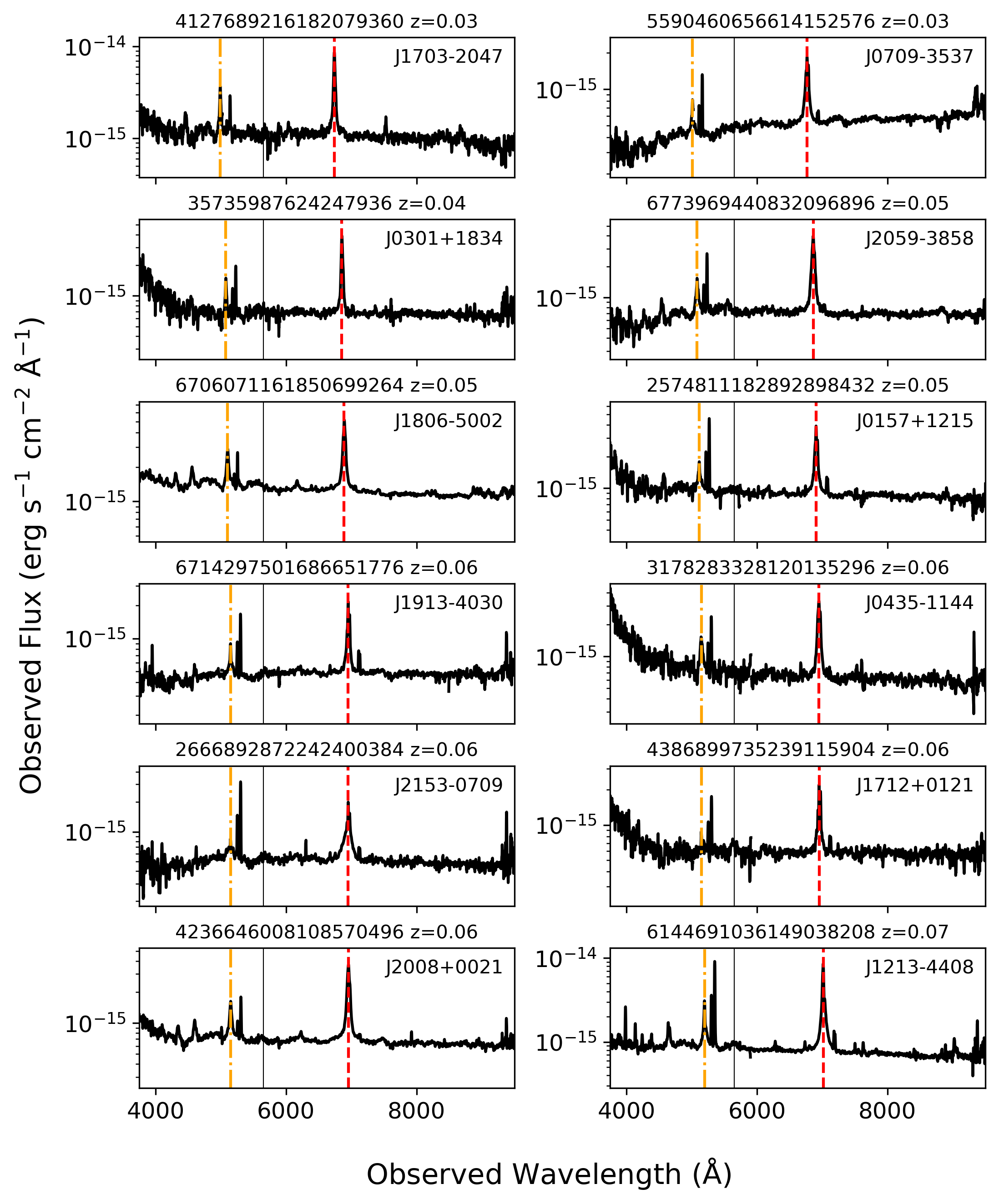} 
    \caption{WiFeS spectra of quasars newly identified by AllBRICQS, ordered by redshift. The y-axis shows the flux on a logarithmic scale, normalised using Gaia DR3 photometry. Vertical lines indicate the positions of various quasar emissions lines: H$\alpha$ (red, dashed), H$\beta$ (orange, dot-dashed), \ion{Mg}{ii} (green, dotted), \ion{C}{iii}] (blue, solid), \ion{C}{iv} (purple, dashed). The thin, black vertical line at 5650~\AA\ indicates where the blue and red arms of the spectrograph are spliced together.}
    \label{fig:spec1}
\end{figure*}
\clearpage 
\begin{figure*}[p]
    \centering \ContinuedFloat
    \includegraphics[width=\columnwidth]{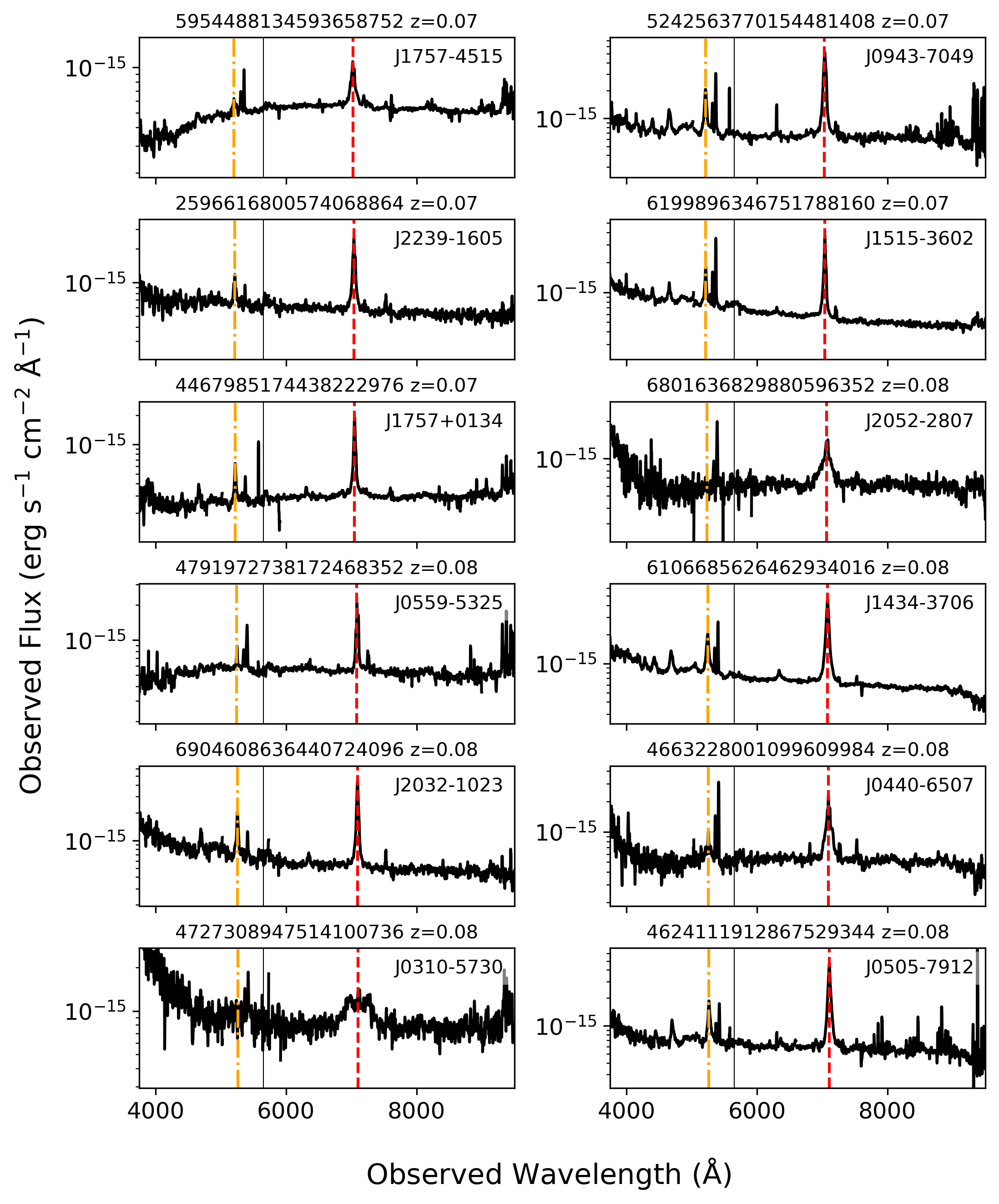} 
    \caption{ (continued). }
\end{figure*}
\clearpage 
\begin{figure*}[p]
    \centering \ContinuedFloat
    \includegraphics[width=\columnwidth]{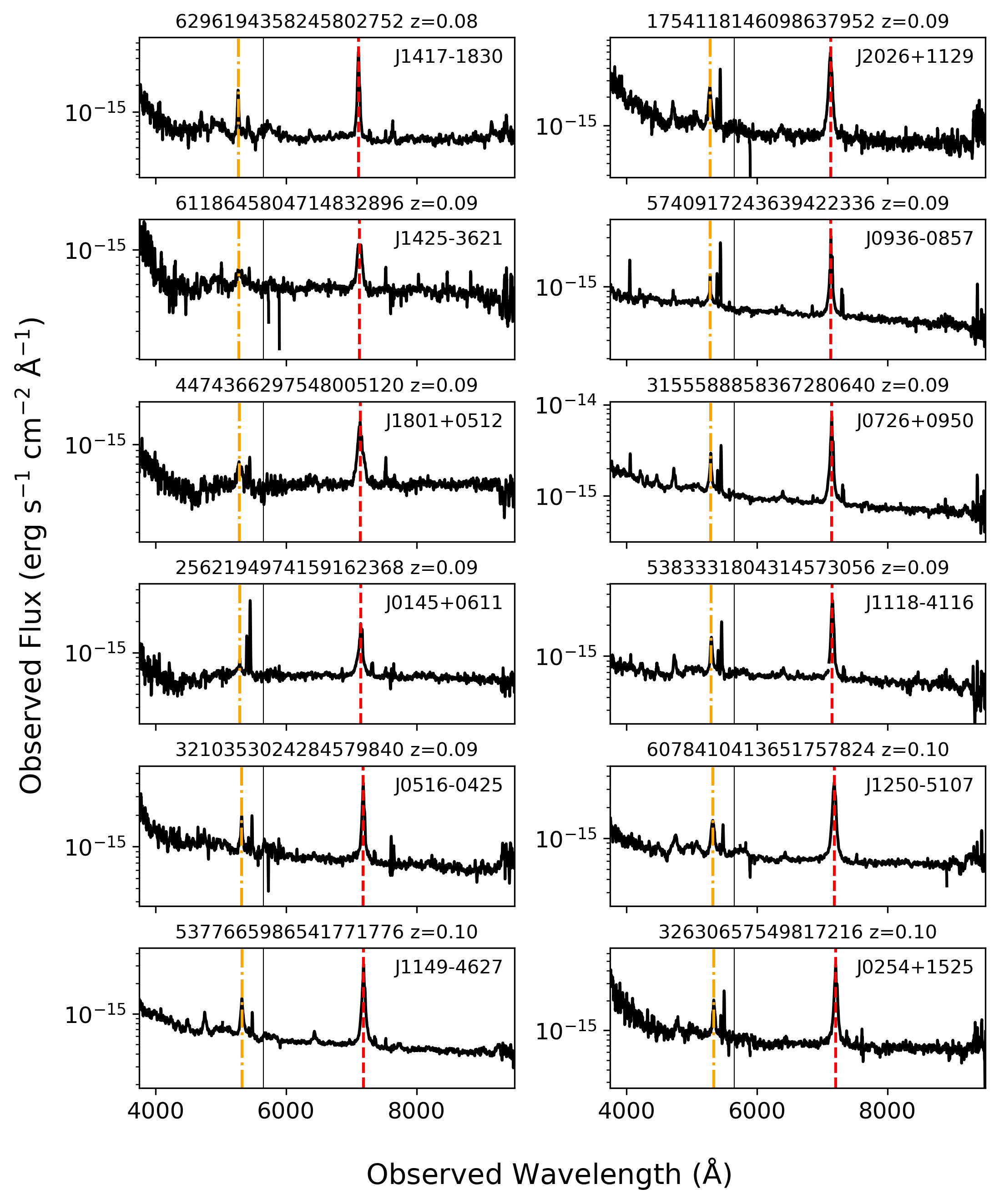} 
    \caption{ (continued). }
\end{figure*}
\clearpage 
\begin{figure*}[p]
    \centering \ContinuedFloat
    \includegraphics[width=\columnwidth]{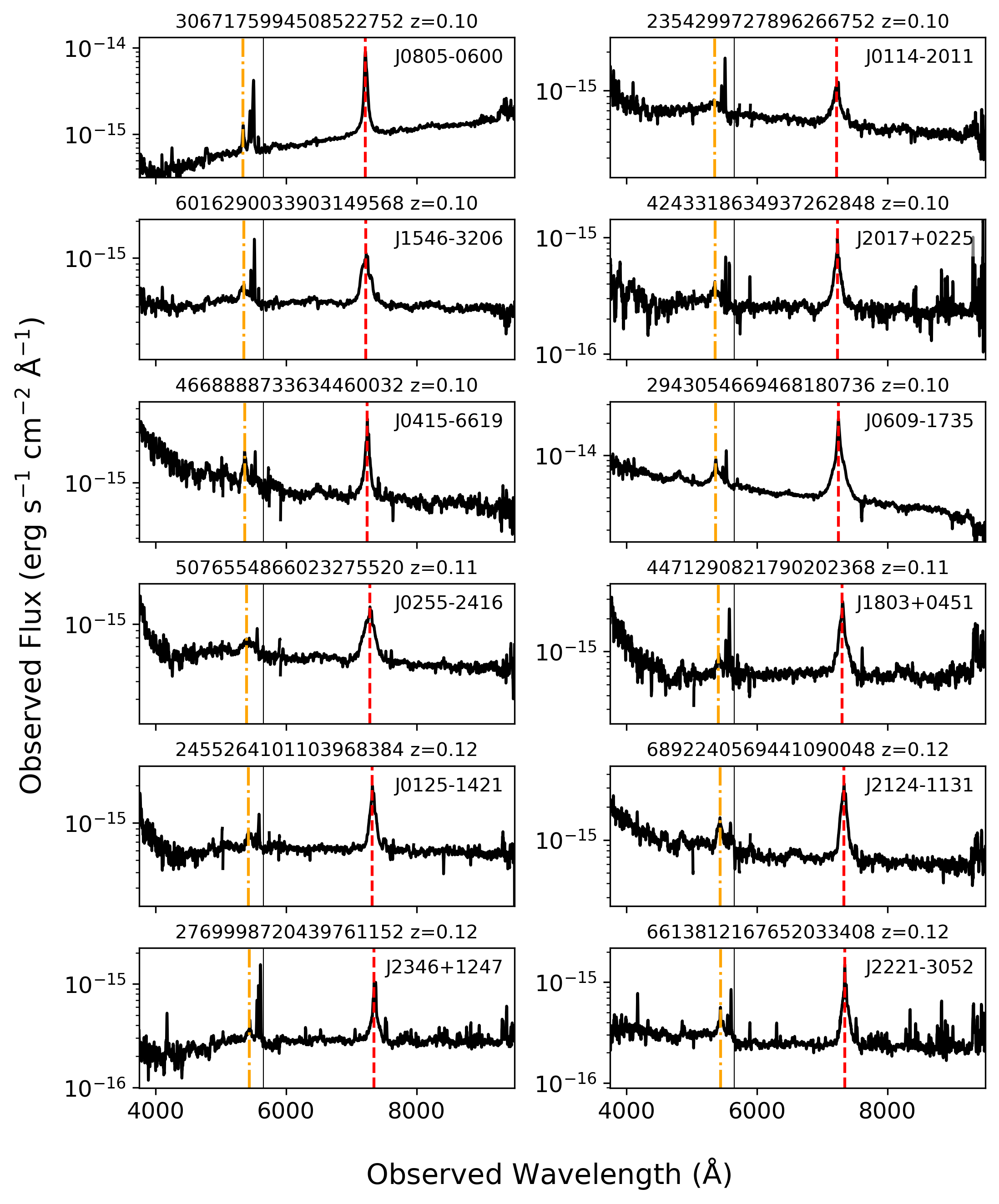} 
    \caption{ (continued). }
\end{figure*}
\clearpage 
\begin{figure*}[p]
    \centering \ContinuedFloat
    \includegraphics[width=\columnwidth]{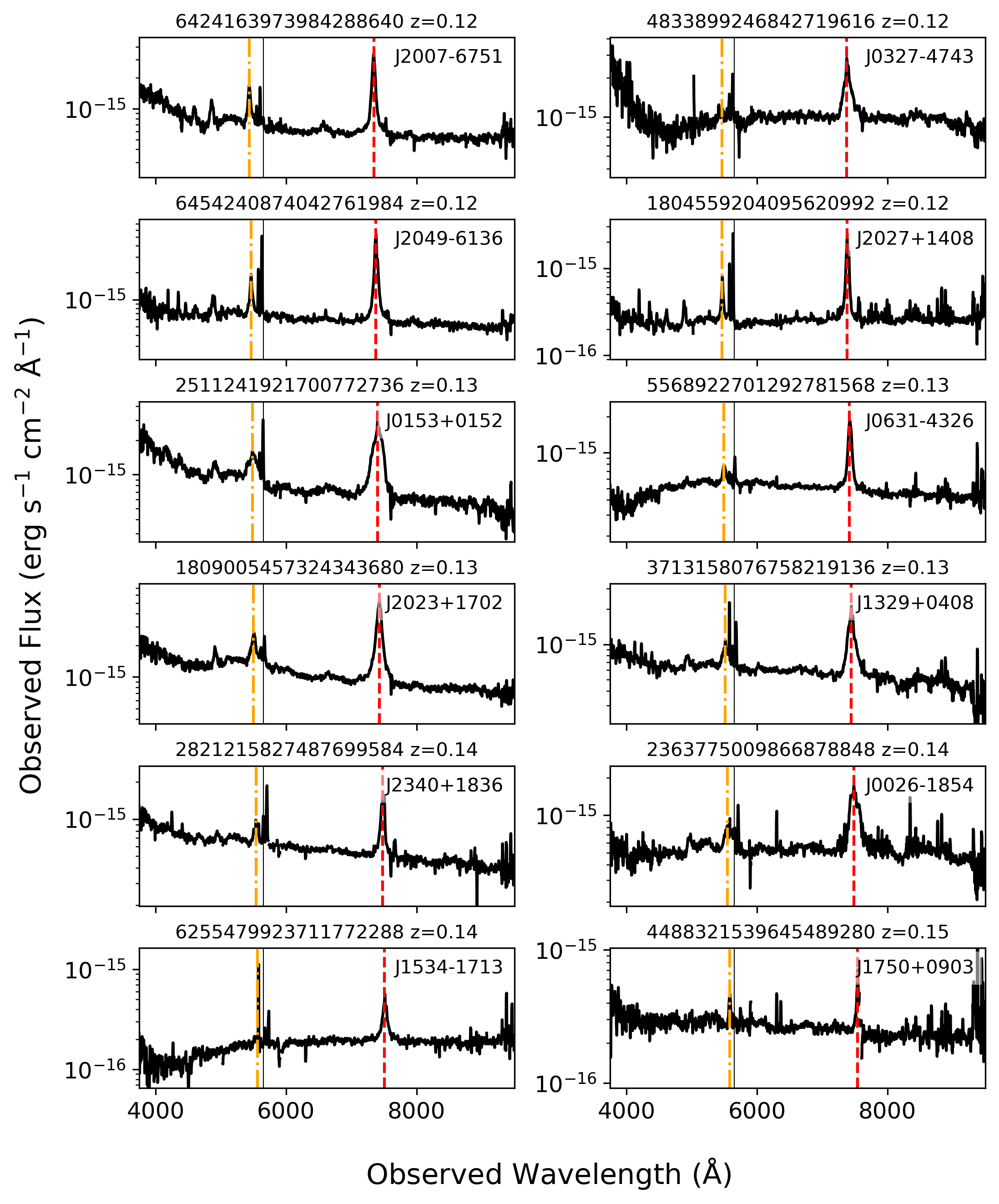} 
    \caption{ (continued). }
\end{figure*}
\clearpage 
\begin{figure*}[p]
    \centering \ContinuedFloat
    \includegraphics[width=\columnwidth]{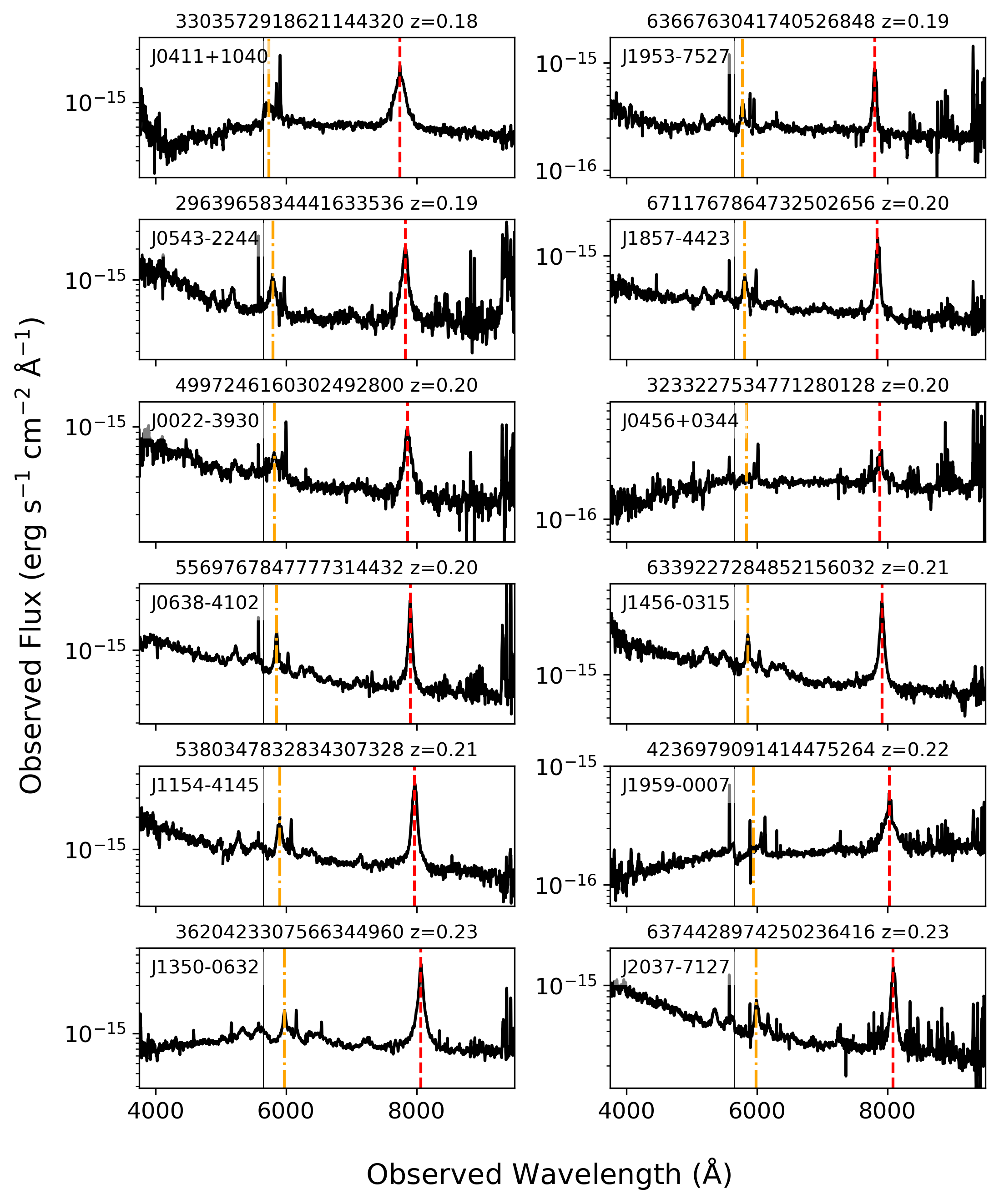} 
    \caption{ (continued). }
\end{figure*}
\clearpage 
\begin{figure*}[p]
    \centering \ContinuedFloat
    \includegraphics[width=\columnwidth]{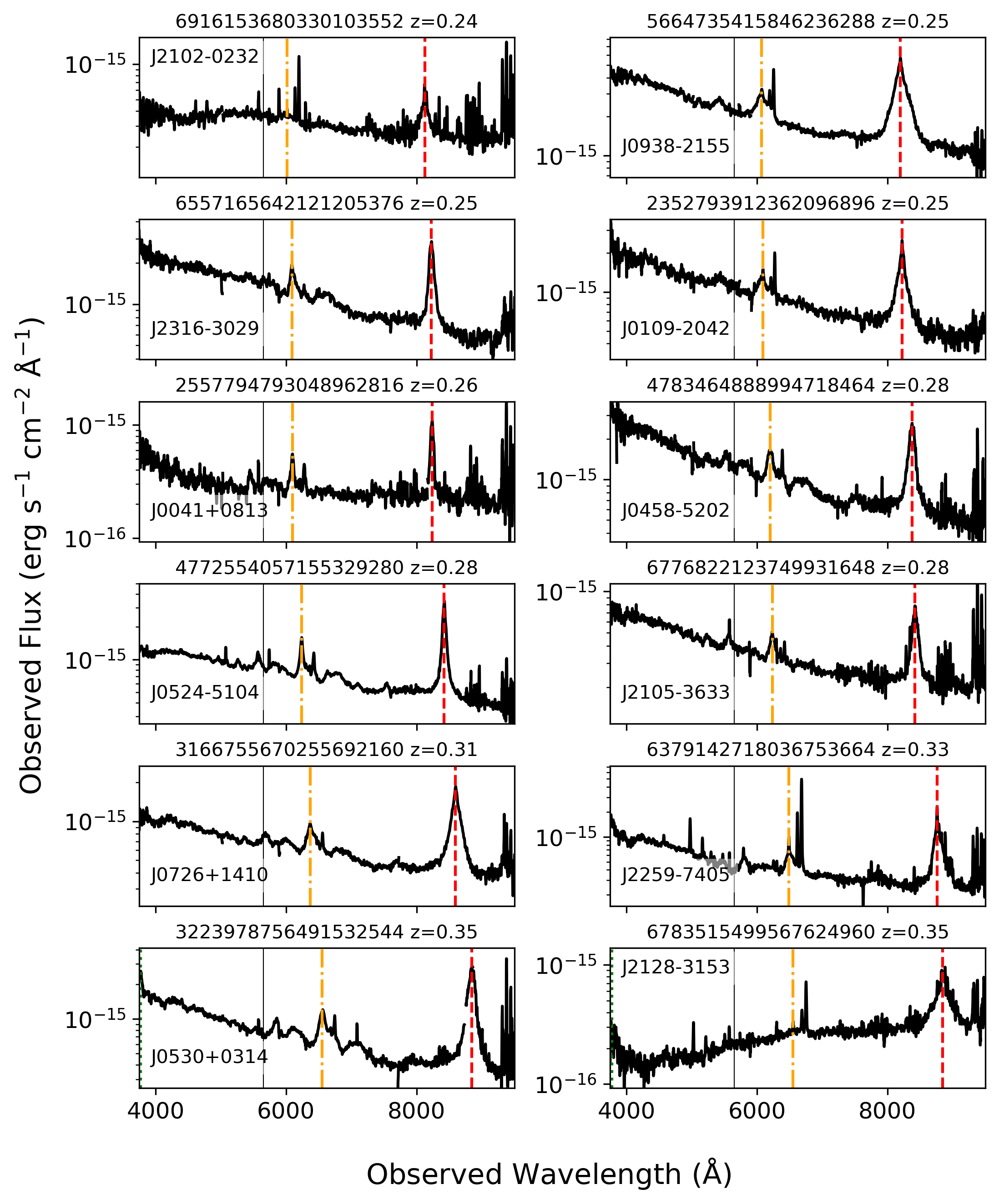} 
    \caption{ (continued). }
\end{figure*}
\clearpage 
\begin{figure*}[p]
    \centering \ContinuedFloat
    \includegraphics[width=\columnwidth]{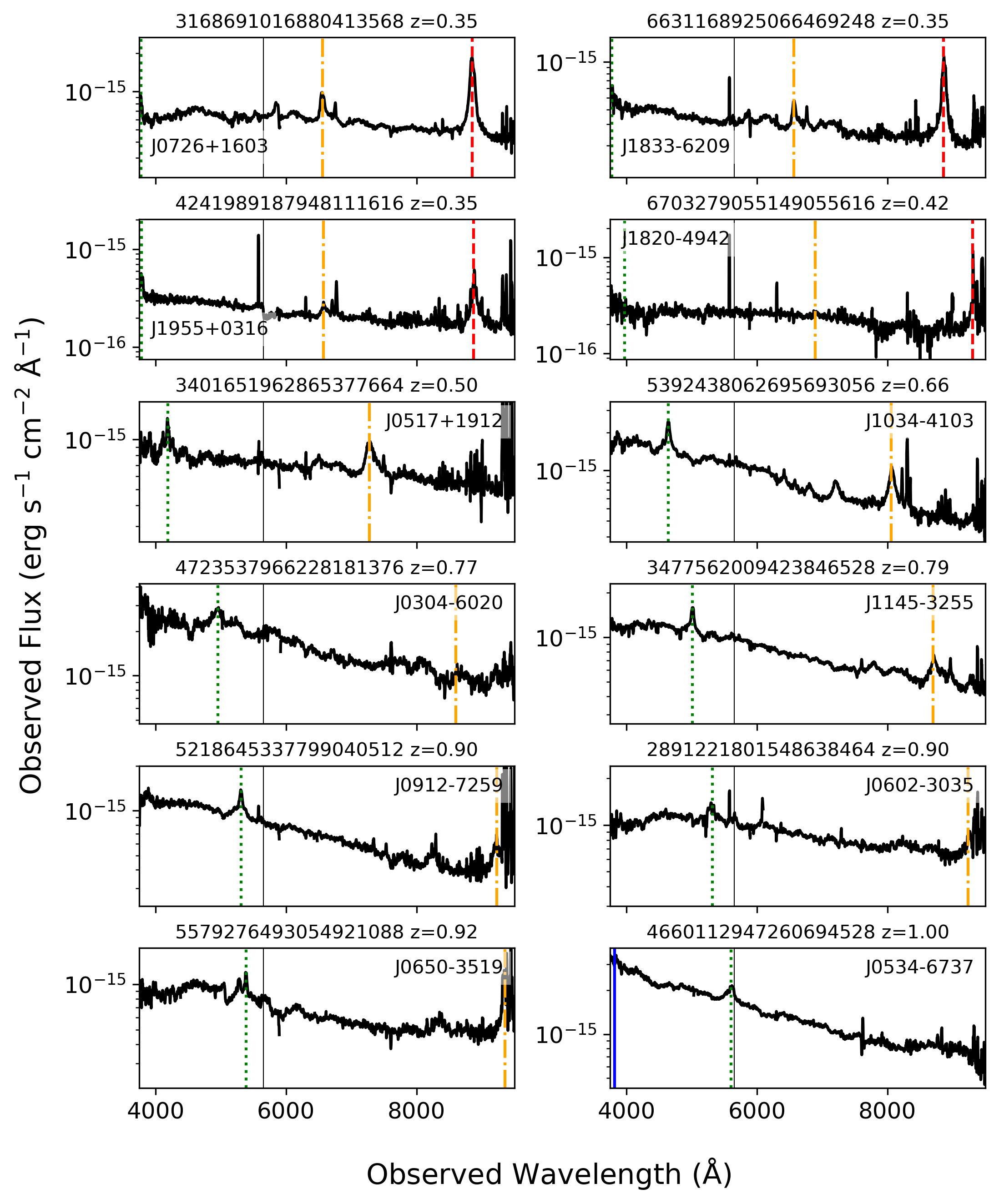} 
    \caption{ (continued). }
\end{figure*}
\clearpage 
\begin{figure*}[p]
    \centering \ContinuedFloat
    \includegraphics[width=\columnwidth]{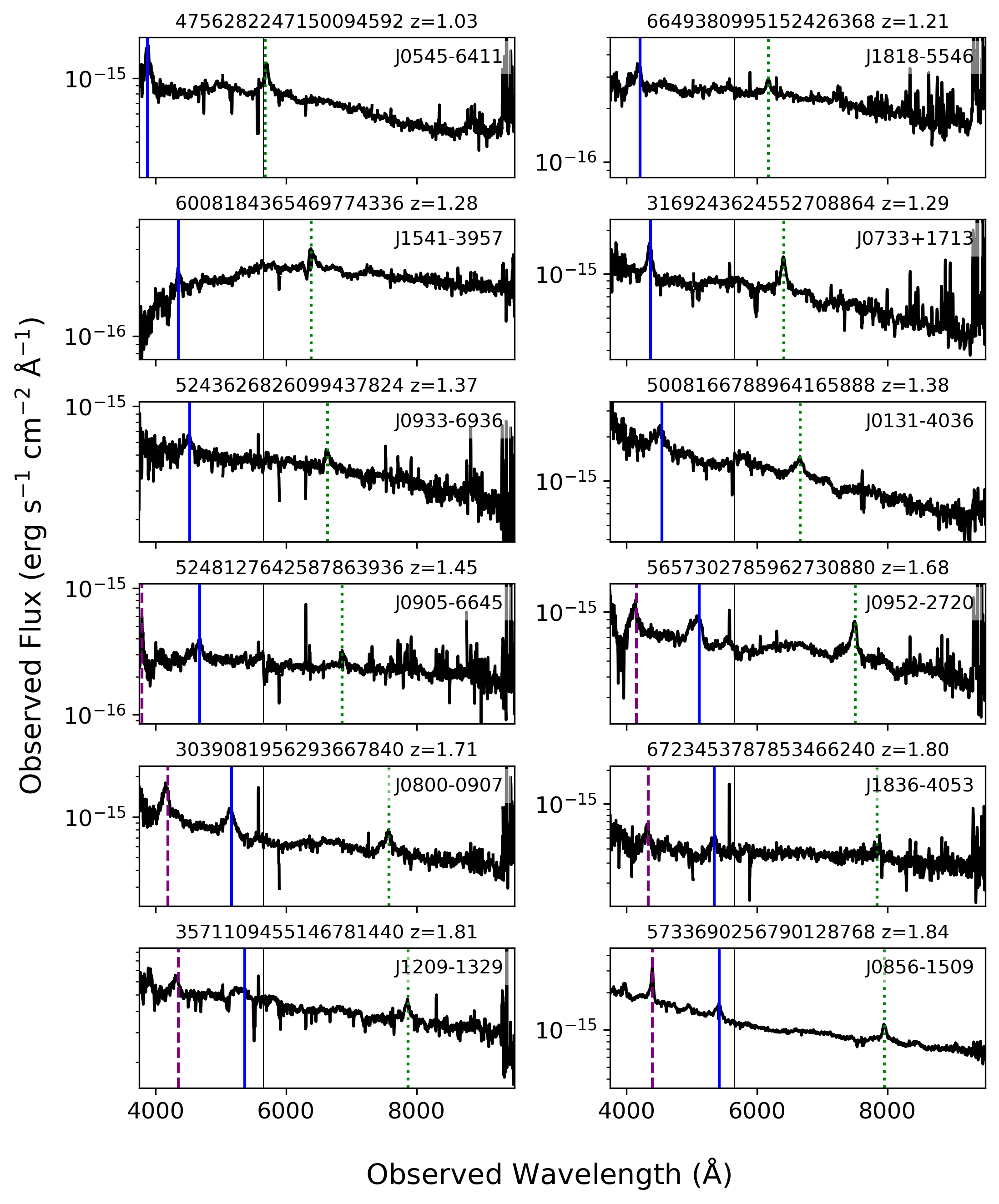} 
    \caption{ (continued). }
\end{figure*}
\clearpage 
\begin{figure*}[p]
    \centering \ContinuedFloat
    \includegraphics[width=\columnwidth]{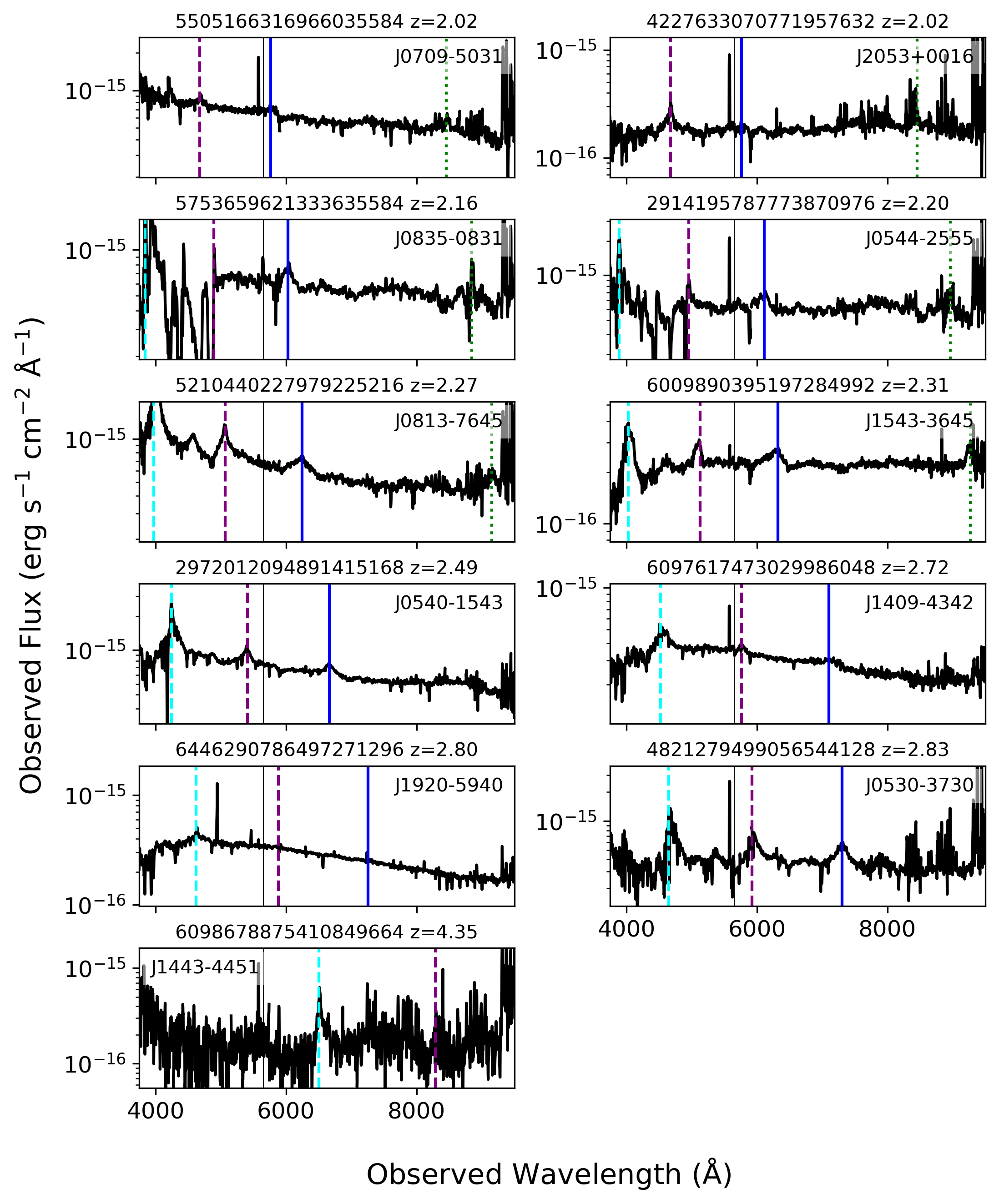} 
    \caption{ (last page). }
\end{figure*}

\begin{figure}[p]
    \centering
    \includegraphics[width=\columnwidth]{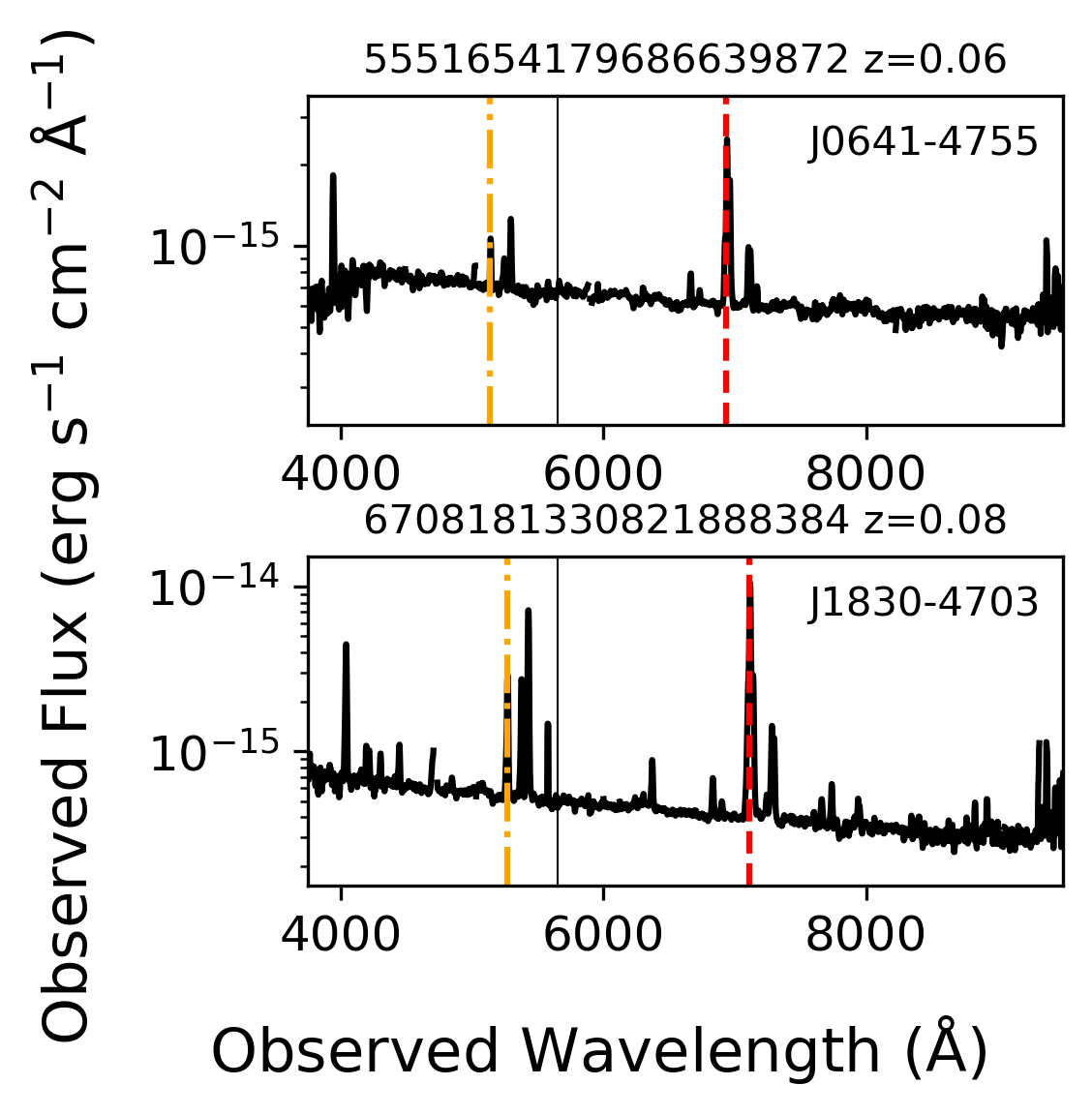} 
    \caption{WiFeS spectra of AllBRICQS quasar candidates found to be NLAGN (format as in Figure~\ref{fig:spec1}. These two candidates had Gaia-WISE colours from a decade ago that are more typical for broad-line AGN, and their Gaia XP spectra show strong emission signatures. They could thus also be turn-off CLAGN without a high-resolution on-stage spectrum. }
    \label{fig:spec2}
\end{figure}

\end{document}